\documentclass[prx, reprint, notitlepage,
nofootinbib, 
superscriptaddress,
floatfix]{revtex4-1}
\usepackage{bbold}
\usepackage{gpkmac}

\newcommand{\TLI}{\raisebox{-2pt}{\includegraphics[height=10pt]{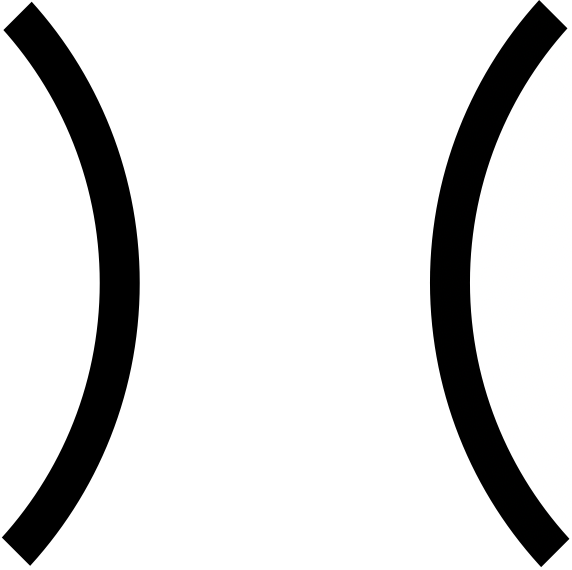}}}
\newcommand{\TLE}{\raisebox{-2pt}{\includegraphics[height=10pt]{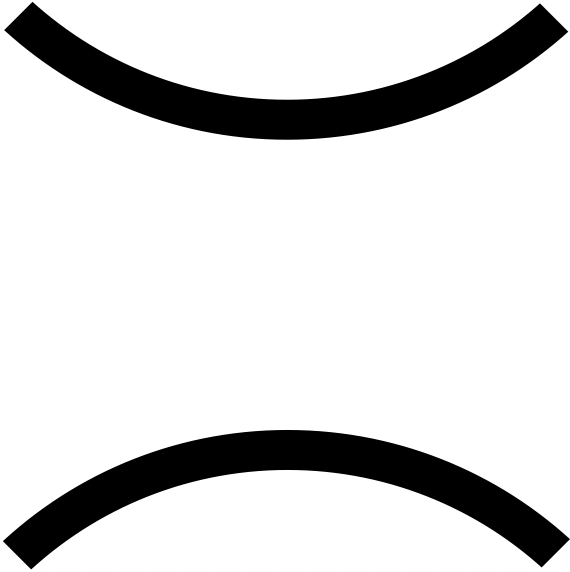}}}

\begin{document}


\title{Decodable hybrid dynamics of open quantum systems with $\mb{Z}_2$ symmetry}
\author{Yaodong Li}
\affiliation{Department of Physics, 
University of California, Santa Barbara, CA 93106}
\affiliation{Department of Physics, Stanford University, Stanford, CA 94305}
\author{Matthew P. A. Fisher}
\affiliation{Department of Physics, University of California, Santa Barbara, CA 93106}
\date{November 14, 2023}

\begin{abstract}

We explore a class of ``open'' quantum circuit {models} with local decoherence (``noise") and local projective measurements, each respecting a global $\mb{Z}_2$ symmetry.
The model supports a spin glass phase where the $\mb{Z}_2$ symmetry is spontaneously broken (not possible in an equilibrium 1d system), a paramagnetic phase characterized by a divergent susceptibility, and an intermediate ``trivial'' phase.
All three phases are also
stable to $\mb{Z}_2$-symmetric local unitary gates, and the dynamical phase transitions between the phases are in the percolation universality class.
The open circuit dynamics can be purified by explicitly introducing a bath with its own ``scrambling" dynamics, as in [Bao, Choi, Altman, \href{https://arxiv.org/abs/2102.09164}{arXiv:2102.09164}], which does not change any of the universal physics.
Within the spin glass phase the circuit dynamics can be interpreted as a quantum repetition code,
with each stabilizer of the code measured stochastically at a finite rate, and the decoherences as effective  bit-flip errors.
{Motivated by the geometry of the spin glass phase,}
we devise a novel
decoding algorithm for recovering an arbitrary initial qubit state in the code space, assuming knowledge of the history of the measurement outcomes, and the ability of performing local Pauli measurements and gates on the final state. For a circuit with $L^d$ qubits running for time $T$, the time needed to
{execute the decoder} scales as $O(L^d \cdot T)$ (with dimensionality $d$).
With this decoder in hand,
we find that the information of the initial encoded qubit state can be retained (and then recovered) for a time logarithmic in $L$ for a 1d circuit, and for a time at least linear in $L$ in 2d below a finite error threshold.  
For both the repetition and toric codes, we compare and contrast our decoding algorithm with earlier algorithms that map
the error model to {the random bond Ising model}.


\end{abstract}

\maketitle


{\hypersetup{linktocpage} \tableofcontents}


\section{Introduction}

Recent theoretical progress in the dynamics of many-body quantum systems has become increasingly laden with ideas from quantum information theory.
With notions such as the entanglement entropy and the out-of-time-ordered correlator, the process of {information spreading} in a spatially extended system can be succinctly quantified.
Two notable non-equilibrium phases of \emph{closed} system dynamics have emerged, namely a thermal phase~\cite{Deutsch1991, Srednicki1994, Calabrese_2005, Calabrese_2007, rigol2007thermalization, Kim2013, suh2013tsunami, greiner2018quantumthermalization, Mezei2017, Ho_2017, nahum2017KPZ, nahum2018coarsegrained, nahum2018operator, keyserlingk2018operator, prosen2018chaosRMT} where the information can spread across the entire system, and a many-body localized (MBL) phase (in the presence of strong quenched disorder)~\cite{BaskoAleinerAltshuler2006, OganesyanHuse2007, Prosen2008, PalHuse2010, Bardarson2012, Bauer2013, Serbyn2013, HuseNandkishoreOganesyan2014, Nandkishore2015, abanin2018rmp} where information in the initial state is retained locally at long times.
The MBL transition~\cite{Luitz2015, Kjall2014, Vosk2015, chandran2015stabilizerMBL, Potter2015, Serbyn2015, Khemani2017, bloch2015mbl_optical, Dumitrescu2017, deroeck2017avalanche, Zhang2016, goremykina2018MBLRG, Dumitrescu2018KTscaling} between the two is an ``entanglement phase transition'', marked by sharp changes in the temporal growth and spatial scaling of the entanglement entropy.

A different type of entanglement phase transition occurs in ``hybrid circuits''~\cite{aharonov2000, halpern2017qiqcog, nandkishore2018hybrid, nahum2018hybrid, li2018hybrid, li2019hybrid, choi2019qec, szyniszewski2019measurement, gullans2019purification, andreas2019hybrid, choi2019spin, gullans2019scalable, Tang2019, chamon2020nonuniversal, gorshkov2020classicalmodel, dalmonte2020twoplusonedim, nahum2020alltoall, vasseur2021chargesharpening}, namely a random unitary circuit~\cite{nahum2017KPZ, nahum2018operator, keyserlingk2018operator, chan2017chaos, chan2018spectral, rakovszky2017conservation, zhou2018emergent, khemani2017operator_conservation, friedman2019spectral} interspersed with {monitored} local measurements.
Similar phenomena are also considered in a broader context~\cite{Hayden2016, cao2018monitoring, vasseur2018rtn, harrow2020efficient, chenxiao2020nonunitary, diehl2020trajectory, jian2020fermionRTN, sagar2020volume, ashida2020continuous, vasseur2020mft, danshita2020cold, pal_lunt_2020_mbl_hybrid, szyniszewski2020stroboscopic, iaconis2020automata, schiro2020subradiance, gopalakrishnan2020nonhermitianQM, liu2020nonunitarySYK, diehl2021engineered,  diehl2021sineGordon, turkeshi2021zeroclicks, tang2021freefermion2+1d, buchhold2021freebosonCFT, melko2021trappedion}.
Due to the competition between unitary gates that increase entanglement and measurements that tend to ``disentangle'',
the steady state can be in a volume law phase of entanglement entropy at small measurement rates, or an area law phase with more frequent measurements, separated by a continuous phase transition~\cite{nahum2018hybrid, li2018hybrid, li2019hybrid, vasseur2018rtn, andreas2019hybrid, choi2019spin, li2020cft, huse2019tripartite, zabalo2021ceff}.
Using the language of quantum error correction~\cite{shor1995scheme, steane1996qec, BDSW9604mixedstate, knill_laflamme_1997}, the dynamics in the volume law phase can be viewed as a ``robust encoding'' circuit~\cite{choi2019qec, gullans2019purification} of a 
{quantum memory}, whose information hiding properties prevent infrequent measurements from rapidly collapsing the wavefunction and suppressing the volume law entropy to an area law~\cite{fan2020selforganized, li2020capillary, fidkowski2020forget, li2021dpre}.
The entanglement transition is then interpreted as a transition in the \emph{code rate} or \emph{channel capacity}, which is an information theoretic upper bound of the residual information.

A crucial aspect of the measurement induced transition is that it only occurs along quantum trajectories labelled by the measurement results~\cite{cao2018monitoring, nahum2018hybrid, li2018hybrid}, but not in the (Lindblad) evolution of the mixed state density matrix, suitable for a dynamics where the ``measurements" are not monitored.
As such, an experimental observation of the transition {will presumably need to} make use of the information of the measurement results.
One possible approach is to post-select on the same measurement results, so that one gets multiple copies of the same state; and from these copies, 
\emph{nonlinear} functions of the state -- such as the entanglement entropy or ``squared correlators'' (see for example Eq.~\eqref{eq:chi_SG_def} below) -- can be measured, in principle.
However, the probability that the same measurement history occurs more than once is exponentially surpressed in the system size, making this approach impractical.
A second approach is to ``decode'' the circuit~\cite{li2019hybrid, gullans2019scalable}, namely replicating the state by applying a 
``feedback unitary''
based on the measurement history, so that the resultant state is the same for different measurement histories.
This is demonstrated in a recent  simulation of a hybrid stabilizer circuit on a trapped-ion quantum computer~\cite{monroe2021TrappedIonCliffordTransition}, where an optimal decoding algorithm (saturating the channel capacity) exists.

Another important class of monitored dynamics are ``measurement-only circuits''~\cite{nahum2019majorana, ippoliti2020measurementonly, barkeshli2020symmetric, sang2020protected, barkeshli2020topological, diehl2020trajectory, buechler2020projectiveTFIM, Hafezi2020competing_monitoring}.
These circuits do not have any built-in unitary gates, and the nontrivial dynamics is generated instead by competing local measurements drawn from a finite set.
In {many} such models, one finds ``measurement-protected phases''~\cite{sang2020protected} with {area law} 
entanglement entropies favored by different types of measurements,
and transitions between {these} area law phases {can also be} transitions in the channel capacity from finite to zero.
Because the measurements are prevalent, the measurement-only circuits closely resemble the conventional ``decoding'' dynamics of a stabilizer code~\cite{gottesman1996hamming, gottesman1997thesis, DKLP2001topologicalQmemory, fowler2012surfacecode} -- with competing check operator measurements and interspersing errors -- rather than an encoding circuit.
Namely, we encode one (or a few) logical qubit(s) of quantum information in
the initial state, and view a subset of mutually commuting measurement operators as ``check operators'' defining a certain code space, while treating all other (incompatible) measurements as ``errors''~\cite{buechler2020projectiveTFIM}.
By recording the measurement results of the check operators, one collects information about the errors, and applies an appropriate decoding unitary based on this \emph{classical} information,  
in order to recover the  \emph{quantum} information encoded in the initial state, in spite of the errors.
(See Sec.~\ref{sec:dynamical_rep_code} for a clearer description of this decoding problem.)
Here, the measurement outcomes enter explicitly as input of the decoder. 
The probability of successful decoding requires a finite channel capacity of the circuit, but deducing a specific decoding protocol is complicated by the presence of errors, details of the error model
{and the possible intrinsic circuit compexity.}

In this work, we study the said decoding problem in the simplest measurement-only circuit, namely the one with $ZZ$ measurements on neighboring qubits and $X$ measurements on single qubits~\cite{nahum2019majorana, sang2020protected, buechler2020projectiveTFIM} (see also Refs.~\cite{ippoliti2020measurementonly, barkeshli2020symmetric, sang2020negativity}).
We view the $ZZ$ operators as check operators defining a quantum repetition code, and the $X$ measurements (anticommuting with the checks) as errors.
Motivated by Bao, Choi, and Altman~\cite{bao2021enriched}, we also include another type of error, namely single-qubit {dephasing} channels in the $X$ direction, making the circuit an \emph{open} system subject to decoherence and generally driving the circuit into a mixed state.
The two types of $X$ errors both become probabilistic bit-flips under check operator measurements, whereas phase-flip errors never occur due to the global $\mb{Z}_2$ symmetry that we impose. 

We define our ``baseline circuit'' in Sec.~\ref{sec:simple_z2_circuit}, and outline in Sec.~\ref{sec:mapping_to_perc} a mapping of the circuit dynamics to bond percolation, which allows us to solve for the phase diagram and for critical properties of the dynamical phase transitions.
(The details are left for Appendix~\ref{app:details_mapping_percolation}.)
Notably, besides the ``paramagnetic/non-percolating'' and the ``spin glass/percolating'' phases~\cite{nahum2019majorana, sang2020protected, buechler2020projectiveTFIM}, a third ``trivial'' phase with neither order is enabled by decoherence, and is smoothly connected to the infinite temperature fixed point~\cite{bao2021enriched}.
{Furthermore, we follow Ref.~\cite{bao2021enriched} and introduce a ``bath'', whose coupling with the ``system'' replaces the decoherence (Sec.~\ref{sec:intro_bath}).
When the bath dynamics is thermalizing, we find that the system-bath coupling does act like decoherence, and the universal physics is identical with the baseline circuit. }

In Sec.~\ref{sec:dynamical_rep_code}, we focus on the percolating phase~\cite{sang2020protected} of the baseline circuit, and devise a {polynomial time} decoding algorithm (or simply a ``decoder'') for the repetition code based on ``error-avoiding spanning paths'' in the percolating lattice.
The performance of the decoder defines a ``decoding phase'', which coincides with the percolating phase when the locations of the $X$ errors are known to the decoder.
With unlocated $X$ errors, the decoder finds a zero error threshold for the (1+1)d repetition code, and a finite error threshold for the (2+1)d repetition code. 
In the latter case, the decoding phase is strictly within the percolating phase, and the ``decoding transition'' is in a distinct universality class than three-dimensional critical percolation.
{In Sec.~\ref{sec:faulty_meas}, we confirm the robustness of the decoder against faulty measurements.}
In Sec.~\ref{sec:MWPM_numerics}, we compare our results with previous work~\cite{DKLP2001topologicalQmemory, preskill2002MWPMnumerics},  in which the decoding problem of the repetition code {in one and two dimensions} is
{mapped to a Minimal Weight Perfect Matching (MWPM) problem of Ising defects in random bond Ising models (RBIM) in two and three dimensions, respectively.
Notably, for the three dimensional RBIM, the matching problem is NP-hard, whereas our decoder runs in polynomial time.}

In Sec.~\ref{sec:outlook}, we discuss the general relation between decoding and phases of entanglement in hybrid circuits, and possible future directions.

In Appendix~\ref{app:details_mapping_percolation}, we describe in detail the mapping of the baseline circuit to bond percolation, and derive the critical exponents, confirmed by numerical results.

In Appendix~\ref{app:decoding_alg_proof}, we prove the correctness of the decoder using error-avoiding spanning paths.
The formulation we adopt here slightly generalizes the stabilizer formalism (i.e. treating Clifford circuits with an superposition of stabilizer initial states), and may be of independent interest.
In Appendix~\ref{app:dynamical_toric_code}, we drop the $\mb{Z}_2$ symmetry and consider the toric code in (2+1)d, where both bit-flip and phase-flip errors are allowed. 
For the toric code, the decoder can be generalized, and relies on the presence of two dimensional ``error-avoiding spanning membranes'' in the percolating phase of check operator measurements.
However, with unlocated 
errors, 
{implementing our decoder by summing over a subset of spanning membranes} gives a zero error threshold, 
whereas the MWPM decoder (using a mapping to a (2+1)d statistical mechanics model, generalizing the RBIM) {does give a finite error threshold~\cite{DKLP2001topologicalQmemory, preskill2002MWPMnumerics}.}

In Appendix~\ref{sec:generic_phases}, we move away from the decodable baseline circuit, and study the generic phase diagram when various $\mb{Z}_2$ symmetric perturbations are included.
In particular, 
we consider cooling the ``bath'' (in Sec.~\ref{sec:intro_bath}) 
by making local measurements, and observe that the bath-system coupling ceases to act as decoherence.
In this case, the decoherence-induced trivial phase is replaced by a ``critically entangled phase'' exhibiting (super-)logarithmic scaling of entanglement entropy, similar to the phase found by Sang and Hsieh~\cite{sang2020protected}.












\section{Simplest ``baseline'' open quantum circuit with $\mb{Z}_2$ symmetry in (1+1)-dimensions
\label{sec:simple_z2_circuit}
}

\begin{figure}[b]
    \centering
    \includegraphics[width=.45\textwidth]{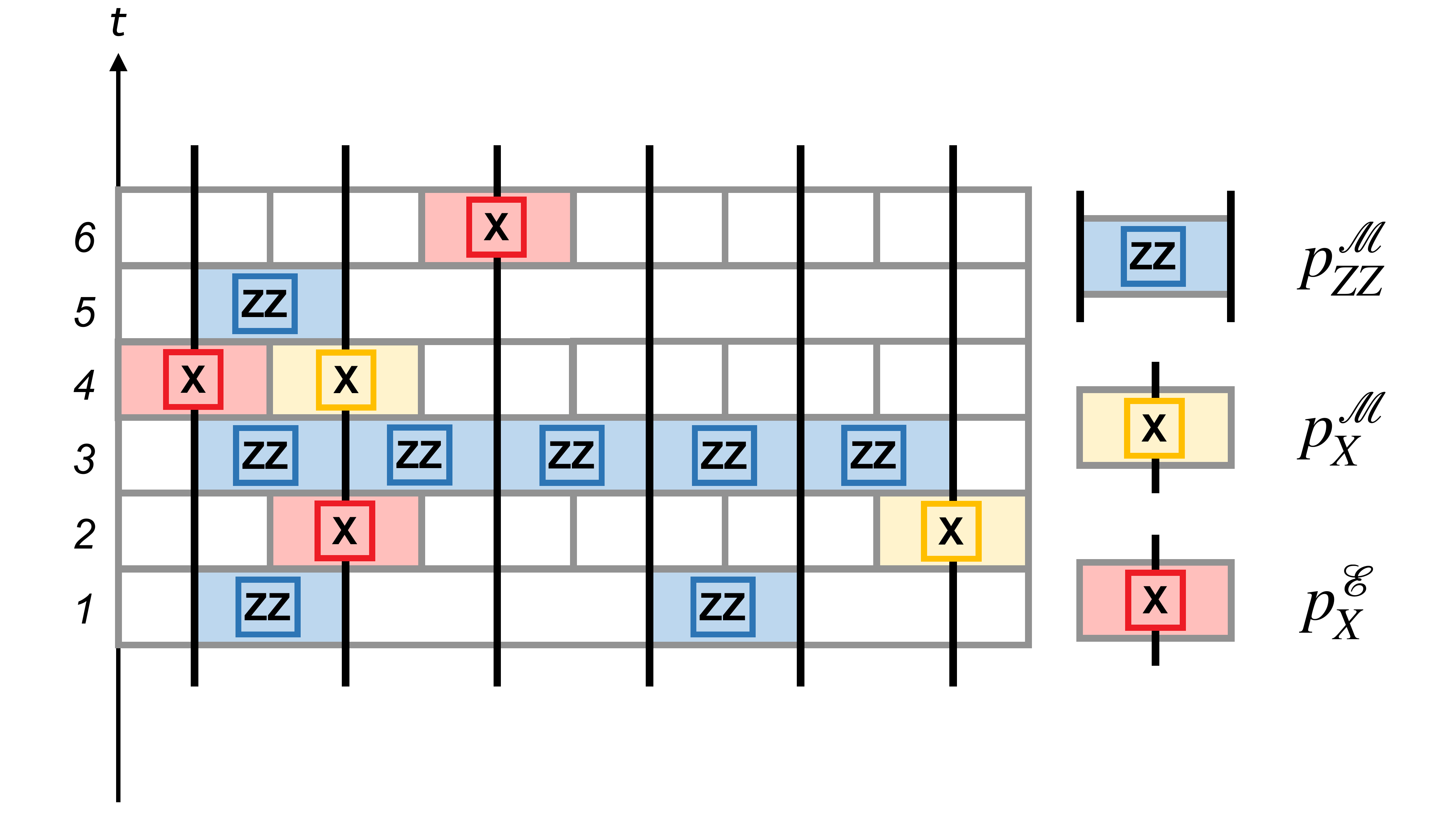}
\caption{
The ``baseline'' open $\mb{Z}_2$ circuit.
There are three types of gates, namely $ZZ$ measurements at odd time steps, and $X$ measurements and $X$ decoherence at even time steps, each occuring with a finite probability.
This may be
obtained from the measurement-only circuit in Ref.~\cite{nahum2019majorana, buechler2020projectiveTFIM, sang2020protected} by randomly swapping out a fraction $q$ of all single-qubit $X$ measurements for  single-qubit $X$ dephasing channels.
}
    \label{fig:z2_circuit}
\end{figure}

We start with the simplest open $\mb{Z}_2$ circuit model [Fig.~\ref{fig:z2_circuit}], which contains three types of gates.
This model will subsequently be referred to as the ``baseline circuit model''.
The dynamics at the $t$-th time step alternates with the parity of $t$.

If $t$ is odd, each of the nearest neighbor $Z_j Z_{j+1}$ operators ($Z_1 Z_2$, $Z_2 Z_3$, $Z_3 Z_4$, $\ldots$) is measured independently with probability $p_{ZZ}^{\mathcal{M}}$.
The measurement outcome is random and follows Born's rule, and is thereafter recorded on a classical memory.
The state evolution is thus stochastic and depends on the measurement outcome, where
\begin{align}
\label{eq:state_evo_meas}
    \rho \to \mathcal{M}_{\pm Z_j Z_{j+1}}
    (\rho) = \frac{P_{\pm Z_j Z_{j+1}} \cdot \rho \cdot P_{\pm Z_j Z_{j+1}}}{\mathrm{Tr} [P_{\pm Z_j Z_{j+1}} \cdot \rho]}
\end{align}
with Born probabilities $\mathrm{Tr} [P_{\pm Z_j Z_{j+1}} \cdot \rho ]$, respectively.
Here $P_{\pm Z_j Z_{j+1}} \coloneqq (1 \pm Z_j Z_{j+1})/2$ are the projection operators, correpsonding to the measurement outcome of $Z_j Z_{j+1}$ being $\pm 1$, respectively.

If $t$ is even, for each qubit $j$, we have either
\begin{enumerate}
\item 
With probability $p_X^{\mathcal{M}} = (1-q)(1-p_{ZZ}^{\mathcal{M}})$, a projective measurement of $X_j$.
The state evolution $\mathcal{M}_{\pm X_j}$ is similar to Eq.~\eqref{eq:state_evo_meas},
\begin{align}
    \rho \to \mathcal{M}_{\pm X_j}
    (\rho) = \frac{P_{\pm X_j} \cdot \rho \cdot P_{\pm X_j}}{\mathrm{Tr} [P_{\pm X_j} \cdot \rho]},
\end{align}
but with a different set of projection operators, namely $P_{\pm X_j} = (1 \pm X_j)/2$.

\item 
Or, with probability $p_X^{\mathcal{E}} = q(1-p_{ZZ}^{\mathcal{M}})$, a ``decoherence gate'' -- specifically a dephasing channel in the $X$ direction of qubit $j$,
\begin{align}
    \label{eq:X_dep_channel}
    \rho \to \mathcal{E}_{X_j}
    (\rho) = \frac{1}{2} (\rho + X_j \rho X_j).
\end{align}
\end{enumerate}

It is important that the projection operators of the $ZZ$ and $X$ measurements, as well as the Kraus operators of the dephasing channel in Eq.~\eqref{eq:X_dep_channel}, all commute with the global $\mb{Z}_2$ symmetry of the circuit, $\mathbf{X} = \prod_{j=1, \ldots, L} X_j$.
The circuit dynamics is therefore ``strongly symmetric''~\cite{prosen2012strongsymmetry, gorshkov2020symmetrybreaking}.

These parameters {have been chosen to} satisfy $p_{ZZ}^{\mathcal{M}} + p_{X}^{\mathcal{M}} + p_X^{\mathcal{E}} = 1$.
In the limit $q = p_X^{\mathcal{E}} = 0$, the model
becomes the $ZZ$-$X$ measurment-only circuit~\cite{nahum2019majorana, sang2020protected, buechler2020projectiveTFIM, sang2020negativity}. 
For this model, the circuit dynamics can be mapped exactly to a bond percolation problem on the square lattice, where each bond (regardless of whether horizontal or vertical) is present with probability $p_{ZZ}^{\mathcal{M}}$.
The corresponding percolation threshold is at $p_{ZZ}^{\mathcal{M}} = p_{X}^{\mathcal{M}} = 1/2$~\cite{nahum2019majorana}.
When $q > 0$ and $p_X^{\mathcal{E}} > 0$,
with this particular gate set and parameterization, the model can still be solved analytically by a slight generalization of the above mapping to percolation, as we describe in Sec.~\ref{sec:mapping_to_perc}.
The phases and the critical properties are apparently universal and do not depend on these details,
as long as (strong) $\mb{Z}_2$ symmetry is preserved by the measurements and the decoherence.
This is discussed in detail in Appendix~\ref{sec:perturb_unitary}, where we perturb the circuit with random local $\mb{Z}_2$ unitary gates for which the dynamics cannot be exactly solved.

\begin{figure*}[t]
    \centering
    \includegraphics[width=.45\textwidth]{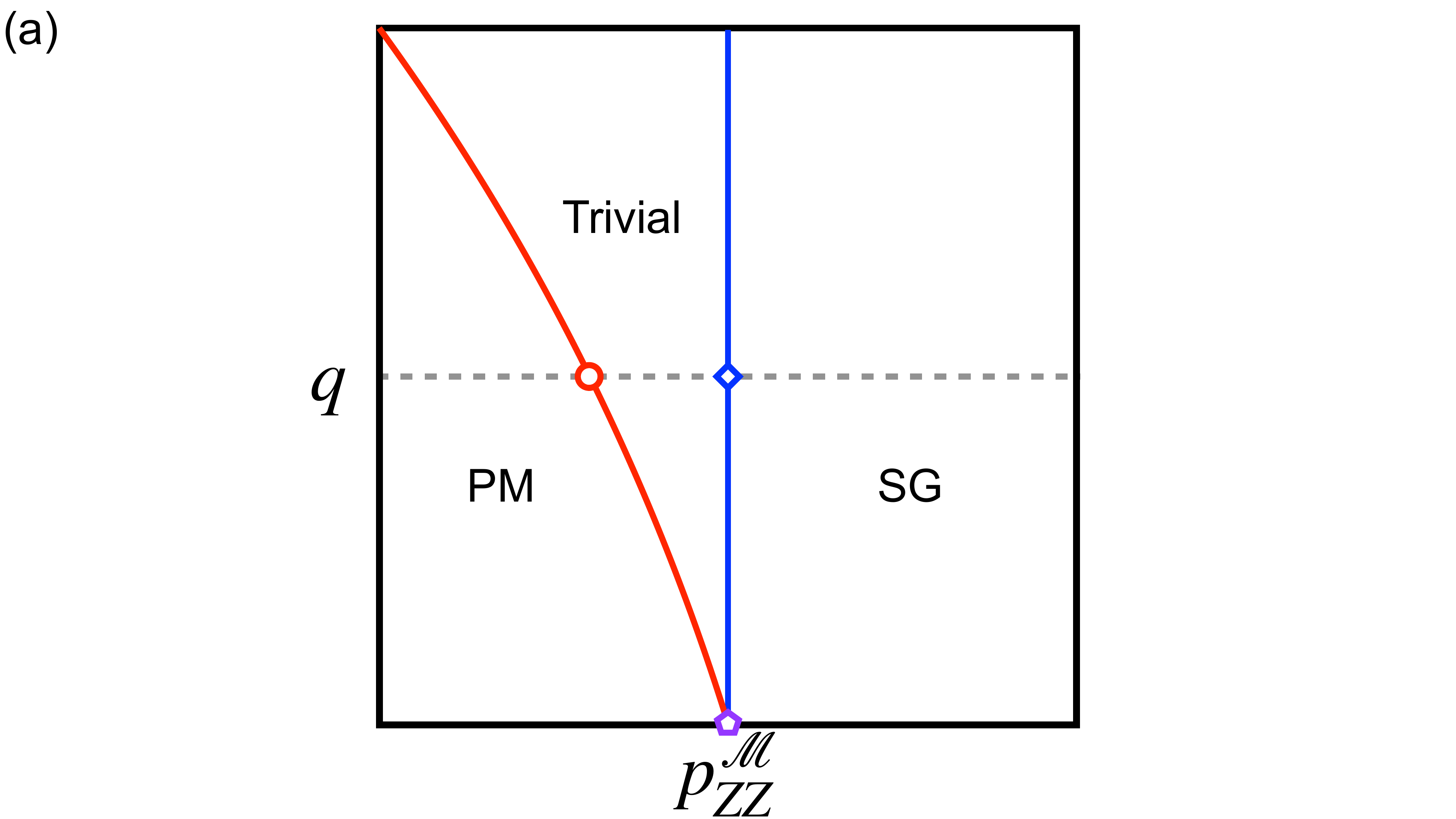}
    \includegraphics[width=.45\textwidth]{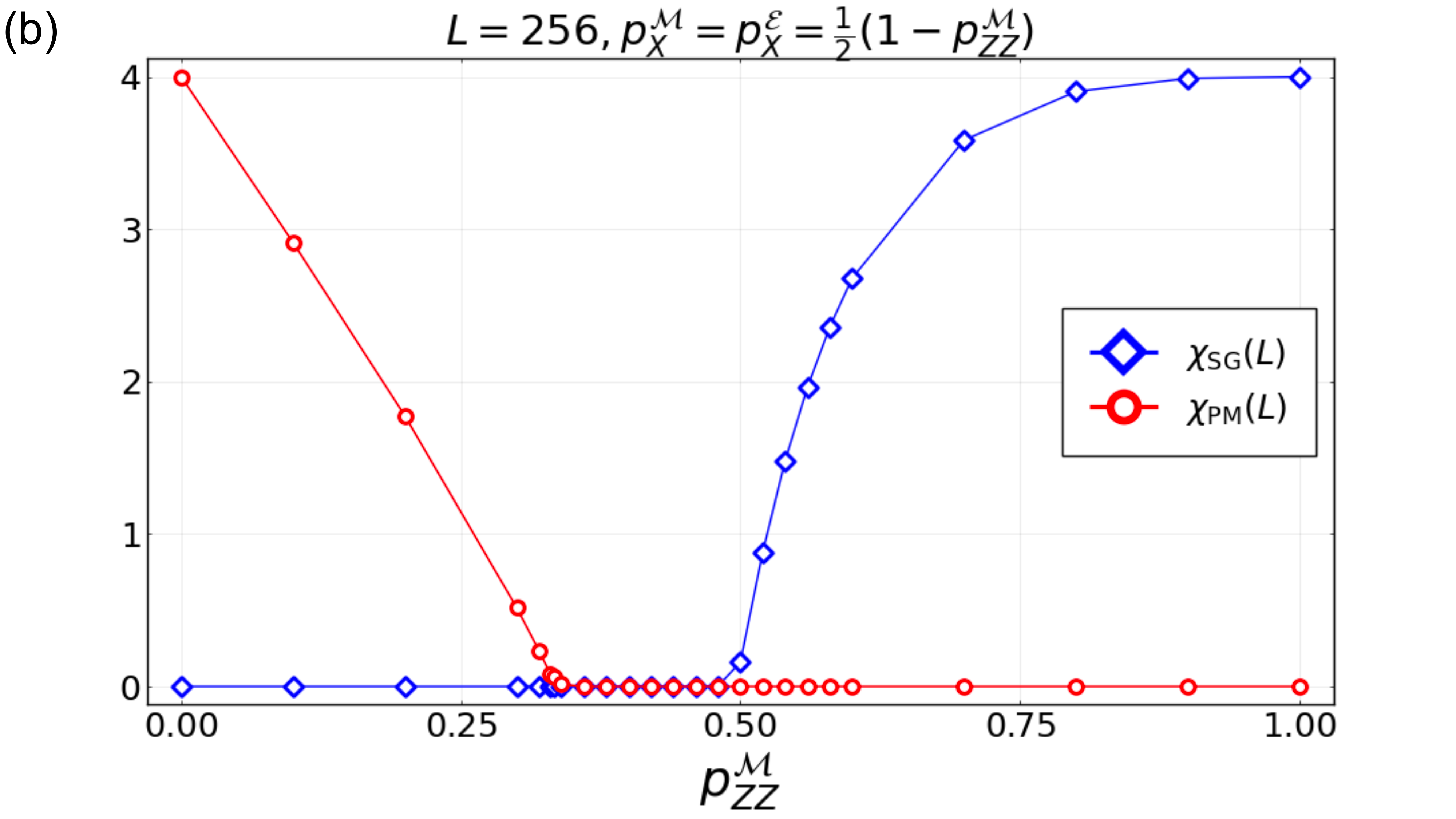}
    \includegraphics[width=.45\textwidth]{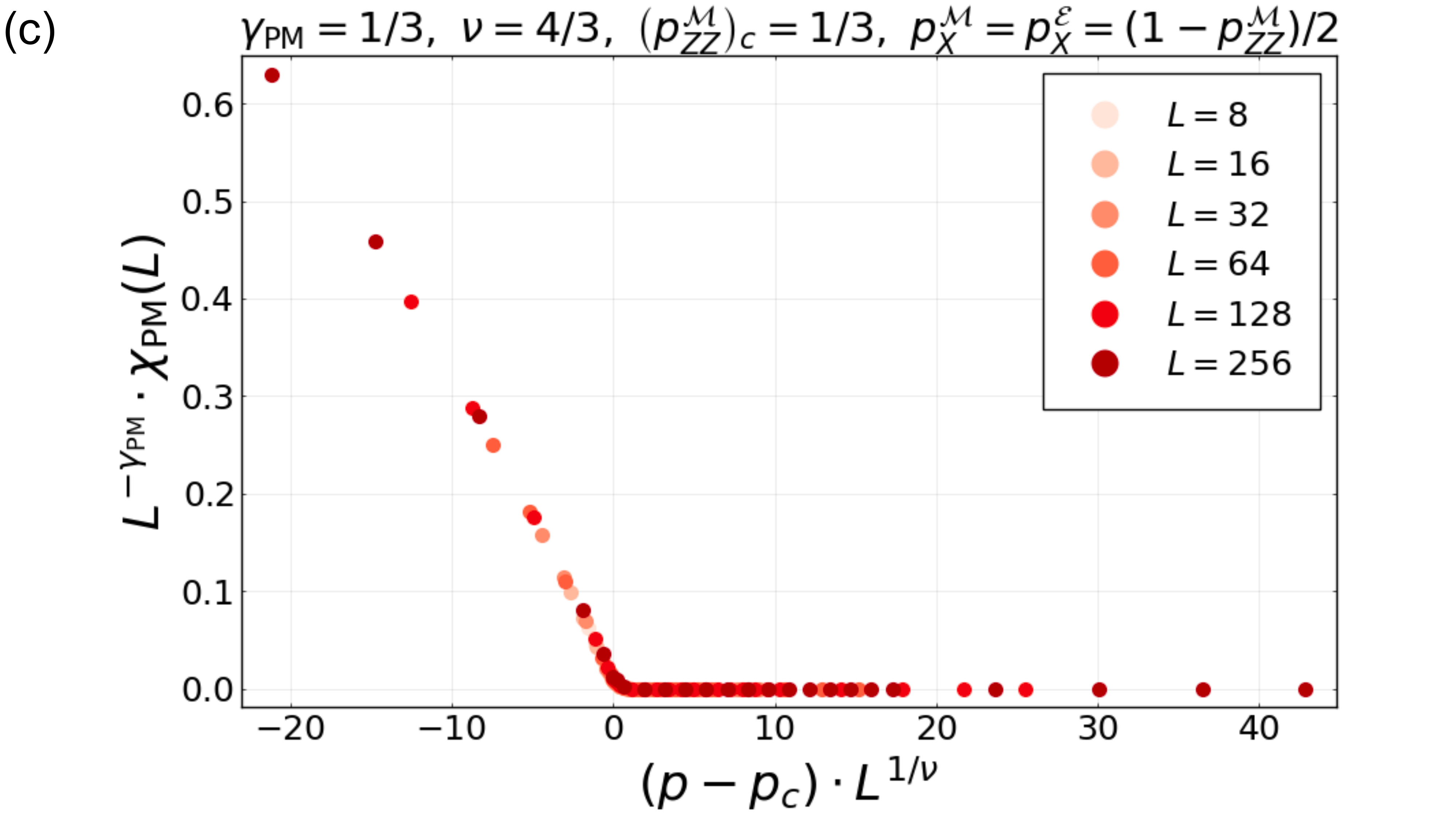}
    \includegraphics[width=.45\textwidth]{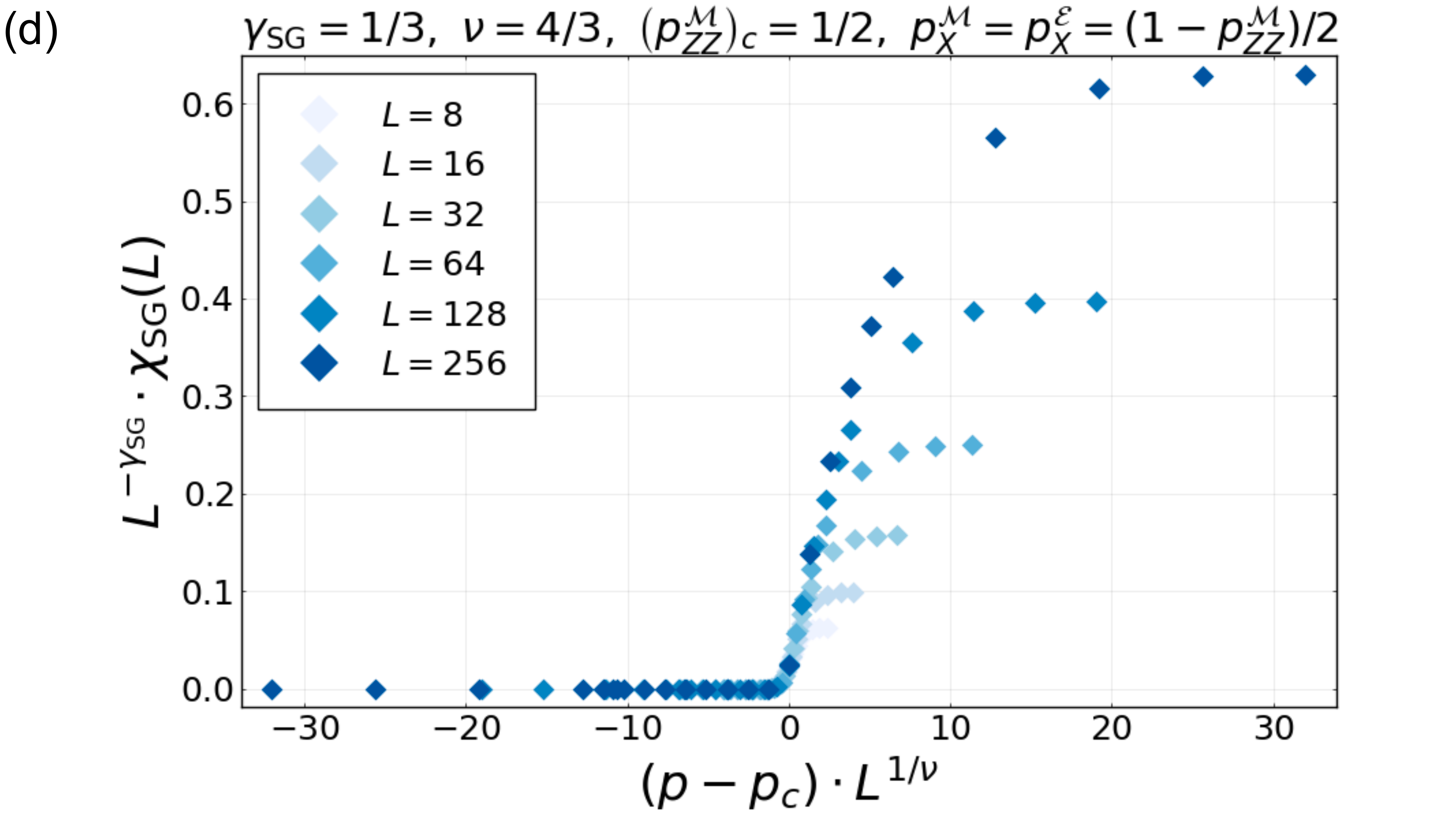}
\caption{
(a) The phase diagram of the baseline circuit [Fig.~\ref{fig:z2_circuit}], and (b,c,d) numerical results for this model along the dashed cross section in (a).
In (b), we see nonvanishing $\chi_{\rm SG}$ and $\chi_{\rm PM}$ at large and small values of $p_{ZZ}^\mc{M}$, respectively.
In between the SG and PM phases, we see an intermediate phase with neither order.
Near the critical points, $\chi_{\rm PM}$ and $\chi_{\rm SG}$ for different system sizes are collapsed against the scaling forms in Eqs.~(\ref{eq:chi_PM_collapse_perc}, \ref{eq:chi_SG_collapse_perc}), as shown in (c) and (d), respectively.
The boundary between {the trivial} and SG {phases} is at $p_{ZZ}^{\mathcal{M}}=1/2$,
and that between {the trivial} and PM {phases} is on {the line given by} $p_{ZZ}^{\mathcal{M}} = p_X^{\mathcal{M}}$.
The phase transitions here are all in the critical percolation universality class, but {generally} the same physical observable can map to different correlation functions for transitions marked by different colors.
}
    \label{fig:phase_diagram}
\end{figure*}

\begin{figure*}[t]
    \centering
    \subfigure[]{
        \includegraphics[width=.45\textwidth]{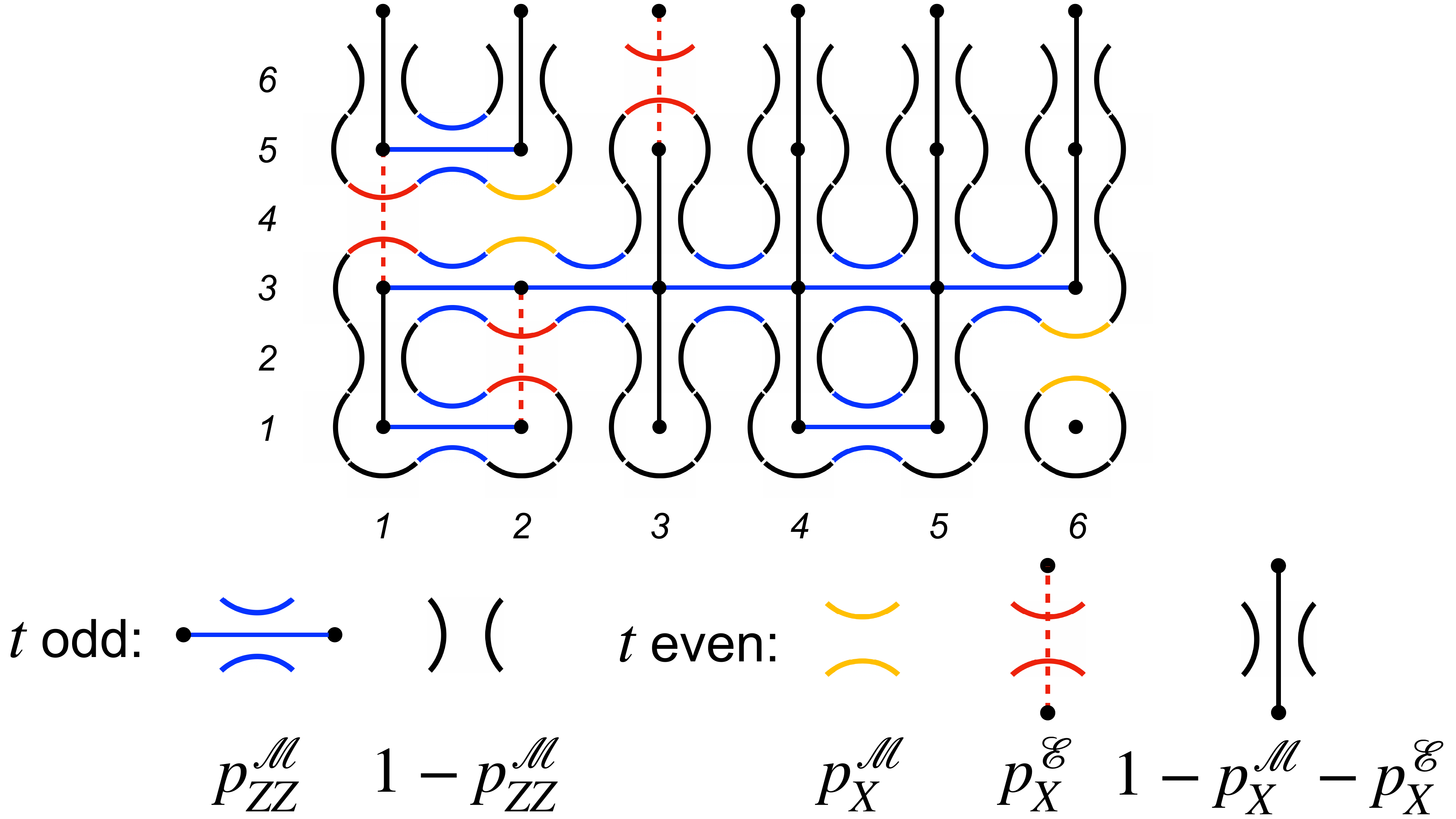}
    }
    \subfigure[]{
        \includegraphics[width=.45\textwidth]{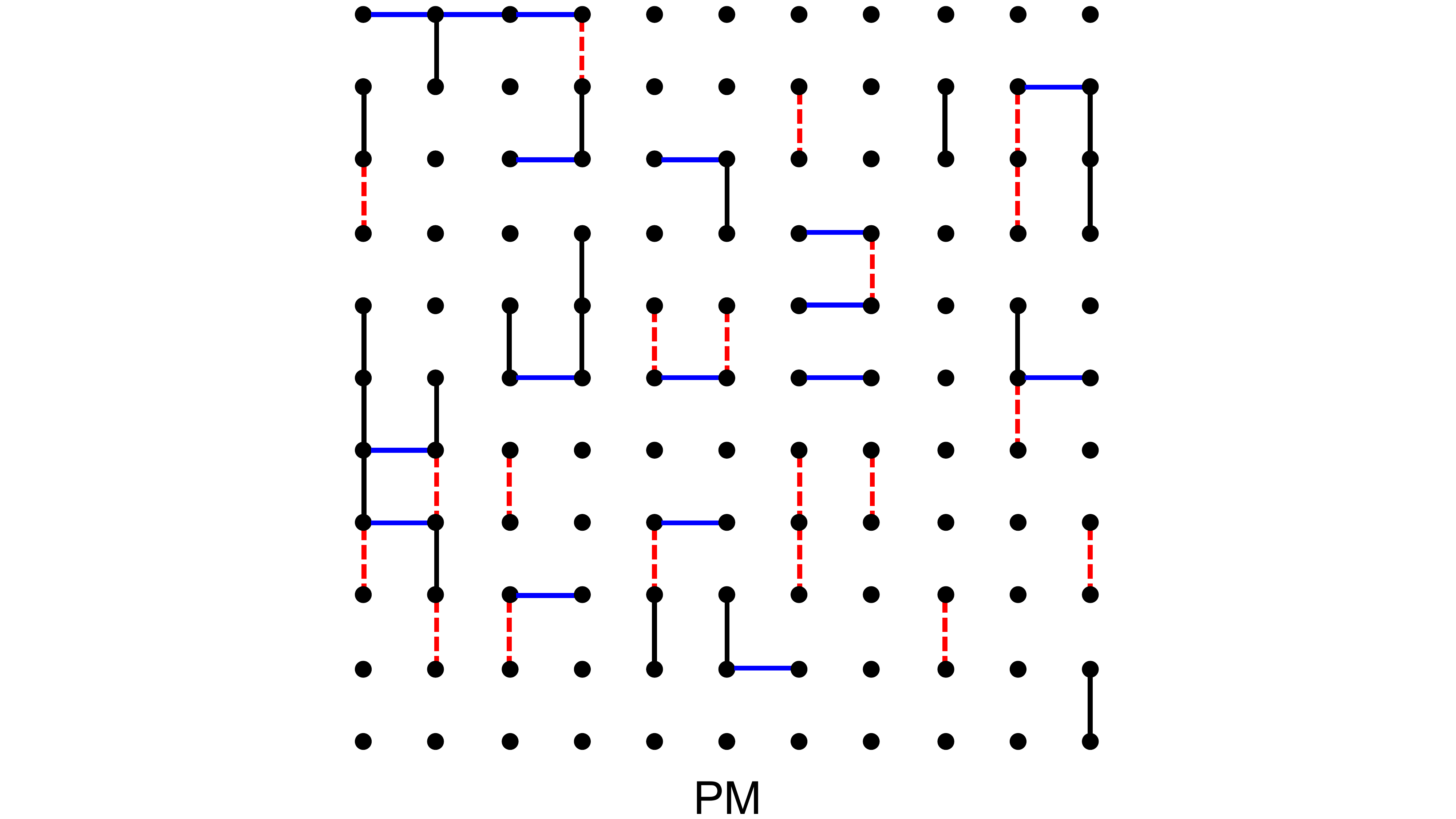}
    }
    \subfigure[]{
        \includegraphics[width=.45\textwidth]{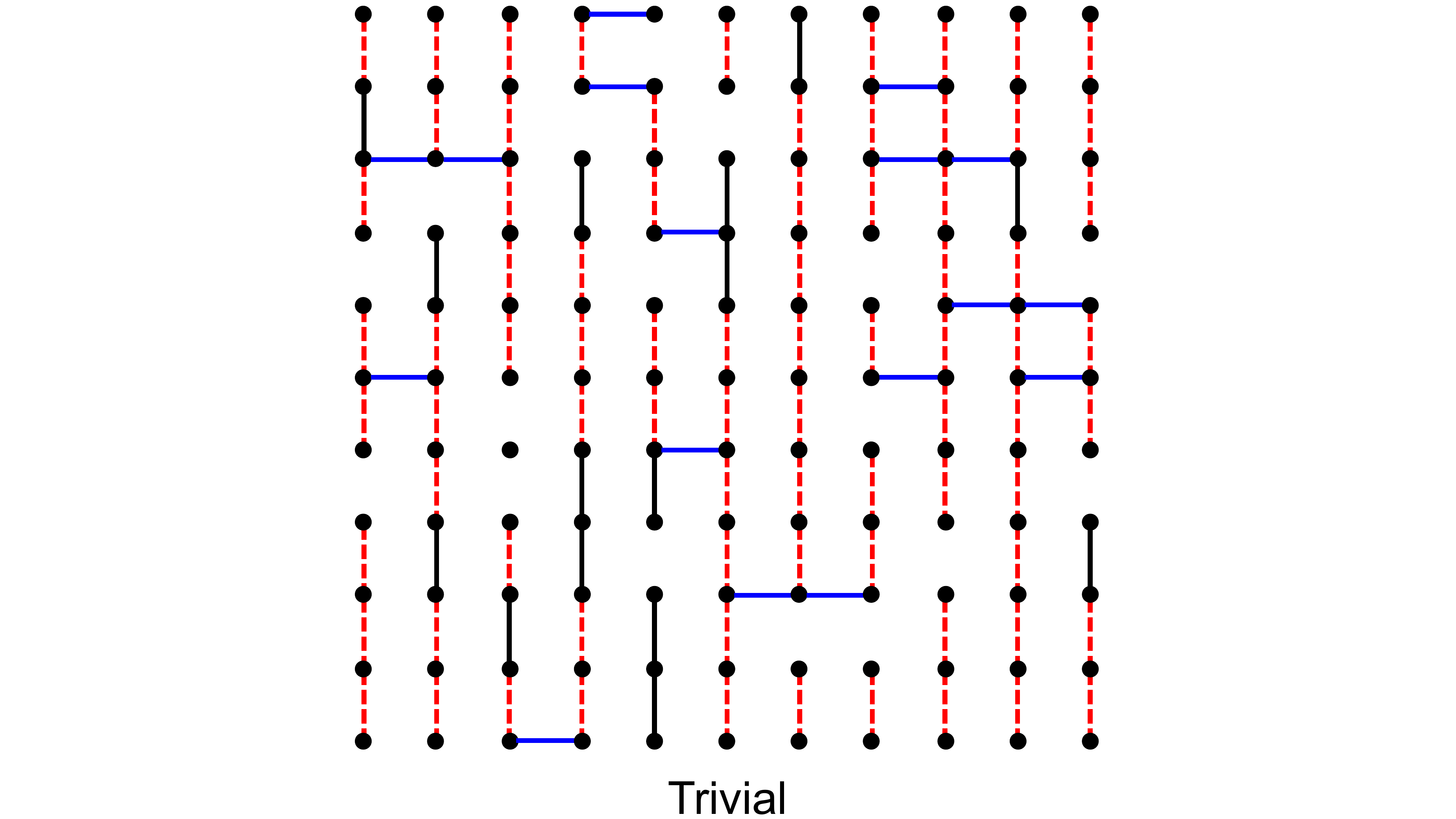}
    }
    \subfigure[]{
        \includegraphics[width=.45\textwidth]{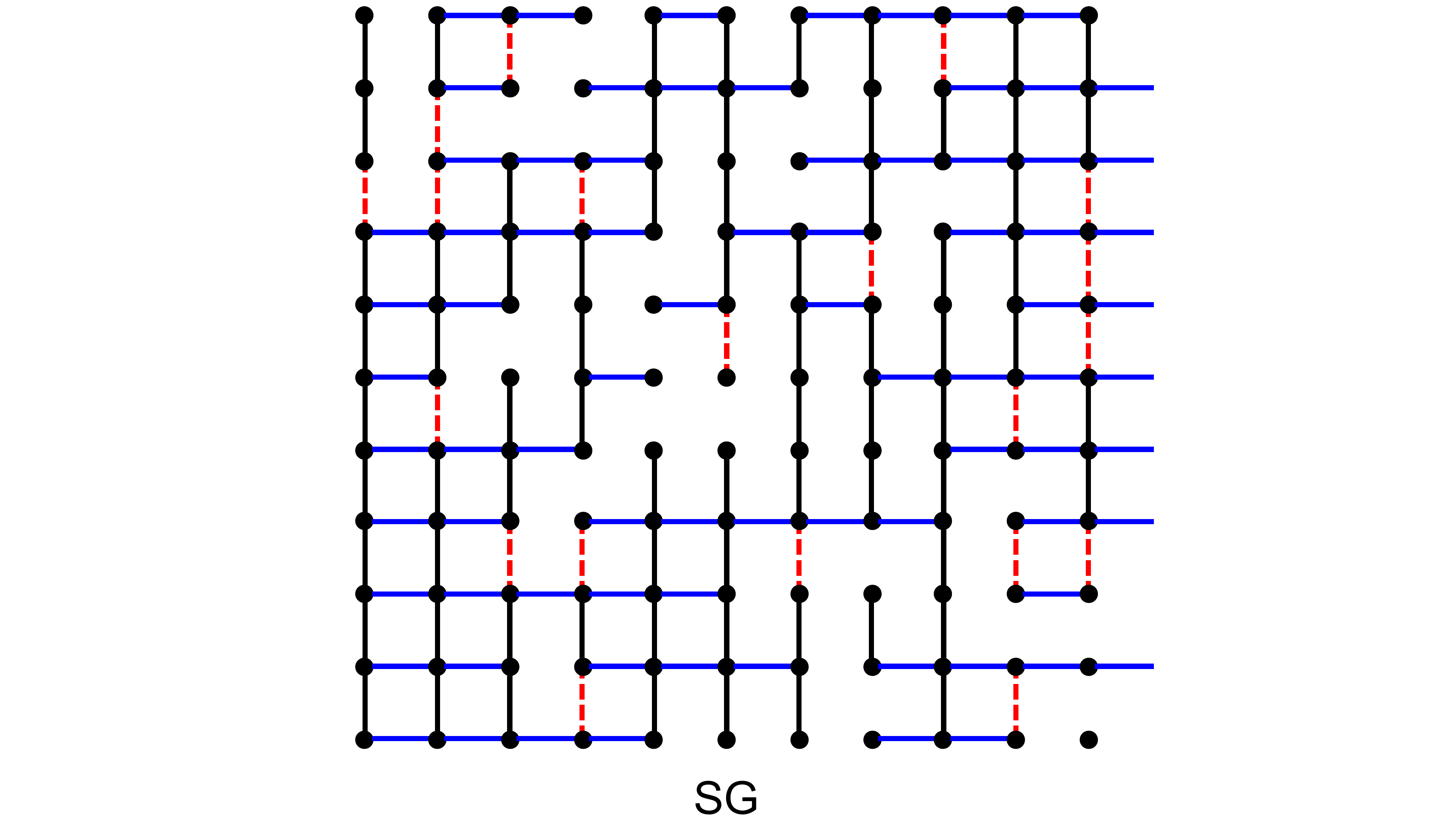}
    }
\caption{
(a)
Mapping from the circuit geometry to a bond percolation problem, for the example in Fig.~\ref{fig:z2_circuit}.
(b,c,d) Typical configurations of the square lattice in the PM, Trivial, and SG phases, respectively.
In the PM phase (b), the broken bonds percolate.
In the SG phase (d), the connected bonds (blue and black) percolate.
In the Trivial phase (c), the third type of decorated bonds (red, dashed) preclude the percolation of the other two types of bonds.
}
    \label{fig:perc_config}
\end{figure*}

For concreteness, we will mostly take the initial state a product of $\ket{+}_j = \frac{1}{\sqrt{2}}(\ket{0}_j + \ket{1}_j)$ on each qubit $j$ in our numerics.
This is a ``stabilizer state'', so that the circuit can be simulated efficiently using the Gottesman-Knill theorem~\cite{gottesman1997thesis, aaronson2004chp}.
This choice of initial state does not affect the phase diagram~\cite{buechler2020projectiveTFIM}, and we will return to the issue of a non-stabilizer initial state in Sec.~\ref{sec:dynamical_rep_code}.
We also take periodic boundary condition in the numerics, and focus on the steady state {with times} $T \gg L$,
{where we have $L$ qubits}.

\subsection{Phase diagram and critical properties \label{sec:mapping_to_perc}}

The model has three phases, shown in  Fig.~\ref{fig:phase_diagram}(a).
In the spin glass (SG) phase where $ZZ$ measurements dominate, the $\mb{Z}_2$ symmetry is spontaneously broken, and we have an extensive ``susceptibility'' as defined below~\cite{sang2020protected},
\env{align}{
    \label{eq:chi_SG_def}
    \chi_{\rm SG}
    =&\
    \frac{1}{L} \overline{
     \sum_{i \in A, j \in B} \lv \avg{Z_i Z_j} \rv^2
    }.
}
Here $A$ and $B$ are antipodal regions each of size $L/8$, $\avg{\ldots}$ denotes the expectation value in the steady state, and the overline denotes an ensemble average {over circuit realizations}.
The paramagnetic (PM) phase is where $X$ measurements dominate, and is similarly characterized by an extensive susceptiblity~\cite{bao2021enriched},
\env{align}{
    \label{eq:chi_PM_def}
    \chi_{\rm PM}
    =&\
    \frac{1}{L} \overline{
     \sum_{i \in A, j \in B} \lv \avg{X_i X_{i+1} \ldots X_{j-1} X_j} \rv^2
    }.
}
These two order parameters are related to each other by a Kramers-Wannier duality.
The SG and PM phases are separated by an intermediate trivial phase
~\cite{bao2021enriched}
where both $\chi_{\rm SG}$ and $\chi_{\rm PM}$ vanish, which is only present with decoherence, $q > 0, p_{X}^{\mc{E}} > 0$.
{Despite being in one-dimension, the symmetry breaking SG phase nevertheless survives decoherence (``noise"), not possible in an equilibrium system}.

We provide numerical evidence for this phase diagram along the line $q = 1/2$ [Fig.~\ref{fig:phase_diagram}(b)], where we calculate both susceptibilities and {indeed} find the above three phases. 
In particular, we have the PM phase when $p_{ZZ}^\mc{M} < 1/3$, the SG phase when $p_{ZZ}^\mc{M} > 1/2$, and the trivial phase in between.
Near the PM-Trivial transition at $(p_{ZZ}^\mc{M})_c^{\rm PM} = 1/3$, we collapse $\chi_{\rm PM}(L)$ for different system sizes against the following scaling form,
\begin{align}
    \label{eq:chi_PM_collapse_perc}
    \chi_{\rm PM}(L) = L^{\gamma_{\rm PM}}
    F[ \big(p-(p_{ZZ}^\mc{M})_c^{\rm PM} \big) \cdot L^{1/\nu} ],
\end{align}
where we take $\gamma_{\rm PM} = 1/3$ and $\nu = \nu^{\rm perc}(d=2) = 4/3$. {Here, and in the caption of Fig.~\ref{fig:phase_diagram}, we have used the notation $p \equiv p_{ZZ}^\mc{M}$.}
Similarly, near the SG-Trivial transition at $(p_{ZZ}^\mc{M})_c^{\rm SG} = 1/2$,
we collapse $\chi_{\rm SG}(L)$ for different system sizes against the following scaling form,
\begin{align}
    \label{eq:chi_SG_collapse_perc}
    \chi_{\rm SG}(L) = L^{\gamma_{\rm SG}}
    {\tilde{F}}[ \big(p-(p_{ZZ}^\mc{M})_c^{\rm SG}\big) \cdot L^{1/\nu} ],
\end{align}
where we take $\gamma_{\rm SG} = 1/3$ and $\nu = \nu^{\rm perc}(d=2) = 4/3$.  {From Kramers-Wannier duality we expect $F(X)=\tilde{F}(-X)$}.
The results are plotted in Fig.~\ref{fig:phase_diagram}(c,d), which clearly suggest continuous phase transitions with exponents consistent with two-dimensional critical percolation~\cite{cardy1991finite, cardy2001lecture}.


The phase diagram and the critical properties follow from a mapping of the circuit dynamics to a bond percolation problem on a two dimensional square lattice~\cite{nahum2019majorana, sang2020protected, buechler2020projectiveTFIM, sang2020negativity}, summarized in Fig.~\ref{fig:perc_config}(a).
In this mapping, gates at odd $t$ map to horizontal bonds, and gates at even $t$ map to vertical bonds.
In particular, at odd $t$, 
a projective measurement of $ZZ$ maps to a \emph{connected} horizontal bond, and the absence of a $ZZ$ measurement maps to a \emph{broken} horizontal bond.
At even $t$,
a projective measurement of $X$ maps to a \emph{broken} vertical bond, and the absence of an $X$ gate -- either an $X$ measurement or an $X$ dephasing channel -- maps to a \emph{connected} vertical bond.
The $X$ dephasing channel represents a third type of \emph{decorated} bonds, which we highlight with red color and dashed line.
With details in Appendix~\ref{app:details_mapping_percolation}, we show that the SG phase is the percolating phase of the \emph{connected} bonds alone -- that is, excluding broken bonds and decorated bonds [Fig.~\ref{fig:perc_config}(d)] -- and, the PM phase is the percolating phase of the \emph{broken} bonds alone, excluding connected bonds and decorated bonds [Fig.~\ref{fig:perc_config}(b)].
The intermediate {trivial} phase is where neither type percolates [Fig.~\ref{fig:perc_config}(c)], and is present only when the decorated bonds (decoherence) take place ($q>0$).
On the square lattice, we know exactly that the boudary between the SG and the trivial phases is at $p_{ZZ}^\mc{M} = 1/2$, and the boudary between the PM and the trivial phases is {on the line given by} $p_{ZZ}^\mc{M} = p_X^\mc{M}$.

\begin{table}[t]
\centering
\begin{tabular}{c || c | c | c | c}
 ~ &
 $\chi_{\rm SG}$
 & 
 $\chi_{\rm PM}$ 
 &
 $\frac{1}{4} I_{A=[0,L/2], \overline{A}}$ 
 &
 $I_{A=[0, \varepsilon], B=[\frac{L}{2}, \frac{L}{2}+\varepsilon]}$
 \\
\hline \hline
 PM
 & 0
 & $\propto L$
 & area law
 & 0
 \\ \hline
 Trivial
 & 0
 & 0
 & area law
 & 0
 \\ \hline
 SG
 & $\propto L$
 & 0
 & area law
 & 1
 \\ \hline
 PM-Trivial 
 & 0
 & $\propto L^{1/3}$
 & $\frac{\sqrt{3}\ln 2}{8\pi} \ln L$ 
 & $L^{-4}$
 \\ \hline
 SG-Trivial 
 & $\propto L^{1/3}$
 & 0
 & $\frac{\sqrt{3}\ln 2}{8\pi} \ln L$ 
 & $L^{-2/3}$
 \\ \hline
 PM-SG
 & $\propto L^{1/3}$
 & $\propto L^{1/3}$
 & $\frac{\sqrt{3}\ln 2}{4\pi} \ln L$ 
 & $L^{-2/3}$
\end{tabular}
\caption{
A summary of the scaling of the order parameters and entanglement properties in each of the 3 phases and at the phase transitions.
These results can be derived from the mapping in Fig.~\ref{fig:perc_config}, as we detail in Appendix~\ref{app:details_mapping_percolation}.
}
\label{table:critical_exponents}
\end{table}

We summarize in Table~\ref{table:critical_exponents} the phases and several critical exponents, and leave the details and further numerical results to  Appendix~\ref{app:details_mapping_percolation}.
The system is area law entangled inside the phases, and logarithmically entangled on the phase boundaries.
Note that although the phase transitions between the phases are clearly in the universality class of two dimensional critical percolation, the same observable can map to different correlation functions in the critical field theory at different transitions.

In Appendix~\ref{sec:ZZ_dep_channel} and Fig.~\ref{fig:phase_diagram_3D} we discuss a generalization of this model, where a fraction of the $ZZ$ measurements are replaced by $ZZ$ dephasing channels.
We find that this is an irrelevant perturbation, which does not change the universality class of the phase transitions.

\subsection{Introduction of an explicit bath \label{sec:intro_bath}}


\begin{figure}[t]
    \centering
    \includegraphics[width=\columnwidth]{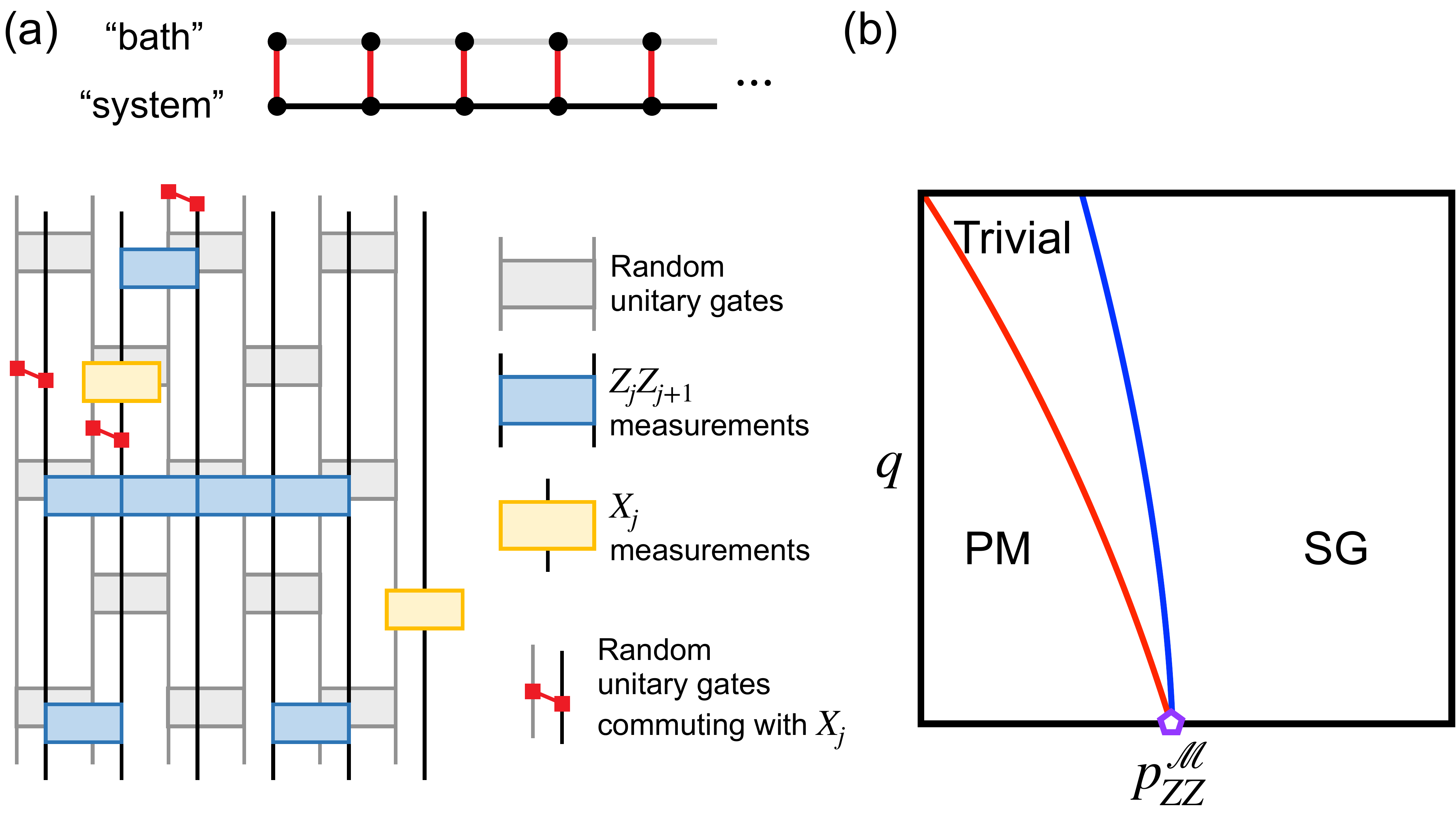}
    \caption{(a) The ``two-leg ladder'' circuit model introduced in Ref.~\cite{bao2021enriched}, and (b) its phase diagram.
    The ``system'' chain is similar to the baseline circuit model [Fig.~\ref{fig:z2_circuit}], but with the $X$ decoherence replaced by $\mb{Z}_2$ symmetric coupling to a random unitary circuit, which acts as a ``bath''.
    The physics of the phases and the phase transitions are  identical to the baseline circuit (compare Fig.~\ref{fig:phase_diagram}).
    }
    \label{fig:phase_diagram_twochain}
\end{figure}

We now introduce an explicit physical model of decoherence by coupling the ``system'' (i.e. the qubit chain in Fig.~\ref{fig:z2_circuit}) unitarily to a ``thermal bath''.
The bath is a collection of new degrees of freedom that has its own unitary dynamics.
The system and the bath together now form a closed system.

We posit two key conditions that must be satisfied by the system-bath coupling, namely
\begin{enumerate}
\item
``Triviality'': when the bath is traced out, the coupling can be generically captured by a local, $\mb{Z}_2$-symmetric quantum channel.
Thus, the coupling should be local, and should commute with the $\mb{Z}_2$ symmetry in the system.
\item
``Markovianity'': the $\mb{Z}_2$ channels induced by the coupling must appear short-range correlated in both the temporal and the spatial directions, despite the internal unitary dynamics of the bath.
To satisfy this condition, the bath in general needs to be extensive, and its internal dynamics needs to be ``scrambling''. 
\end{enumerate}
It is only when both conditions are met can the baseline circuit capture the universal dynamics of the system, after the bath is traced out.




An example satisfying the above conditions is a circuit with the geometry of a two-leg ladder (having $2 \times L$ qubits in total), first introduced in  Ref.~\cite{bao2021enriched}.
On one leg (``system'') of the ladder $Z_j Z_{j+1}$ and $X_j$ operators are measured with finite rates (defined as $p_{ZZ}^{\mc{M}}$ and $p_X^{\mc{M}} = (1-p_{ZZ}^{\mc{M}})(1-q)$, respectively), in a similar fashion as in Fig.~\ref{fig:z2_circuit}.
The other leg is another one dimensional chain undergoing random unitary dynamics, which is thermalizing (on its own) and acts as a ``bath''.
The two legs are coupled via random two-site unitary gates commuting with $\mathbf{X}$ in the system, on the rungs of the ladder, at a rate $p_X^{\mc{I}} = (1-p_{ZZ}^{\mc{M}})q$ (this quantity is similar to $p_X^{\mc{E}}$).
The phase diagram (previously found in Ref.~\cite{bao2021enriched}, and shown in Fig.~\ref{fig:phase_diagram_twochain}) is very much like Fig.~\ref{fig:phase_diagram}, up to reparametrizations and/or phase boundaries.

We further confirm numerically (see Appendix~\ref{app:twochain_numerics}) that all the critical properties in the two-leg ladder circuit model are again fully consistent with Table~\ref{table:critical_exponents}.
Thus, the two-leg ladder circuit in Ref.~\cite{bao2021enriched} behaves qualitatively the same as the model in Fig.~\ref{fig:z2_circuit}, and the difference between a random unitary circuit and an infinite bath is irrelevant, as far as the phase diagram of the system is concerned.
{We note that all three phases of Fig.~\ref{fig:phase_diagram_twochain}(b) coexist with a volume law entanglement of the bath~\cite{bao2021enriched}.}

{In Appendix~\ref{sec:bath_transition}, we investigate an extension of the two-leg ladder circuit, where there are also local measurements on the bath.
Upon driving the bath through a volume law to area law entanglement transition, the ``trivial" phase is replaced by a ``critical'' phase with logarithmic entanglement.}


\section{The $\mb{Z}_2$ circuit as a partially monitored dynamical memory \label{sec:dynamical_rep_code}}

In this section, we take a different view on the circuit in Fig.~\ref{fig:z2_circuit}, namely as a model of error and error-correction for the quantum repetition code.
Similar problems have been considered previously, as we discuss in Sec.~\ref{sec:MWPM_numerics}.

Recall that 
the repetition code is a stabilizer code with stabilizers $\{ Z_1 Z_2, Z_2 Z_3, \ldots, Z_{L-1} Z_L \}$, whose code space is two-dimensional (i.e. one logical qubit) and has a basis,
\env{align}{
    \{ \ket{\mathbf{0}} = \ket{000\ldots}, \ket{\mathbf{1}} = \ket{111\ldots} \}.
}
The logical $X$ operator can be chosen to be the $\mb{Z}_2$ symmetry operator $\mathbf{X} = X_1 \ldots X_L$, and the logical $Z$ operator (denoted $\mathbf{Z}$) can be chosen to be $Z_j$ for any $1 \leq j \leq L$.
In the circuit, we take the initial state to be the encoding of an arbitrary one-qubit state, 
\env{align}{
    \label{eq:psi_encoded}
    \ket{\psi} = \alpha \ket{0} + \beta \ket{1}
    \,
    \xrightarrow[]{\text{encoding}}
    \,
    \ket{\Psi} = \alpha \ket{\mathbf{0}} + \beta \ket{\mathbf{1}}.
}
This \emph{encoded logical state} $\ket{\Psi}$ is then subject to gates in the circuit.
Here, the \emph{errors} are the $X$ measurements and $X$ decoherences that bring the state outside the code space while also reducing the purity of the state.
These types of errors are occuring at a finite rate ($p_X^{\mc{M}}$ and $p_X^{\mc{E}}$) in time.
{Meanwhile}, 
stabilizer measurements of each $Z_j Z_{j+1}$ are also made at a finite rate $p_{ZZ}^{\mc{M}}$, that are trying to project the state back into the code space.

As the circuit time grows, the errors will accumulate, and the information encoded in the initial state will eventually be lost if the errors prevail (e.g. when they are the \emph{only} processes in the dynamics).
However, by making stabilizer measurements concurrently with the errors at a finite rate $p_{ZZ}^\mc{M} < 1$, one can hope to obtain clues about the history of the errors without destroying the encoded state, thereby preserving that information.
The question here is whether by making $O(TL)$ measurements as the circuit evolves, {is it} 
possible \emph{in principle} to  reverse these errors and reliably recover the initial state $\ket{\Psi}$; and when this is indeed possible, what is an efficient \emph{decoding algorithm} (or simply a ``decoder'') in practice.
Here, the decoding algorithm refers to a sequence of quantum operations (i.e. measurements and unitaries) on the final state, aided by the information about the circuit bulk -- namely the $ZZ$ measurement locations and the corresponding outcomes.

We discuss two versions of this question, depending on whether or not we assume knowledge of the locations of the errors (i.e. the $X$ gates) in the decoding, in the following two subsections, respectively.
When the errors are {``located"}, there exists an exact decoding algorithm if and only if the information is still encoded in the final state.
As we will see, this condition is related to a geometrical property of the underlying circuit, and the ``encoding phase'' coincides with the spin glass phase.
When the errors are {``unlocated"}, we devise a heuristic decoder, whose probability of success in the thermodynamic limit defines a ``decoding phase''.

\subsection{Exact decoding algorithm with located errors in the spin glass ``encoding'' phase \label{sec:located_error}}

\begin{figure}[t]
    \centering
    \includegraphics[width=.5\textwidth]{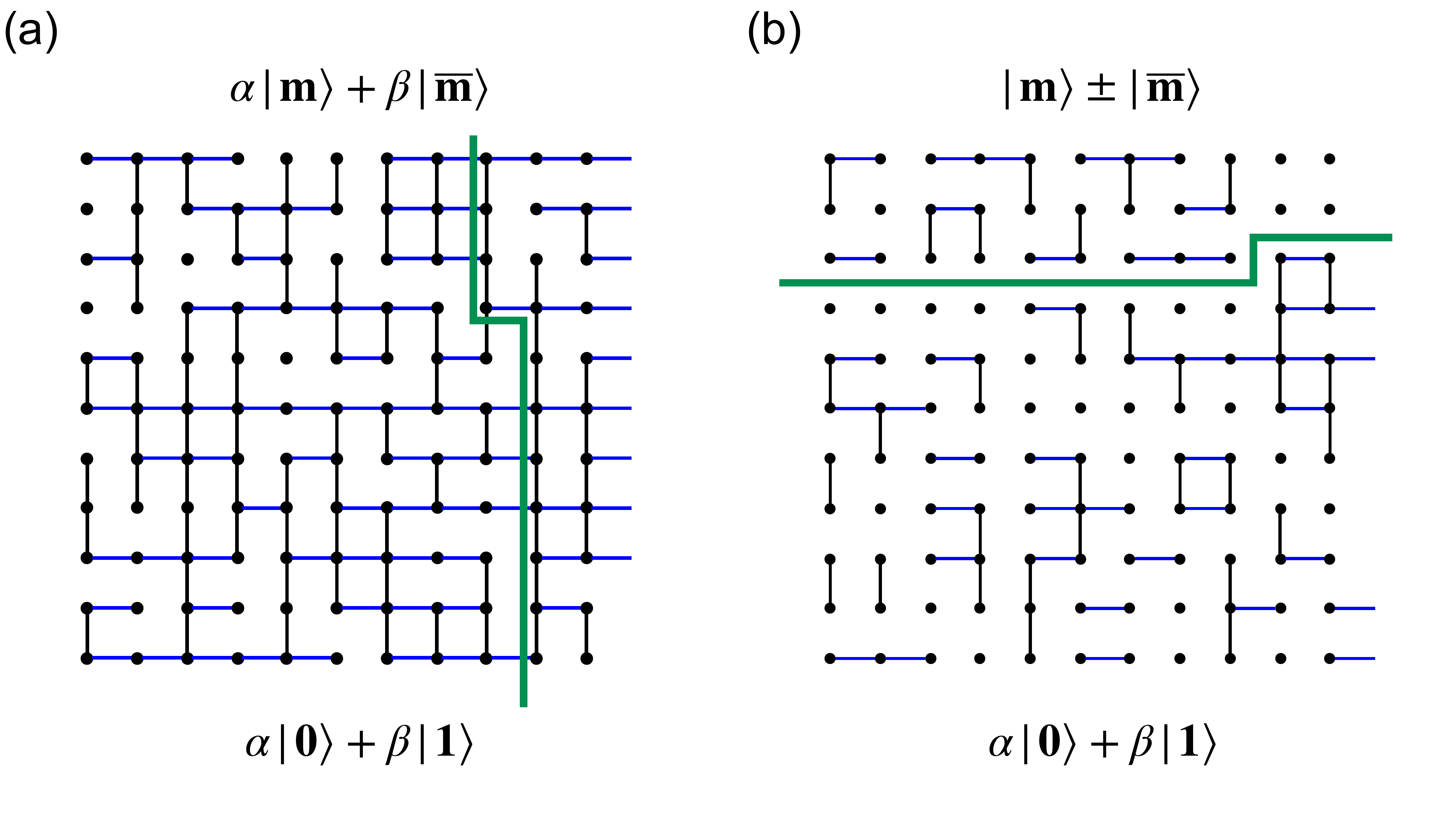}
    \includegraphics[width=.5\textwidth]{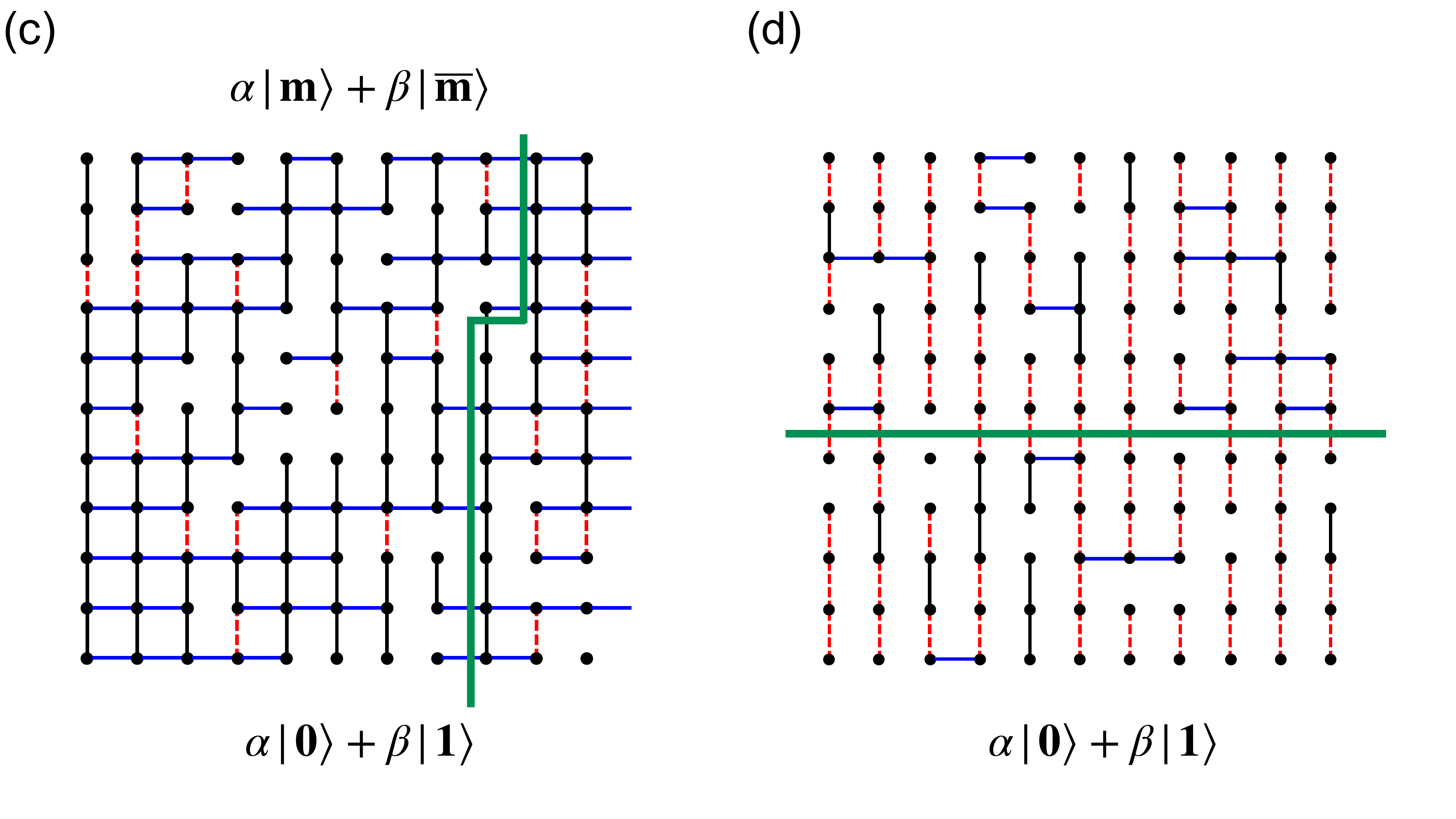}
    \caption{
    (a, b)
    With only $X$ measurement errors but no $X$ dephasing errors, 
    we illustrate in (a) an error-avoiding path in the SG phase connecting the upper and lower boundaries, and in (b) a ``cut'' in the PM phase precluding an error-avoiding path.
    In (a), the final state is a coherent code state containing the initial information (albeit in encoded form), whereas in (b) the final state is a classical mixture of $\ket{\mathbf{m}} \pm \ket{\ovl{\mathbf{m}}}$, and the initial information has been lost~\cite{buechler2020projectiveTFIM}.
    (c, d)
    Similarly, in the presence of $X$ decoherence, we  illustrate in (c) a ``path''  when in the SG phase, and in (d) a ``cut''  when outside the SG phase.
    In the SG phase (c), the final state is again a coherent code state, while outside the SG phase (d) the final state is in general mixed.
    }
    \label{fig:lang_buechler}
\end{figure}

The recovery problem with located $X$ measurements (but without $X$ decoherences)
was first discussed by Lang and B\"{u}chler~\cite{buechler2020projectiveTFIM}.
The authors showed that the initial state is \emph{in principle} recoverable from the final state \emph{provided} that the lower and upper boundary of the circuit geometry are connected by an ``error-avoiding spanning path'' consisting of connected bonds \emph{only} [Fig.~\ref{fig:lang_buechler}(a)].
In particular, after measuring $Z_j Z_{j+1}$ stabilizers for all $1 \leq j < L$ (which is a quantum operation denoted $\mc{M}$) on the final state (which we denote as $\mathcal{E}\ket{\Psi}$, where $\mathcal{E}$ represents the circuit evolution), the resultant state takes the form,
\env{align}{
    \label{eq:lang_buechler}
    \ket{\Phi} \coloneqq (\mathcal{M} \circ \mathcal{E}) \ket{\Psi} = \alpha \ket{\mathbf{m}} + \beta \ket{\ovl{\mathbf{m}}}.
}
Here, $\mathbf{m} = m_1 m_2 \ldots m_L$, {with $m_j=\pm1$},
and thus $\ket{\mathbf{m}}$ is a state in the computational basis. Here $\ket{\ovl{\mathbf{m}}} = \mathbf{X} \ket{\mathbf{m}}$.
In this case, the \emph{quantum} information is still stored in the system, and can be recovered given the classical information of $\mathbf{m}$.
On the {other} hand, when such a path does not exist [Fig.~\ref{fig:lang_buechler}(b)], the resultant state is a probabilistic mixture of $\ket{\mathbf{m}} + \ket{\ovl{\mathbf{m}}}$ and $\ket{\mathbf{m}} - \ket{\ovl{\mathbf{m}}}$, and the quantum information is irretrievably lost~\cite{buechler2020projectiveTFIM}.
We rederive their results using a slightly different approach in Appendix~\ref{app:decoding_alg_proof}.

Clearly, the existence of the spanning path is related to the geometry of the circuit.
In the spin glass phase where the connected bonds percolate, such a path exists with probability $1$, and the information can be preserved for any time $T$ polynomial in $L$.
{The ``encoding phase'' thus coincides with the spin glass phase.}

We now describe an explicit algorithm for the recovery of the state $\ket{\Psi}$,
starting from the state $\ket{\Phi} = (\mathcal{M} \circ \mathcal{E}) \ket{\Psi}$.
From the stabilizer measurement outcomes in $\mc{M}$, we already know the following products,
\env{align}{
    (-1)^{m_j + m_{j+1}} = 
    (-1)^{\ovl{m}_j + \ovl{m}_{j+1}} = Z_{j} Z_{j+1}.
}
Thus, we will know $\mathbf{m}$ -- and consequently which bits one should flip in $\ket{\Phi}$ such that the result is $\ket{\Psi}$ -- \emph{provided} that we know the value of $m_j$ for any single $j$.
This bit of information must be inferred from a spanning path through the circuit history, but not from the state $\ket{\Phi}$, for which the bitstring $\mathbf{m}$ looks completely random at long times.
We claim that, for any spanning path $\pi$ with endpoints at sites $(u, t=0)$ and $(v, t=T)$ (denoted $(u,0) \xrightarrow[]{\pi} (v,T)$) we have
\env{align}{
    \label{eq:Z_cum_def}
    (-1)^{m_v} = Z_{\rm cum}(\pi) \coloneqq \prod_{ \{(j,t), (j+1,t) \} \in \pi} \lz Z_j  Z_{j+1}\rz(t) .
}
Here, the product is taken over all horizontal bonds on $\pi$, and $\lz Z_j  Z_{j+1}\rz(t)$ denotes the measurement outcome on the bond $\{(j,t), (j+1, t)\}$, of $Z_j  Z_{j+1}$ at time $t$.
We prove this in Appendix~\ref{app:decoding_alg_proof}.
When there are multiple spanning paths ending at the upper boundary (as will almost always be the case in the spin glass phase), any one of the paths would do the job, 
since they will result in the same $\mathbf{m}$.


With the inclusion of $X$ decoherence -- in addition to $X$ measurements -- the same decoding algorithm is still correct.
As in Fig.~\ref{fig:lang_buechler}(a,b), we have a similar graph, but now with decorated bonds representing the decoherence [Fig.~\ref{fig:lang_buechler}(c,d)].
Knowing where the $X$ measurements and $X$ decoherence are in the graph,  we can {still} find a spanning path $\pi$ with endpoints at sites $(u, t=0)$ and $(v, t=T)$, assuming connectivity between the upper and lower edges, {which is valid in the SG phase}.
The path $\pi$ uses only connected bonds, and avoids all $X$ measurements or $X$ decoherences.
When such a spanning path exists,
the resultant state $(\mathcal{M} \circ \mathcal{E} ) \ket{\Psi}$ takes the same form as $\ket{\Phi}$ in Eq.~\eqref{eq:lang_buechler}, despite the $X$ decoherences in the circuit history (see Appendix~\ref{app:decoding_alg_proof}).\footnote{Here, the {final} measurements of $Z_j Z_{j+1}$ for all $j$ are necessary.
{Before} these measurements, the final state is either disconnected due to the $X$ measurements, or mixed due to the $X$ dephasing channels.
The effects of these two types of errors are illustrated in Appendix~\ref{app:details_mapping_percolation}.
{We are also assuming the final measurements are perfect, so that the initial state can be recovered with fidelity $1$, which is the condition we use below to define the success probability.
In a less idealized situation, the output of the decoder will always have a fidelity smaller than $1$.
}
}
Again, we can decode the state 
knowing $(-1)^{m_v} = Z_{\rm cum}(\pi)$.







To summarize, with the locations of all the $X$ errors known, the information encoded in the initial state is decodable from the final state \emph{if and only if} the circuit is in the spin glass phase.
Here, in addition to the knowledge of the locations of the errors, we also assume the knowledge of the $O(TL)$ stabilizer $Z_j Z_{j+1}$ measurement outcomes in the bulk, as well as the ability of performing reliable local quantum operations (i.e. measurements and unitaries) on the final state.

\subsection{Decoding with unlocated errors in (1+1)-dimensions \label{sec:unlocated_error}}

\begin{figure}[b]
    \centering
    \includegraphics[width=.5\textwidth]{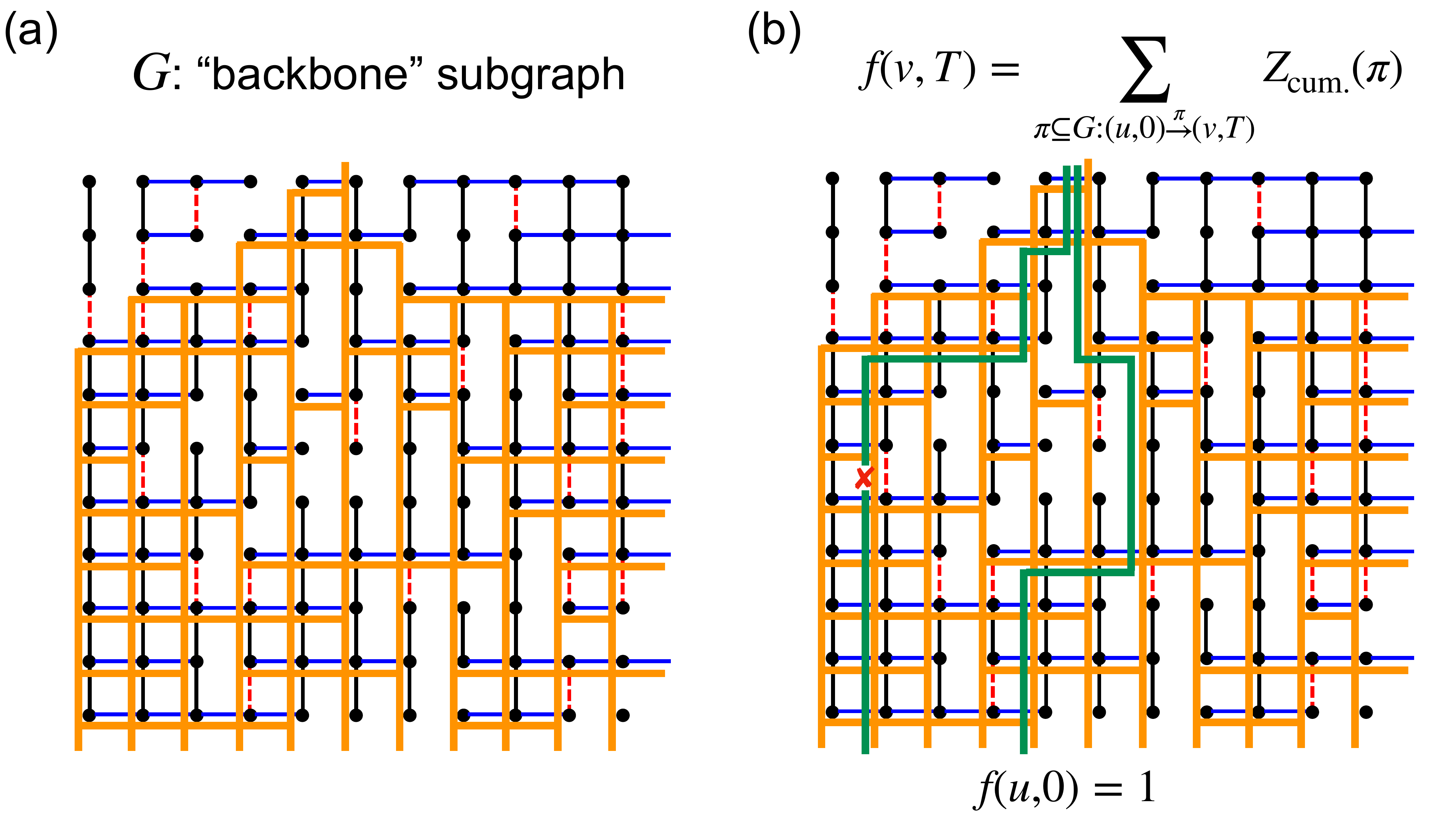}
    \caption{(a)
    The backbone subgraph $G$ (highlighed in orange) constructed out of stabilizer measurements on the horizontal bonds, and by treating all the vertical bonds as if connected (i.e. assuming no error occured on the vertical bonds). 
    (b)
    On $G$ there are two types of paths (highlighted in green).
    The ``good'' paths do not use any errored vertical bonds, and contribute a correct sign of $Z_{\rm cum}(\pi)$ to the summation in $f(v,T)$.
    The ``bad'' paths use one or more vertical bonds that contain errors (denoted by a red cross), and will contribute to the summation in $f(v,T)$ as ``noise''.
    }
    \label{fig:path_sum}
\end{figure}

When the $X$ measurements and $X$ decoherences are viewed {as errors}, 
it is usually the case that these errors are not controlled, and in particular their spacetime locations are unknown to the experimenter.
In conventional quantum error correction, the error locations need to be inferred from the stabilizer measurements (i.e. the syndromes).
In this subsection, we discuss the possibility of information recovery \emph{without} knowing the locations of the errors, or trying to locate the errors explicitly.
However, we still assume locations and outcomes of the stabilizer measurements in the bulk are known to the experimenter, 
who actively performed these measurements and dutifully recorded the results.

Given the circuit final state $\mc{E} \ket{\Psi}$, we can still follow the measurements $\mc{M}$ and obtain a state as in Eq.~\eqref{eq:lang_buechler}. 
However, to successfully decode, we need to know
$Z_{\rm cum}(\pi)$,
which is defined on a path $\pi$ that avoids all errors, {but} 
without the knowledge of the location of the errors it is impossible to calculate this quantity exactly.
We instead try to estimate $Z_{\rm cum}(\pi)$ with whatever information available, by summing over all paths (see Eq.~\eqref{eq:path_sum}), and will declare victory if our method can produce the correct result most of the time -- to be defined more precisely below.

After mapping to the square lattice percolation, what remains available to us is then the information of the horizontal bonds (i.e. whether they are broken or connected, and the measurement outcome on the bond if connected), but not of the information of the vertical bonds.
With these, we construct a ``backbone subgraph'' $G$ for each site $(v,T)$ on the upper boundary [Fig.~\ref{fig:path_sum}(a)].
The vertices of $G$ are all the sites that can be reached from $(v,T)$, using connected horizontal bonds, and any type of vertical bonds (connected, broken, or decorated).
In other words, without knowing what the vertical bonds are, we treat all of them as if connected.
For simplicity, we require that the vertical bonds can only be travelled downwards, so that all paths are directed in the temporal direction.
The edges of $G$ are the connected horizontal bonds, as well as all vertical bonds (as if they are connected) between pairs of vertices in $G$.

We define the following quantity on $G$ for each $(v,T)$,
\begin{align}
    \label{eq:path_sum}
    f(v,T) = 
    \sum_{\pi \subseteq G: (u,0) \xrightarrow[]{\pi}(v,T)}
    Z_{\rm cum}(\pi).
\end{align}
The summation on the RHS is over all paths in $G$ \emph{directed} in the temporal direction, with one endpoint fixed at $(v,T)$, and the other endpoint at any site on the lower boundary, $(u,0)$.
The quantity $Z_{\rm cum}(\pi)$ {has} 
the same definition as in Eq.~\eqref{eq:Z_cum_def}, namely as the product of horizontal $ZZ$ measurement results on the path $\pi$.
This summation can then be evaluted efficiently 
using the following recursion relation,
\env{align}{
    \label{eq:f_i_t_dp}
    &f(i,t) \nn
    =&\ 
    f(i,t-1) \nn
    &+
    \sum_{j \neq i: (j,t) \sim (i,t)}
    f(j,t-1) \cdot
    \prod_{\mathrm{min}\{i,j\} \le k < \mathrm{max}\{i,j\} } [Z_k Z_{k+1}](t).
}
Here $(j,t) \sim (i,t)$ means that $(j,t)$ and $(i,t)$ are connected by an array of consecutive $ZZ$ measurements at time $t$, and the product is over all the measurement outcomes in this consecutive array.
{Provided $
p_{ZZ}^{\mc{M}}<1$, we expect a finite number of}
terms on the RHS of Eq.~\eqref{eq:f_i_t_dp}.

In Eq.~\eqref{eq:path_sum}, there are two types of paths involved in the summation, as we illustrate in Fig.~\ref{fig:path_sum}(b).
There are ``good'' paths that avoids all errors; these will contribute the correct value of $Z_{\rm cum} (=(-1)^{m_v})$ to the sum, constructively.
There are also ``bad'' paths that pass through at least one error.
They will behave as ``noise'' that contribute $\pm Z_{\rm cum}$ at random.
We will get a correct estimate of $Z_{\rm cum}$ from the \emph{sign} of $f(v,T)$ (denoted $\mathrm{sgn} [f(v,T)]$), provided that the ``signal'' can beat the ``noise''.

Under what circumstances does the decoding algorithm defined above give the correct estimate of $Z_{\rm cum}$ from $\mathrm{sgn} [f(v,T)]$?
A necessary condition for the algorithm to produce anything more than noise is that the circuit must be in the spin glass phase with percolating connected bonds.
As we discussed earlier, the quantum information is irretrievably lost when the upper and lower boundaries are disconnected [Fig.~\ref{eq:lang_buechler}] -- that is, no algorithm can recover the initial state.
Meanwhile, the spin glass phase certainly seems to help the decoding.
In this phase, the number of ``good'' directed paths diverges exponentially in the circuit depth $T$, and we can hope to discern the signal if the noise diverges more slowly, although still exponentially in $T$.
This naive observation does not immediately tell us if the spin glass phase alone is sufficient for approximate decoding.
Next, we turn to numerical tests of the decoder.


Here we are only interested in the probability of success of the approximate estimation algorithm described above (rather than faithfully representing the time evolution of the quantum state and access the phase transitions), which should not be too sensitive to the input ensemble (i.e. the ensemble of backbone graphs $G$ as in Fig.~\ref{fig:path_sum}, and measurement results on its horizontal bonds),
we {initially} dispense with a full simulation on the evolution of a quantum state, and instead directly sample the inputs from a classical dynamics, so that 
large system sizes can be accessed.
In Sec.~\ref{app:benchmark} below, we show that the algorithm performs similarly on samples generated from the evolution of stabilizer 
quantum states, for {accessible} system sizes. 
There, we will also see the percolation transition out of the spin glass phase, which is not accessible with the classical dynamics discussed in this section.

The classical dynamics is defined as follows.
We take an array $m_j = 0$ for $1 \le j \le L$ to start with.
At odd time steps we ``measure'' each $(-1)^{m_j + m_{j+1}}$ with probability $p_{ZZ}^{\mc{M}}$, and at even time steps we flip $m_j \to 1-m_j$ with probability $p^{\rm err}$ for each $j$.
The quantity $p^{\rm err}$ is similar to $p_X^{\mc{M}} + p_X^{\mc{E}}$ for the model in Fig.~\ref{fig:z2_circuit}, except that we now vary $p^{\rm err}$ and $p_{ZZ}^{\mc{M}}$ independently (i.e. dropping the constraint $p_{ZZ}^{\mc{M}} + p^{\rm err} = 1$).
At the final step, we also ``measure'' $(-1)^{m_j + m_{j+1}}$ for each $j$.
We then compute $f(v,T)$ using Eq.~\eqref{eq:f_i_t_dp}, and compare its sign with the correct value of $Z_{\rm cum}(\pi_{\rm ESP})$ for any error-avoiding spanning path $\pi_{\rm ESP}$ terminating at $(v,T)$.
In this classical model {the correct value in the final state is} $Z_{\rm cum}(\pi_{\rm ESP}) = (-1)^{m_v}$.

\begin{figure}[t]
    \centering
    \includegraphics[width=.48\textwidth]{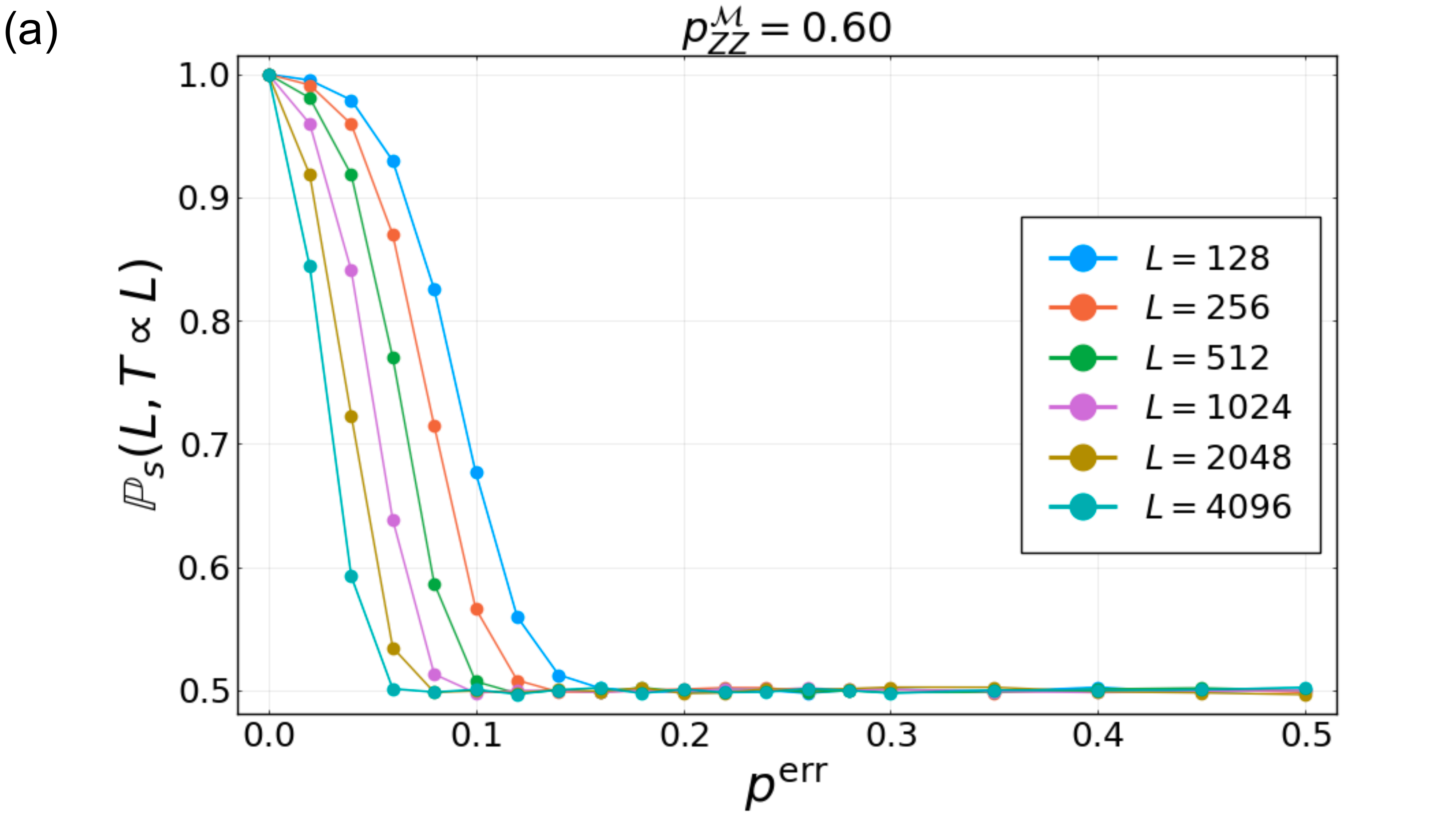}
    \includegraphics[width=.48\textwidth]{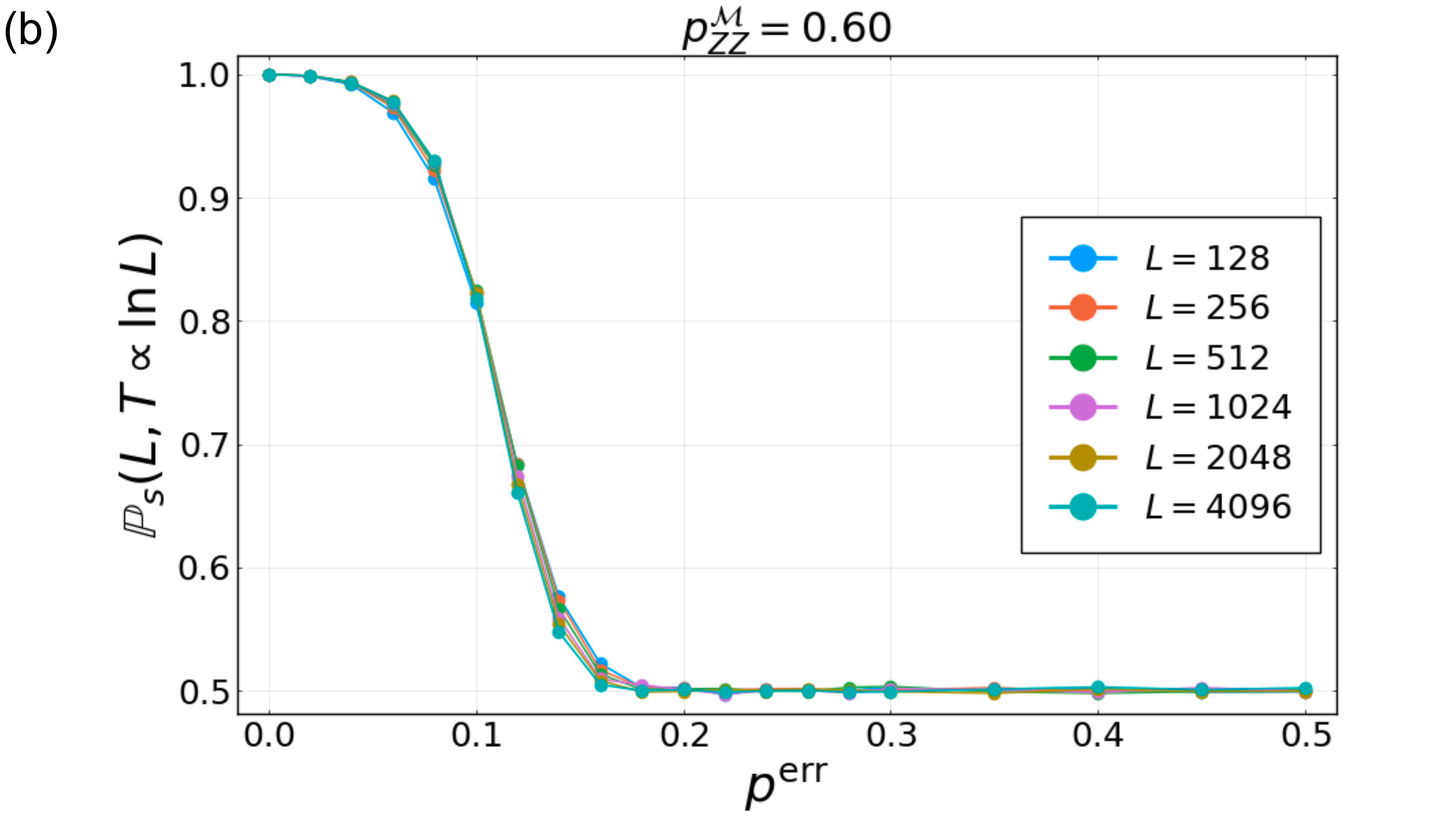}
    \includegraphics[width=.48\textwidth]{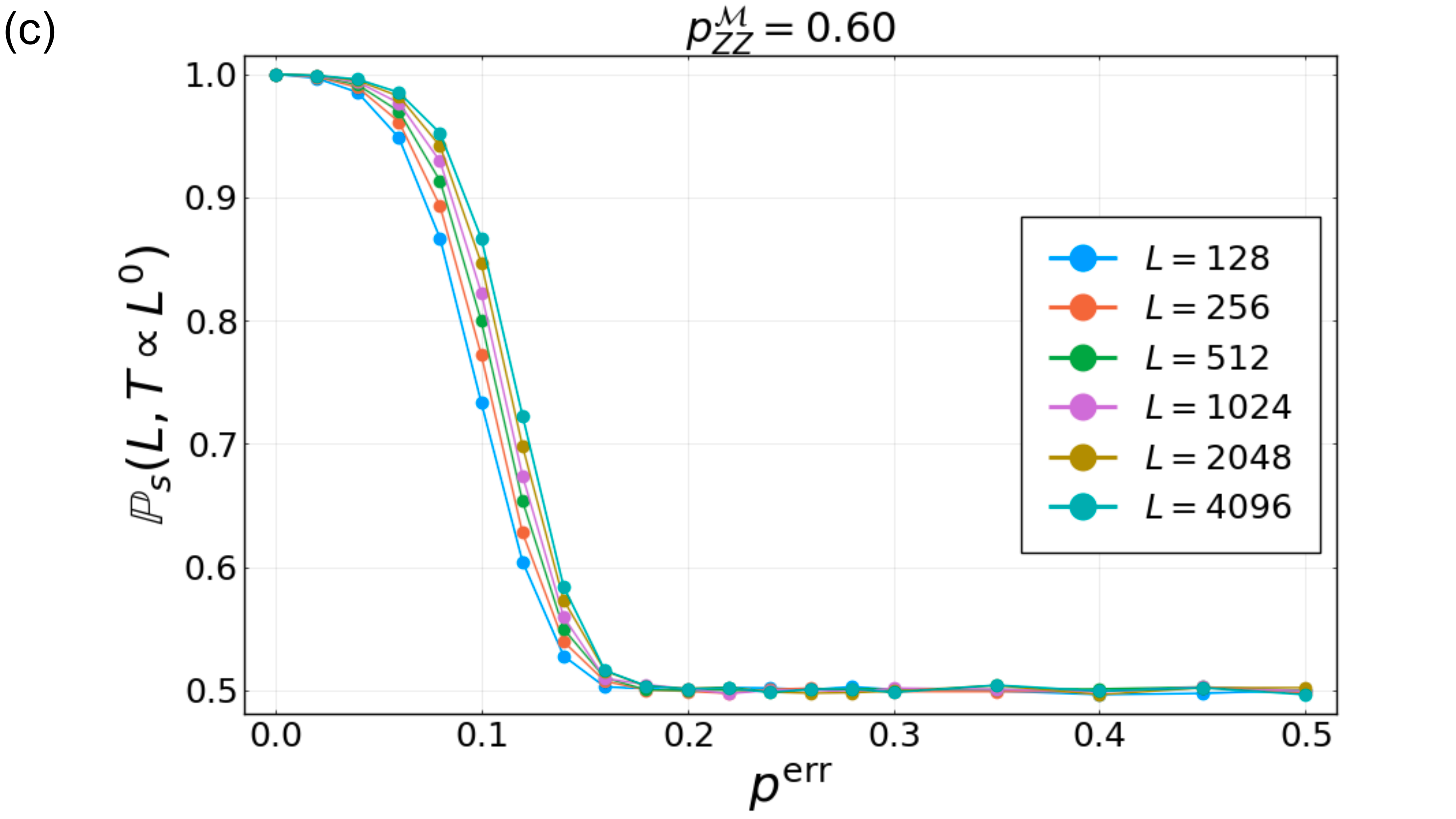}
    \caption{Results of the decoder in Eq.~\eqref{eq:f_i_t_dp} with unlocated errors for the repetition code in (1+1) spacetime dimensions, for (a) $T \propto L$, (b) $T \propto \ln L$, and (c) $T \propto L^0$, respectively.
    Here, the error model is the classical stochastic dynamics generated by random bit flips with probability $p^{\rm err}$.
    The results are summarized in Eqs.~(\ref{eq:P_s_rep_1+1d_a}, \ref{eq:P_s_rep_1+1d_b}), and the paragraphs below.
    }
    \label{fig:decoding_result_1+1d}
\end{figure}

We run this classical stochastic dynamics for various system sizes $L$ and various circuit depths $T$ over an ensemble of $O(10^5)$ samples.
We then compute the success probability of decoding, 
\begin{align}
    \mathbb{P}_s(L,T) \coloneqq \mathrm{Prob} [\mathrm{sgn} [f(v,T)] = Z_{\rm cum}(\pi)].
\end{align}
The numerical results are shown in Fig.~\ref{fig:decoding_result_1+1d} (a,b,c), where we take $T \propto L$, $T \propto \ln L$, 
and $T \propto L^0$,
respectively, and fix $p_{ZZ}^{\mc{M}} = 0.60$.
The results with a given set of $(p_{ZZ}^{\mc{M}}, p^{\rm err})$ in Fig.~\ref{fig:decoding_result_1+1d}(a, b) can be summarized as follows,
\begin{align}
    \label{eq:P_s_rep_1+1d_a}
    \lim_{L\to \infty} \mathbb{P}_s(L,T\propto L) =& \begin{cases} 
        1, & p^{\rm err} = 0 \\
        1/2, & p^{\rm err} > 0
    \end{cases}; \\
    \label{eq:P_s_rep_1+1d_b}
    \lim_{L\to \infty} \mathbb{P}_s(L,T\propto \ln L) =&\ F(p^{\rm err}),  \quad p^{\rm err} \ge 0.
\end{align}
That is, when $T \propto \ln L$, the probability of successful decoding appears to saturate to a smooth function in the thermodynamic limit.    
On the other hand, as long as $T$ grows faster than $\ln L$, there is not a finite error threshold, and the probability of successful decoding equals $1/2$ with any finite error rate.

Less clear is the case
when $T$ grows slower than $\ln L$ [Fig.~\ref{fig:decoding_result_1+1d}(c)].
The data suggest the following scenario: when the error rate is below a threshold, the probability of successful decoding approaches $1$ with increasing $L$, albeit slowly.


These result for decoding  with unlocated errors should be contrasted with the case of exact decoding with located errors, for which the decoding is successful with probability $1$ for $T \propto L$ as long as in the spin glass phase.
Note that the values we took for $p^{\rm err}$ in Fig.~\ref{fig:decoding_result_1+1d} are well below the percolation threshold.

\subsection{Decoding with unlocated errors in (2+1)-dimensions \label{sec:unlocated_error_2plus1d}}

\begin{figure}[b]
    \centering
    \includegraphics[width=.49\textwidth]{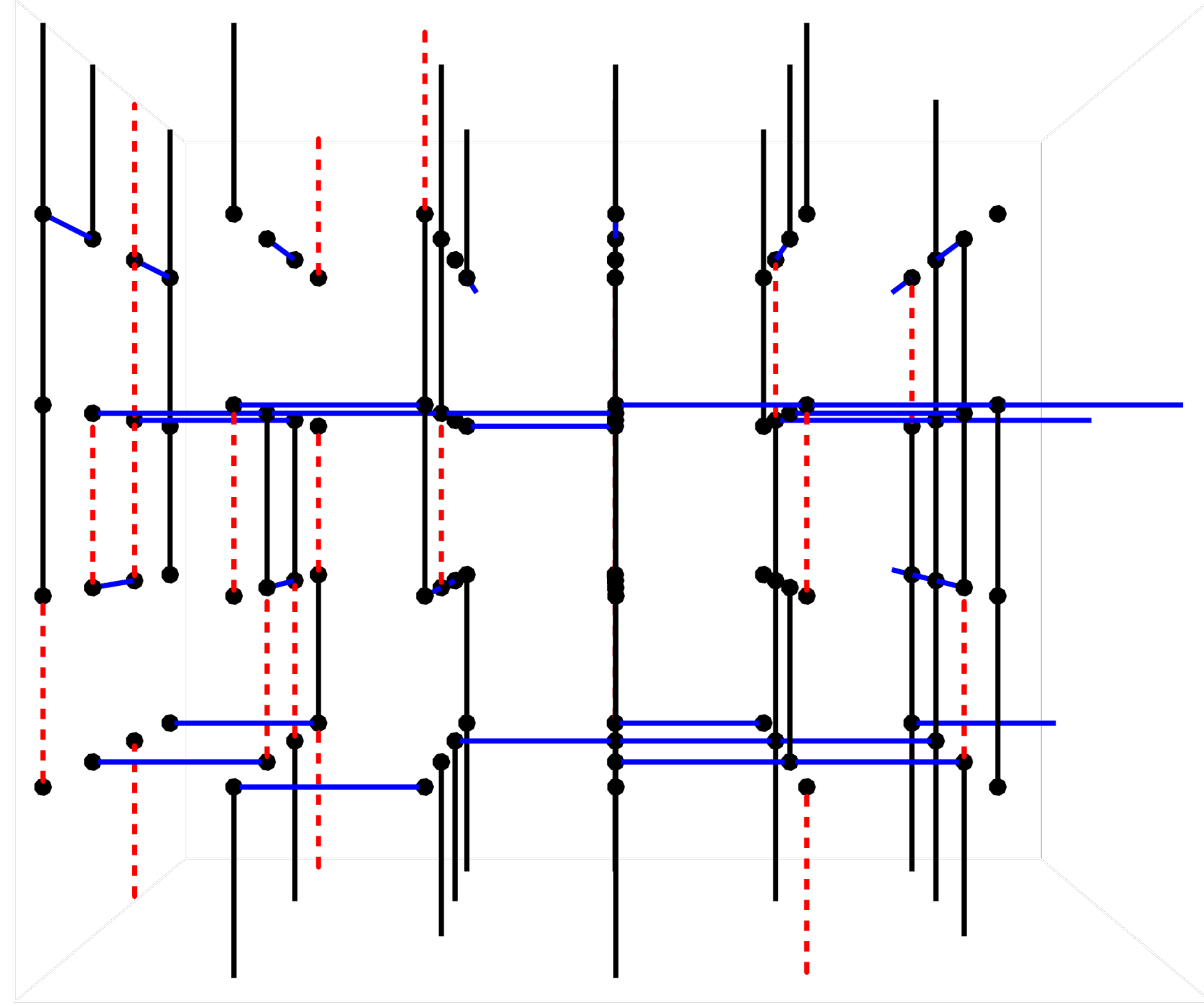}
\caption{
Dynamics of the repetition code in (2+1)d.
Following the convention in Fig.~\ref{fig:perc_config}, we represent measurements of $ZZ$ check operators (of rate $p_{ZZ}^\mc{M}$) by blue bonds in the $xy$-plane, $X$ measurements errors (of rate $p_X^\mc{M}$) by broken (absent) vertical bonds, and $X$ dephasing errors (of rate $p_X^\mc{E}$) by dashed red vertical bonds.
We choose to measure the check operators along $x$- and $y$- directions in alternative time steps.
}
    \label{fig:perc_config_2plus1d}
\end{figure}

In this subsection, we discuss the decoding algorithm for the repetition code in one higher spatial dimension.

For concreteness, we consider the circuit geometry of errors and stabilizer measurements in Fig.~\ref{fig:perc_config_2plus1d}.
Each ``period'' of the circuit consists of $4$ time steps.
In the 1st time step, we measure the $ZZ$ stabilizer on each bond in the $x$-direction, with probability $p_{ZZ}^{\mc{M}}$.
Similarly, in the 3rd time step, we measure bonds in the $y$-direction with the same probability.
At the 2nd and the 4th time step, the $X$ measurement and decoherence errors occur on each qubit with probabilities $p_{X}^{\mc{M}}$ and $p_{X}^{\mc{E}}$, respectively.

We briefly discuss the phase diagram.
The mapping from the circuit model to bond percolation introduced in Fig.~\ref{fig:perc_config} can be directly carried over, with the $ZZ$ measurements mapped to bonds in the $xy$-plane, $X$ measurements to broken bonds in the $t$-direction, and $X$ decoherence to decorated bonds in the $t$-direction -- a convention we already adopted in Fig.~\ref{fig:perc_config_2plus1d}.
As its (1+1)d counterpart, the spin glass and paramagnetic ``susceptibilities'' $\chi_{\rm SG}$ and $\chi_{\rm PM}$ can also be defined [Eqs.~(\ref{eq:chi_SG_def}, \ref{eq:chi_PM_def})], and will be extensive when the connected/broken bonds percolate, corresponding to the spin glass and paramagnetic phases, respectively.
When the decoherence is strong, there is similarly an intermediate phase where both susceptibilities vanish.
However, due to the larger dimensionality, there is now a phase where both susceptibilities can be extensive when the decoherence is weak.
The transitions between these phases should all be in the universality class of three-dimensional percolation.

We will henceforth focus on the question of decoding the quantum information in the initial state, in the spin glass phase.
With located errors, the decoding follows Sec.~\ref{sec:located_error}, which entails 
finding a spanning path $\pi_{\rm ESP}$ that avoids all errors and calculating $Z_{\rm cum}(\pi_{\rm ESP})$.
Again, the spin glass phase is identical to the encoding phase.

With unlocated errors, we can similarly compute the sum $f(\mathbf{v},T)$ [Eqs.~(\ref{eq:path_sum}, \ref{eq:f_i_t_dp})] for all directed paths ending at $(\mathbf{v},T)$ on the upper boundary of the circuit -- whether they have errored vertical bonds or not -- and compare $\mathrm{sgn}[f(\mathbf{v},T)]$ with the correct value of $Z_{\rm cum}(\pi)$.

\begin{figure}
    \centering
    \includegraphics[width=.5\textwidth]{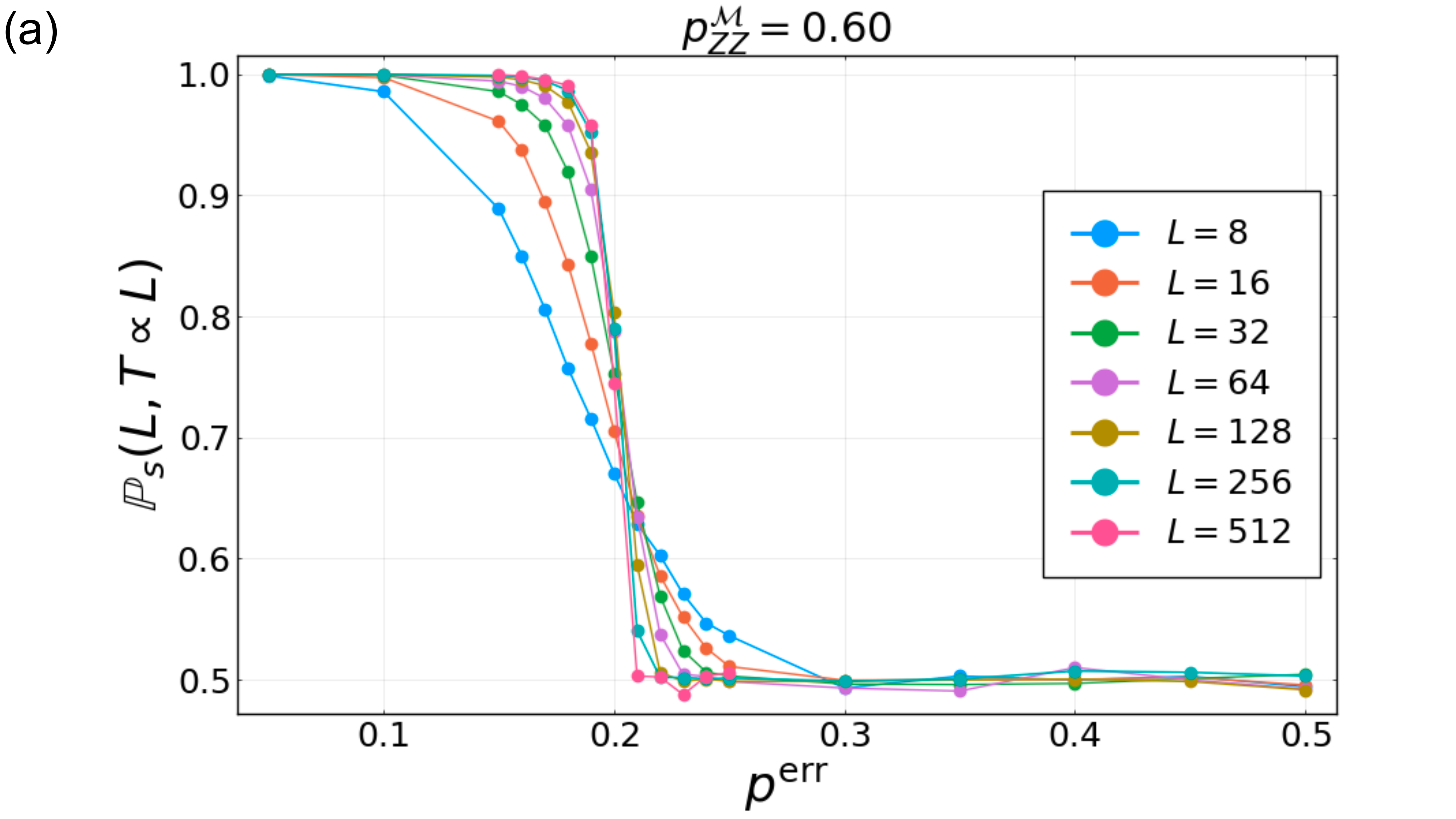}
    \includegraphics[width=.5\textwidth]{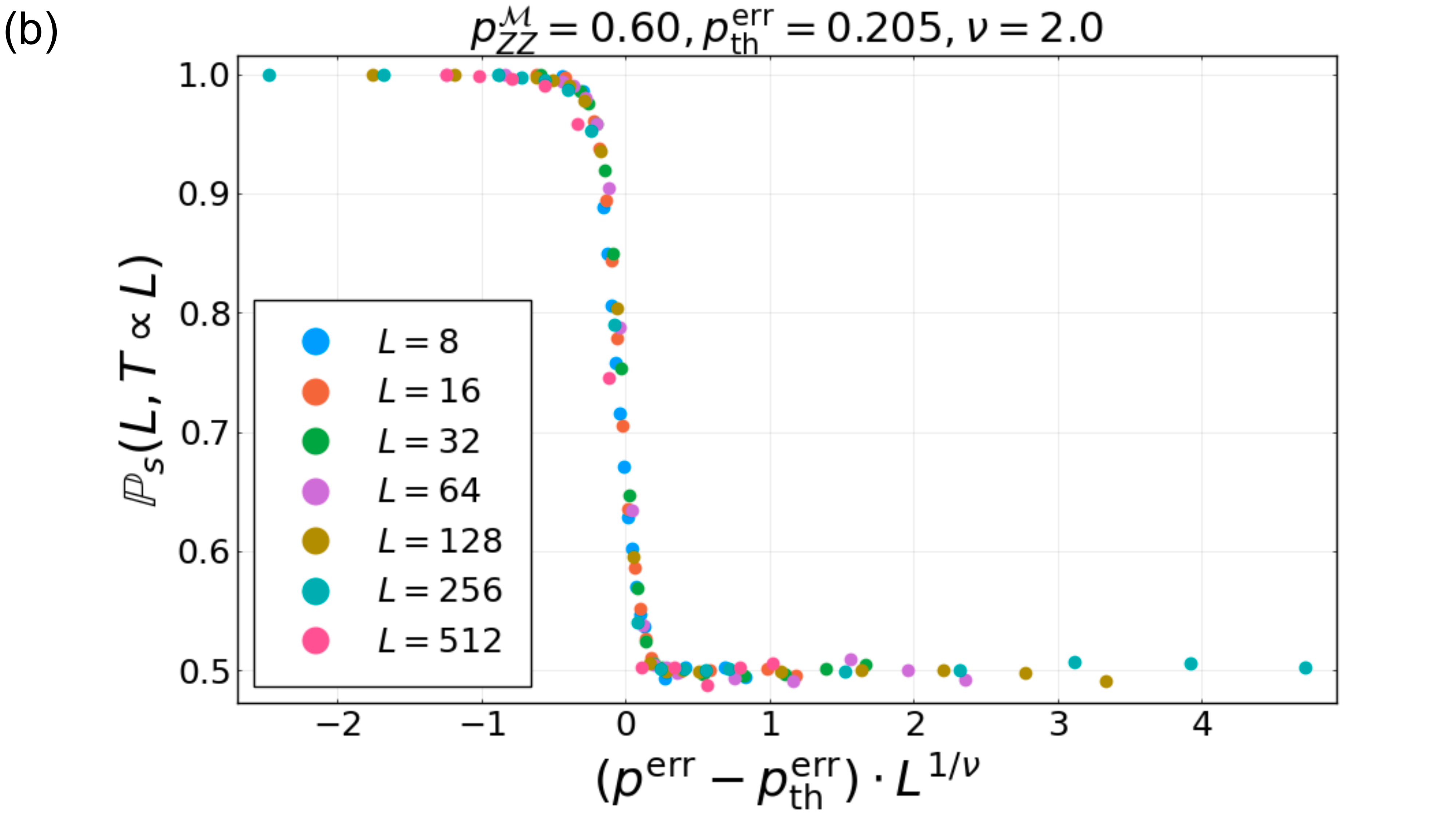}
    \caption{
    Results of the decoder in Eq.~\eqref{eq:f_i_t_dp} with unlocated errors for the repetition code in (2+1) spacetime dimensions.
    Here, the error model is the classical stochastic dynamics generated by random bit flips with probability $p^{\rm err}$.
    The results are summarized in Eqs.~(\ref{eq:P_s_rep_2+1d_collapse}, \ref{eq:P_s_rep_2+1d_scalingfcn}).
    }
    \label{fig:decoding_result_2+1d}
\end{figure}

As before in Sec.~\ref{sec:unlocated_error}, we introduce the parameter $p^{\rm err} = p_{X}^{\mc{M}} + p_{X}^{\mc{E}}$, and compute the probability of successful decoding on a ensemble of input graphs generated by the classical dynamics with parameters $p_{ZZ}^{\mc{M}}$ and $p^{\rm err}$ taken to be independent.
The numerical results are shown in Fig.~\ref{fig:decoding_result_2+1d}(a), where we focus on $L_x = L_y = L$ and $T \propto L$, and take $p_{ZZ}^{\mc{M}} = 0.6$.
The data with increasing system sizes suggest a second order phase transition at a finite threshold $p^{\rm err}_{\rm th} \approx 0.205$, from a phase where the decoding is successful with probability $1$, to another phase in which the decoding is no better than a coin flip.
In Fig.~\ref{fig:decoding_result_2+1d}(b), we attempt the following finite-size scaling form near $p^{\rm err}_{\rm th}$ and find good collapse of the data for different $L$,
\begin{align}
\label{eq:P_s_rep_2+1d_collapse}
    \mathbb{P}_s(L, T \propto L) = F \lz (p^{\rm err} - p^{\rm err}_{\rm th}) \cdot L^{1/\nu} \rz,
\end{align}
where $F$ is a scaling function with the following asymptotics,
\begin{align}
\label{eq:P_s_rep_2+1d_scalingfcn}
    F(x) = \begin{cases}
        1, & x \to -\infty \\
        1/2, & x \to +\infty
    \end{cases},
\end{align}
and $\nu \approx 2.0 \pm 0.5$ is a ``correlation length exponent'' obtained from the fitting.
Thus, in (2+1)d, we have a phase where the information can be recovered with unlocated errors, which sits deep within the spin glass phase.

We note that the (three-dimensional) percolation threshold for the circuit in Fig.~\ref{fig:decoding_result_2+1d} at $p_{ZZ}^{\mc{M}} = 0.60$ is approximately $p^{\rm err}_c  \approx 0.74$ (from numerics), and the correlation length exponent is $\nu^{\rm perc}(d=3) \approx 0.88$~\cite{dengyoujin2014threeDpercolation}.
The percolation transition and the ``decoding transition'' are well separated and, {as expected}, appear to be in different universality classes.



\subsection{Performance of the decoder with faulty measurements on the (2+1)d repetition code \label{sec:faulty_meas}}

\begin{figure}[t]
    \centering
    \includegraphics[width=\columnwidth]{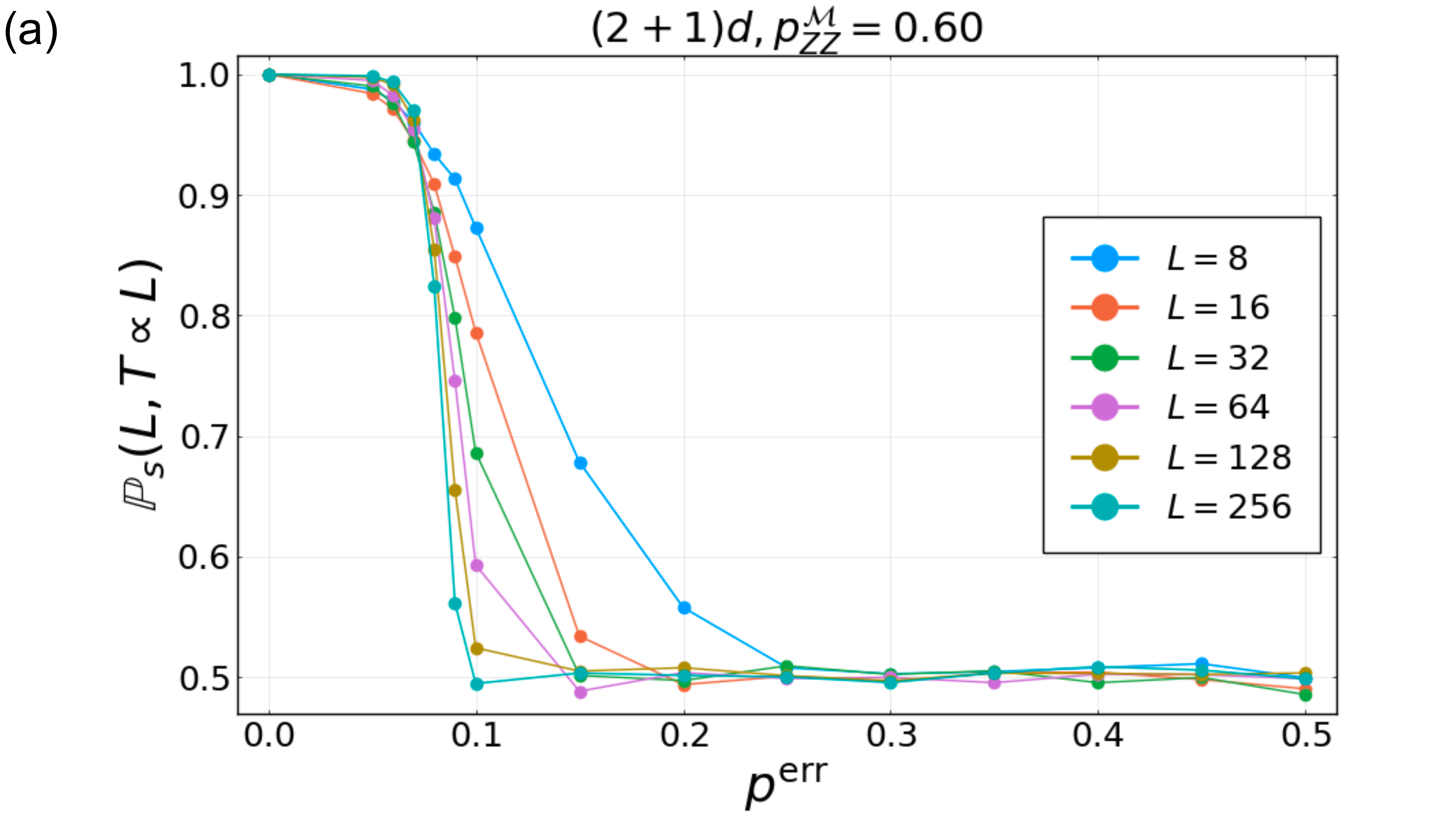}
    \includegraphics[width=\columnwidth]{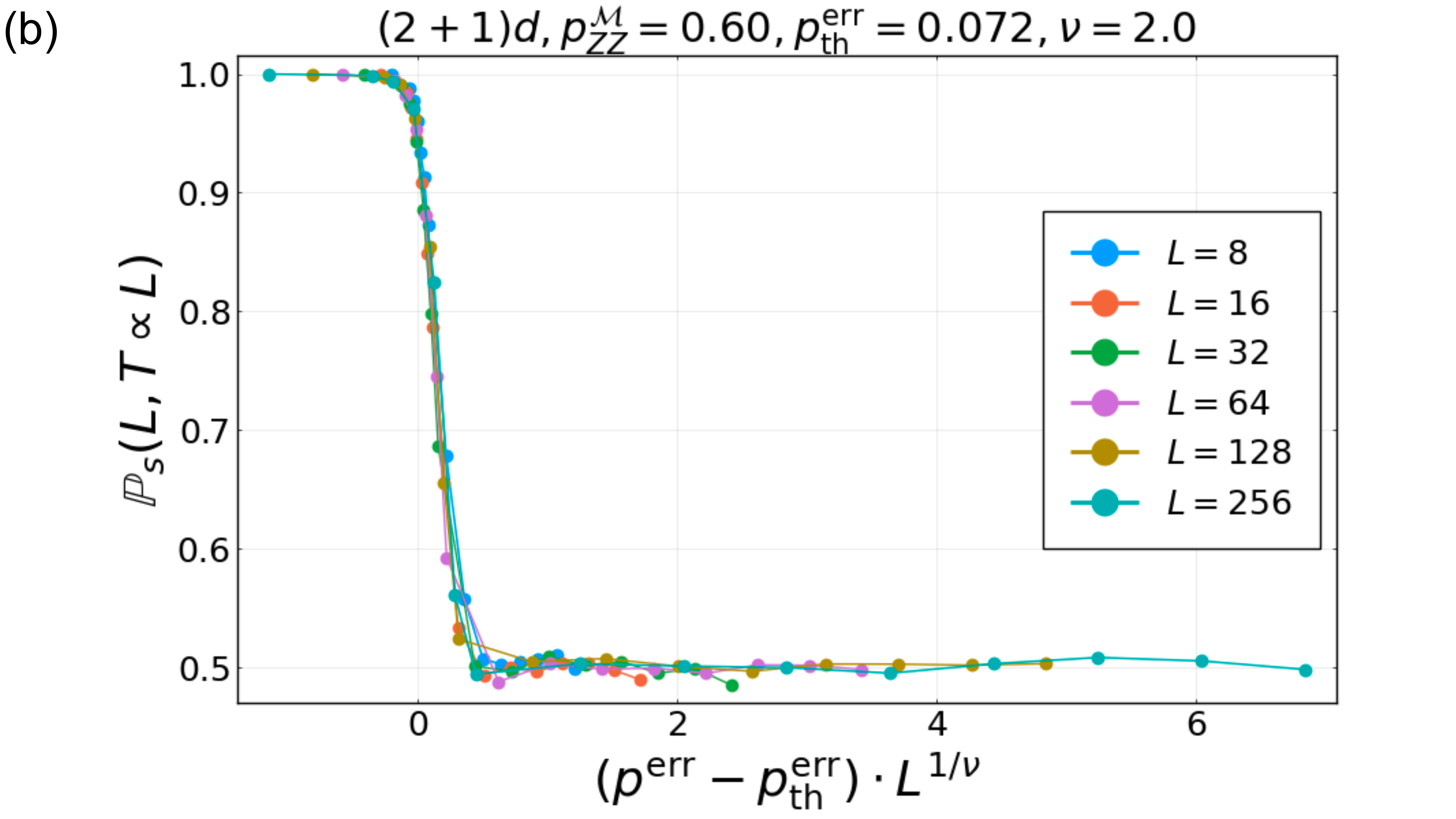}
    \caption{Numerical results for the decoder in Sec.~\ref{sec:unlocated_error_2plus1d}, for the (2+1)d repetition code with both bit-flip errors and faulty measurements.
    Here, the error model is the classical stochastic dynamics generated by random bit flips with probability $p^{\rm err}$, as well as random flips of measurement results, also with probability $p^{\rm err}$.
    The results are similar to those with perfect measurements (see Fig.~\ref{fig:decoding_result_2+1d}), except for a smaller $(p^{\rm err})_{\rm th}$.
    }
    \label{fig:P_decode_T_L_2+1d_Faulty}
\end{figure}

Here, we test the ``robustness'' of the decoder in Sec.~\ref{sec:unlocated_error_2plus1d}, by allowing a finite fraction of the check operator measurements in the bulk to be ``faulty''.\footnote{We still assume the measurements on the final state are perfect; see the previous footnote.}
We focus on (2+1)d, where a finite threshold was found for perfect measurements (see Fig.~\ref{fig:decoding_result_2+1d}).

To access larger system sizes, we again consider the classical bit-flip dynamics.
As before, we denote the probability of a bit-flip to be $p^{\rm err}$.
For simplicity, we take the probability of a ``faulty'' measurement to be also $p^{\rm err}$.

The numerical results are shown in
Fig.~\ref{fig:P_decode_T_L_2+1d_Faulty}, where we similarly find a finite $(p^{\rm err})_{\rm th}$, and a consistent fit for $\nu$ as compared to Fig.~\ref{fig:decoding_result_2+1d}.
Thus, the effect of faulty measurements seems ``irrelevant'' -- it reduces the error threshold, but does not change the universality class of the ``decoding transition''.

\subsection{Comparing the classical error model with Clifford circuits \label{app:benchmark}}

\begin{figure}[h]
    \centering
    \includegraphics[width=\columnwidth]{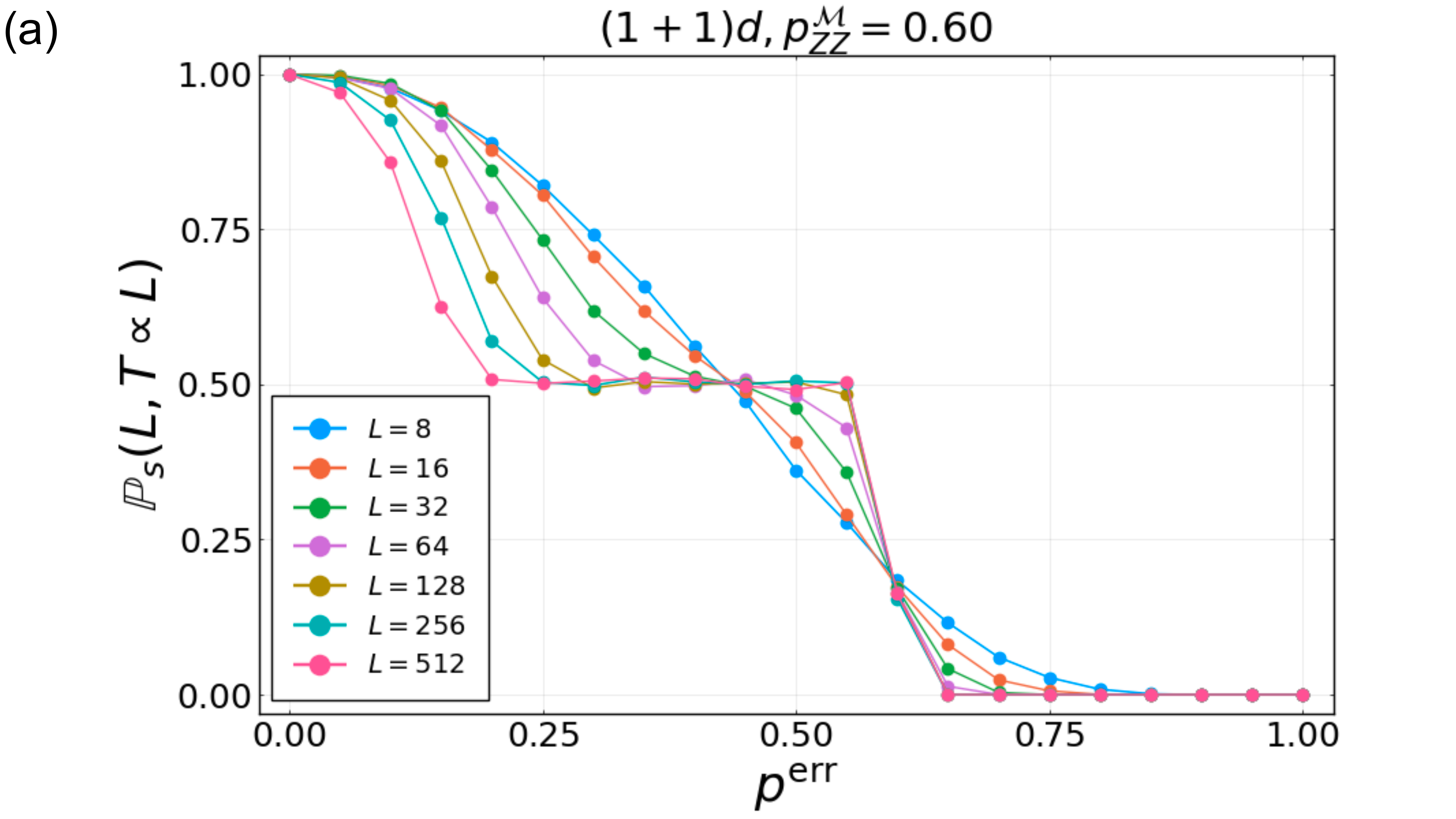}
    \includegraphics[width=\columnwidth]{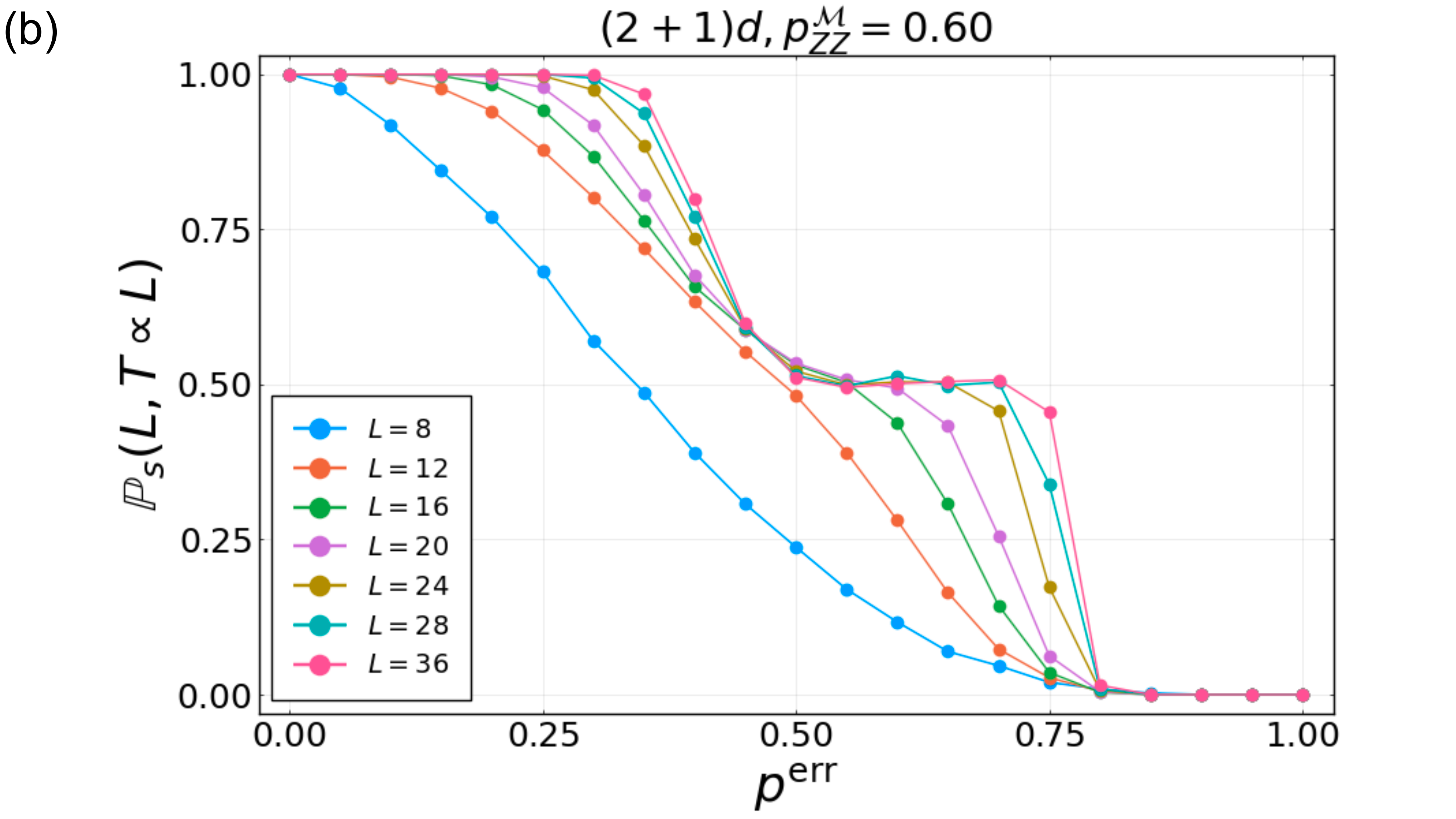}
    \caption{Performance of the decoder in Sec.~\ref{sec:unlocated_error} and Sec.~\ref{sec:unlocated_error_2plus1d} with unlocated errors, on measurement histories generated from the baseline circuit [Fig.~\ref{fig:z2_circuit}], for the repetition code (a) in (1+1) dimensions and (b) in (2+1) dimensions.
    Here the quantum time evolution takes the form of a Clifford circuit, which we simulate using the stabilizer formalism.
    They should be compared with Figs.~\ref{fig:decoding_result_1+1d}, \ref{fig:decoding_result_2+1d}, for which the measurement histories are generated by a classical bit-flip dynamics.
    As far as the success probability of decoding $\mb{P}_s$ is concerned, the baseline circuit dynamics and the classical bit-flip dynamics are very similar.
    However, for the circuit dynamics here, we also see evidences of the percolation transition, and when $p^{\rm err} > (p^{\rm err})_c$, the initial state can never be recovered, and $\lim_{L \to \infty} \mb{P}_s(L, T \propto L) \to 0$.  
    }
    \label{fig:P_decode_CLF}
\end{figure}

In Sec.~\ref{sec:unlocated_error} and Sec.~\ref{sec:unlocated_error_2plus1d}, we tested the performance of the decoder by generating the measurement history by a classical dynamics (see Figs.~\ref{fig:decoding_result_1+1d}, \ref{fig:decoding_result_2+1d}).
Here, we confirm that the decoder performs similarly when the measurement history is taken from running a {fully quantum} Clifford circuit.

We simulate the baseline circuit [Fig.~\ref{fig:z2_circuit}], and take $\ket{\mathbf{0}}$ to be the initial state.
The parameters are similar to those in Fig.~\ref{fig:z2_circuit}, but here we take, for simplicity, $p^{\rm err} = p_X^\mc{E}$, $p_X^\mc{M} = 0$, and allow $p_{ZZ}^\mc{M}$ and $p^{\rm err}$ to vary independently (dropping the condition $p_{ZZ}^\mc{M} + p^{\rm err} = 1$).
While running the circuit, we collect the results of the $ZZ$ measurements, and place them on the horizontal bonds of a square lattice to obtain the ``backbone subgraph'', as in Fig.~\ref{fig:path_sum}.
At the end of the time evolution, we measure all $ZZ$ operators, and apply a decoding unitary based on the sign of $f(v, T)$, defined in Eq.~\eqref{eq:path_sum}.
We then compare the resultant state with $\ket{\mathbf{0}}$, and declare success if they agree.

Running this simulation many times, we get an estimate for the success probability $\mb{P}_s$, shown in Fig.~\ref{fig:P_decode_CLF}.
We focus on the case where $T \propto L$, where we get results similar to Figs.~\ref{fig:decoding_result_1+1d}, \ref{fig:decoding_result_2+1d}, namely $(p^{\rm err})_{\rm th} = 0$ for the (1+1)d repetition code, and $(p^{\rm err})_{\rm th} > 0$ for the (2+1)d repetition code.
Recall that for the repetition code, $\lim_{L \to \infty} \mb{P}_s(L, T \propto L) \to 1$ when $p^{\rm err} < (p^{\rm err})_{\rm th}$, and $\lim_{L \to \infty} \mb{P}_s(L, T \propto L) \to 1/2$ when $p^{\rm err} > (p^{\rm err})_{\rm th}$.

In addition, we also see the percolation transition at a larger value of $p^{\rm err}$, above which the success probability becomes $0$.
This is because when the check operator measurements are non-percolating, the final state of the circuit will always be mixed, {and thus different from the initial (pure) state}.

We also tested the performance of the decoder on other stabilizer initial states, and also on non-stabilizer initial states for smaller system sizes, where we obtained similar results.


\begin{figure*}[t]
    \centering
    \includegraphics[width=.48\textwidth]{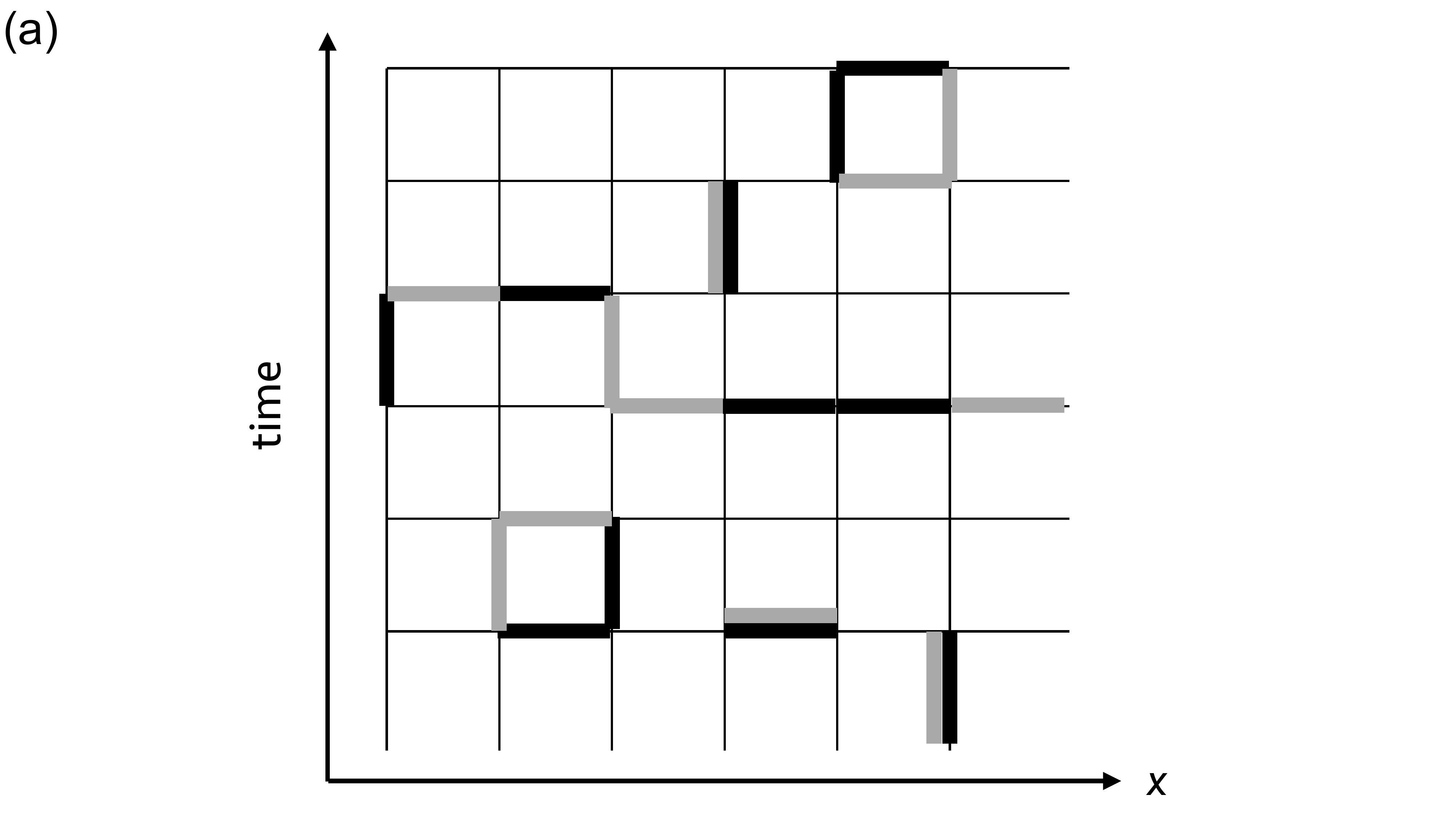}
    \includegraphics[width=.48\textwidth]{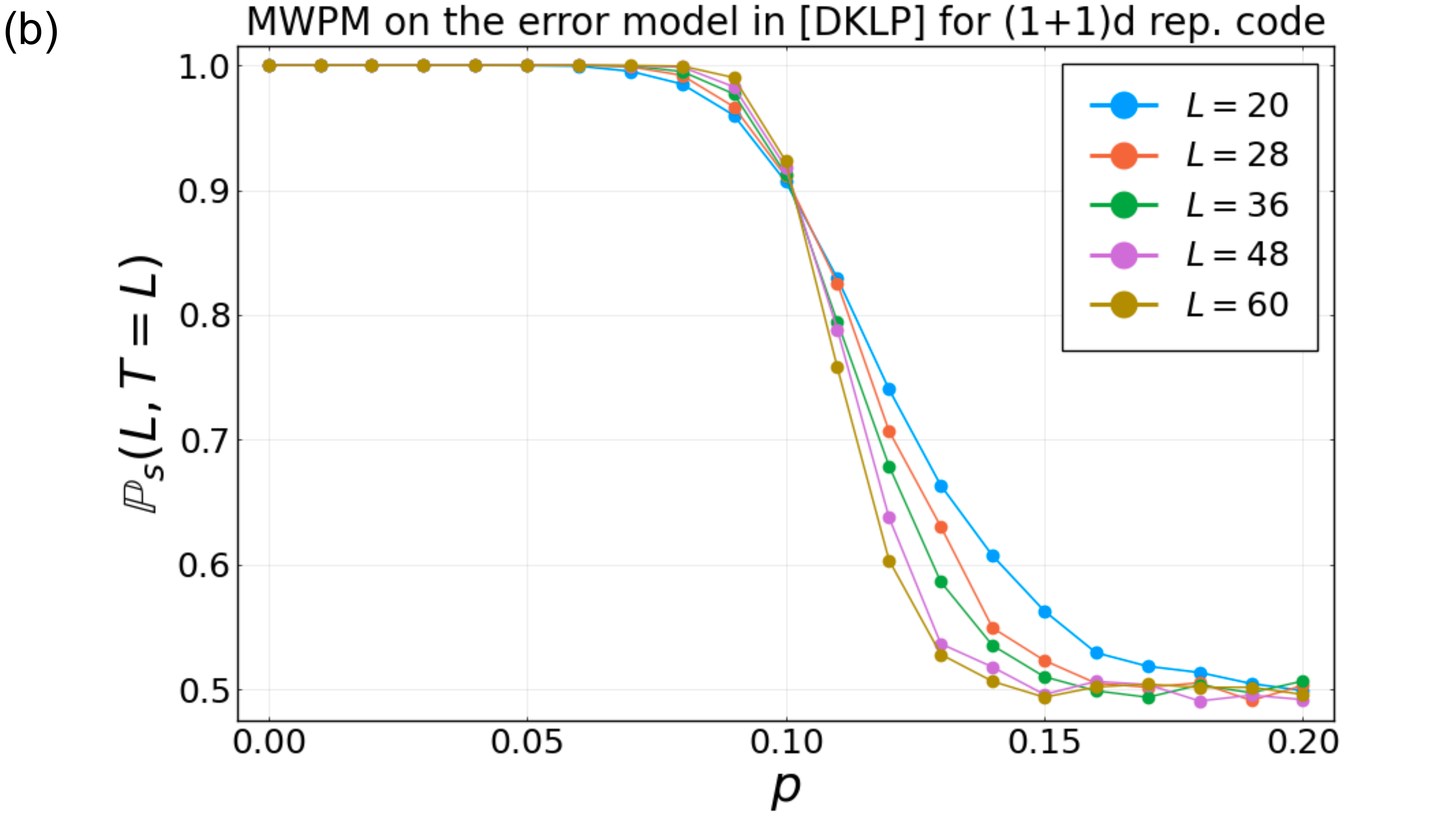}
    \includegraphics[width=.48\textwidth]{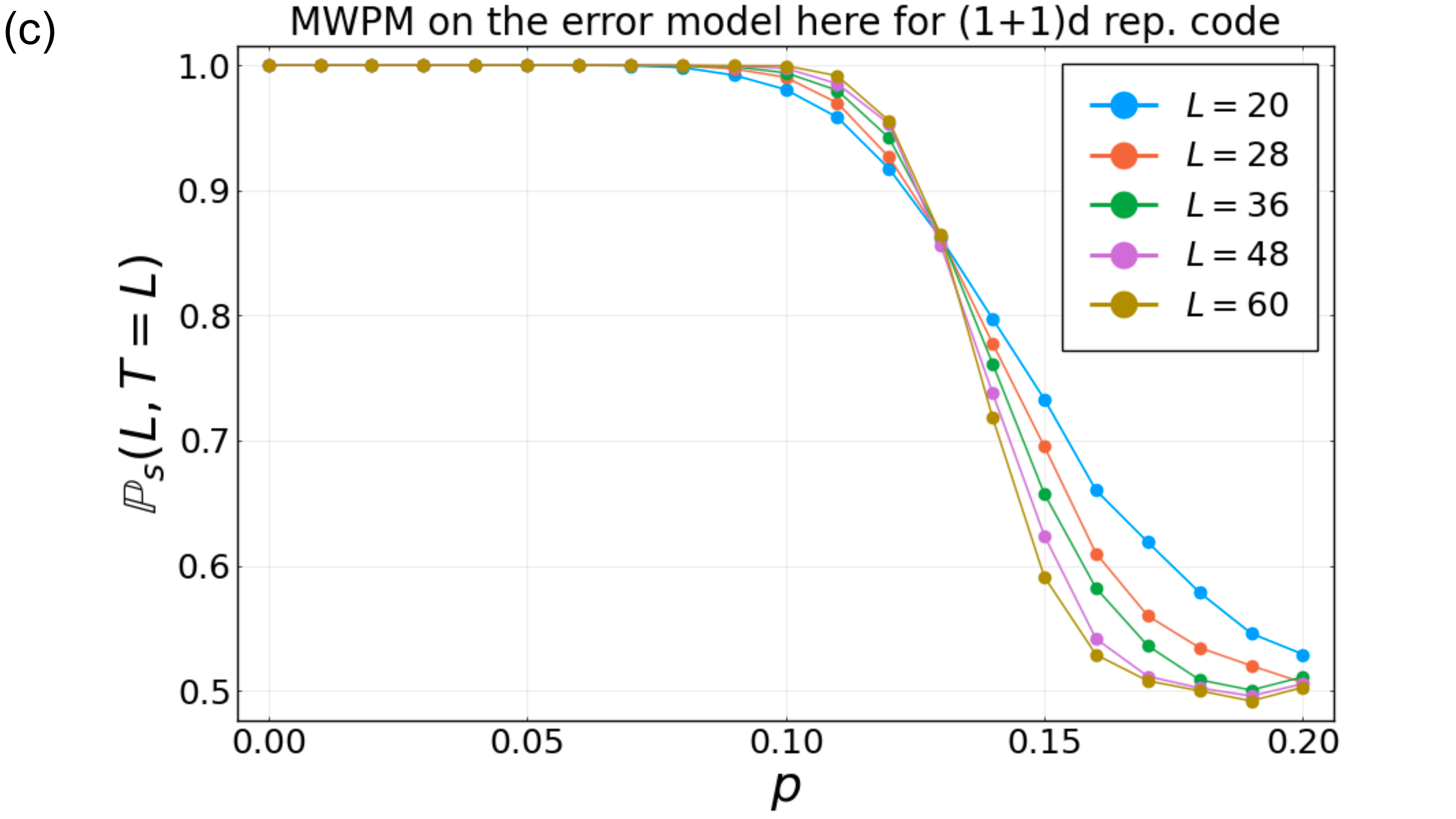}
    \includegraphics[width=.48\textwidth]{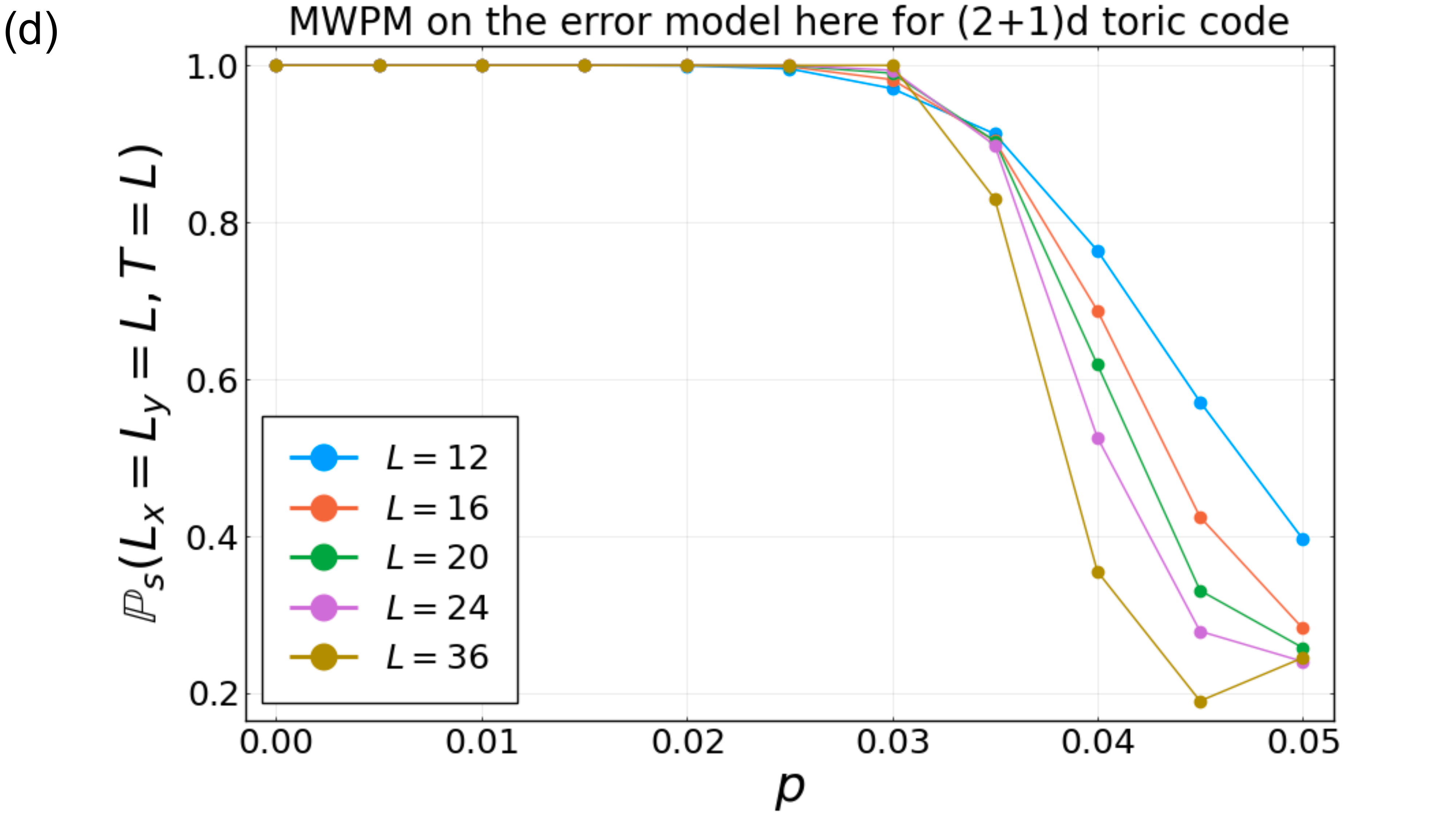}
    \caption{The MWPM decoder.
    (a) A graphical representation of the error model in Ref.~\cite{DKLP2001topologicalQmemory}, by Dennis, Kitaev, Landahl, and Preskill (DKLP).
    The square lattice is periodic in the spatial direction, and has the topology of a cylinder.
    The black bonds form the so-called ``error chains'', which represent the actual error history.
    The qubits now live on each square, the vertical bonds represent faulty measurements, and the horizontal bonds represent bit-flip errors.
    The endpoints of the error chains are Ising vortices that need to be paired up.
    The gray bonds represent a possible such ``perfect matching'', which is also an estimate of the error history, that will give the same syndrome as the actual error history.
    Together, the black and gray bonds form closed loops, and the decoding is successful if the chains wraps the cylinder around for an even number of times.
    Refs.~\cite{DKLP2001topologicalQmemory, preskill2002MWPMnumerics} showed that the algorithm that finds the perfect matching with the minimal weight has success probability $1$ when the error rate is below a finite threshold.
    (b)
    The MWPM decoder on the (1+1)d repetition code with $p_{ZZ}^\mc{M} = 1$, where we take the bit flip error and faulty measurements to both occur at probability $p$, for convenience.
    We also take $T = L$.
    The MWPM gives a finite error threshold, as expected.
    (c)
    The MWPM decoder on the (1+1)d repetition code with $p_{ZZ}^\mc{M} < 1$, where we assume the $ZZ$ check operator measurements are perfect, and the only errors are the bit-flip errors.
    Taking $T = L$ and $p \coloneqq 1-p_{ZZ}^{\mc{M}} = p^{\rm err}$, we find a threhold $p_{\rm th} \approx 0.13$.
    (d)
    The MWPM decoder on the (2+1)d toric code with $p_{\Box}^\mc{M} < 1$, where we assume the plaquet check operator measurements are perfect, and the only errors are the bit-flip errors (see Appendix~\ref{app:dynamical_toric_code}).
     Taking $T = L_x = L_y = L$ and $p \coloneqq 1-p_{\Box}^{\mc{M}} = p^{\rm err}$, we find a threhold $p_{\rm th} \gtrsim 0.03$.
    }
    \label{fig:MWPM}
\end{figure*}

\subsection{
Decoding with Minimal Weight Perfect Matching (MWPM) \label{sec:MWPM_numerics}}


The dynamics of the repetition code in a time duration proportional to the linear size of the code block was previously considered in Refs.~\cite{DKLP2001topologicalQmemory, preskill2002MWPMnumerics}.
There, all of the check operators are measured at each time step, and in addition to bit-flip errors, there are also errors due to faulty measurements.
Using a mapping of the dynamics to a random bond Ising model (RBIM) in (d+1) spatial dimensions, it can be shown that there is a finite error threshold in (1+1)d dimensions, given an optimal decoding algorithm.
The optimal algorithm finds the {most} likely error history\footnote{Or more accurately, its homology class in spacetime,
since multiple error histories can belong to the same ``homology class''.} given the syndrome measurements.
This amounts to finding the homology class with the lowest \emph{free energy} in the statistical mechanics model, which is computationally hard.
The authors instead employed a sub-optimal decoder which finds the configuration with the lowest \emph{energy}, or equivalently a ``minimal weight perfect matching'' (MWPM)~\cite{edmonds_1965_blossom} of point-like Ising vortices, that can be solved in polynomial time.
Despite being sub-optimal, the MWPM decoder can give a finite threshold in (1+1)d, as shown by both analytical estimates~\cite{DKLP2001topologicalQmemory} and numerics~\cite{preskill2002MWPMnumerics}.

Below, we reproduce a numerical result in Ref.~\cite{preskill2002MWPMnumerics} of the MWPM decoder on the repetition code in (1+1)d.
In addition, we adapt the MWPM decoder to our case, namely with check operator measurements at a rate less than $1$, for both the repetition code in (1+1)d and the toric code in (2+1)d.
We find a finite error threshold in both cases, as {we next discuss}.

\subsubsection{MWPM for (1+1)d repetition code}

In this section, we compare the performance of the decoder proposed in Sec.~\ref{sec:unlocated_error} and Sec.~\ref{sec:unlocated_error_2plus1d} with the MWPM decoder.
To start with, we briefly introduce the error model for the repetition code in (1+1)d, shown in Fig.~\ref{fig:MWPM}(a).
We refer the reader to Refs.~\cite{DKLP2001topologicalQmemory, preskill2002MWPMnumerics} for a detailed discussion.
Here, it is convenient to work on the dual lattice of the original square lattice, and place qubits at the center of each square.
Time runs in the vertical direction, upwards.
We will also take periodic spatial boundary condition, so that the lattice has the topology of a cylinder.

At each odd time step, each $ZZ$ check operator on neighboring qubits is measured; that is, $p_{ZZ}^\mc{M} = 1$.
These measurements can be thought of as performed on the vertical bonds.
These measurement can be faulty, with a probability we take to be $p^{\rm err}$.
In Fig.~\ref{fig:MWPM}(a), faulty measuremend are highlighted with black \emph{vertical} bonds.
At each even time step, each qubit can also experience a bit flip error, which occurs with a probability we also take to be $p^{\rm err}$.
The bit flip errors are highlighted with black \emph{horizontal} bonds in Fig.~\ref{fig:MWPM}(a).

The black bonds form one-dimensional ``error chains'' on the square lattice.  {Under a mapping to a 2d random bond Ising model with spins living on the plaquettes, these error chains correspond to negative bonds ~\cite{DKLP2001topologicalQmemory, preskill2002MWPMnumerics}}.
The end points of the ``error chains" can then be identified as Ising ``vortices". 
There is always an even number of such Ising vortices.

{The locations of the Ising vortices can be deduced from the syndrome measurements of the nearest neighbor $ZZ$ check operators.  The collection of non-trivial syndrome measurements with $ZZ=-1$ can be drawn with vertical bonds (not depicted Fig.~\ref{fig:MWPM}(a)) and form ``syndrome chains" running vertically.  The end points of these syndrome chains coincide with the end points of the error chains, both ending at the Ising vortices.}

The MWPM {scheme} finds a perfect matching between pairs of Ising vortices with the minimum weight, where the weight between a pair of Ising vortices is given by their lattice (Manhattan) distance. 
In Fig.~\ref{fig:MWPM}(a), paired Ising vortices are connected by gray bonds.   {In the Ising model the ``gray bond chains" correspond to domain walls, and the MWPM is effectively minimizing the length of the domain walls, i.e. minimizing the energy of the random bond Ising model.}

The gray bonds alone may be understood as the algorithm's estimate of the error history, with horizontal gray bonds being bit-flip errors, and vertical ones faulty measurements, and a decoding can be subsequently carried out based on this estimate. 

Putting the black bonds and the gray bonds together, we obtain closed loops on the cylinder, {which in the Ising model mapping enclose
opposite domains of Ising spins.}
The decoding will be successful if the number of \emph{non-contractible} loops is even.
As can be shown from the mapping to the Ising statistical mechanics model~\cite{DKLP2001topologicalQmemory, preskill2002MWPMnumerics}, the probability of success of the MWPM decoder is $1$ for $p^{\rm err}$ below a finite threshold. {The decodable phase below threshold corresponds to the ferromagnetic phase of the Ising model}.

The MWPM is a well-known problem in graph theory that can be solved by Edmond's algorithm~\cite{edmonds_1965_blossom} in time at least $O((LT)^3 \ln (LT))$~\cite{Kolmogorov09blossomv}. Here, we run the MWPM decoder for the (1+1)d repetition code, and the result is shown in Fig.~\ref{fig:MWPM}(b).  Notice that 
Fig.~\ref{fig:MWPM}(b) shows a finite $p^{\rm err}_{\rm th} \gtrsim 0.10$ {(the crossing point)}, consistent with Ref.~\cite{preskill2002MWPMnumerics}.

The MWPM decoder can also be adapted for the error model considered {in this paper}, with perfect measurements of check operators at a rate $p_{ZZ}^{\mc{M}} < 1$, by simply assuming the unmeasured $ZZ$ check operators all have the outcome $+1$.
This should be equivalent to having {some} faulty measurements, and the MWPM should then succeed for $p_{ZZ}^{\mc{M}}$ sufficiently close to $1$.
We calculate the success probability numerically, where for convenience we take $p \coloneqq 1-p_{ZZ}^{\mc{M}} = p^{\rm err}$ and $T = L$.
The results are shown in Fig.~\ref{fig:MWPM}(c), which also show a finite error threshold.

\subsubsection{MWPM for (2+1)d toric code}

The MWPM can also be used for decoding the toric code~\cite{kitaev1997} in (2+1) dimensions by matching point-like defects, and would similarly give a finite error threshold~\cite{DKLP2001topologicalQmemory, preskill2002MWPMnumerics}.
As for the repetition code, all of the check operators are measured at each time step.
Both bit-flip errors and phase errors are allowed here for the toric code, as well as faulty measurements of check operators.

Here, we test the MWPM on the toric code dynamics in (2+1)d in our error model (see Appendix~\ref{app:dynamical_toric_code}), with perfect measurements of ``plaquette'' check operators at a rate $p_{\Box}^{\mc{M}} < 1$, and with bit-flip errors only .
Phase-flip errors can be tracked by measuring the ``star'' operators, and they can be considered separately from the bit-flip errors~\cite{kitaev1997, DKLP2001topologicalQmemory, calderbank1995good, steane1996multiple}.
{Again, we assume that the unmeasured check operators all give the outcome $+1$.}
We similarly take $p \coloneqq 1-p_{\Box}^{\mc{M}} = p^{\rm err}$ and $T = L_x = L_y = L$.
The results are shown in Fig.~\ref{fig:MWPM}(d).
A finite error threshold is found, as expected~\cite{DKLP2001topologicalQmemory, preskill2002MWPMnumerics}.
Note that above the threshold, the success probability saturates to $1/4$, since there are now 4 different homology classes on the two-dimensional torus.

\subsubsection{Comparison with the MWPM decoder \label{sec:previous_work_preskill}}

Comparing our decoder with Refs.~\cite{DKLP2001topologicalQmemory, preskill2002MWPMnumerics}, there are differences in the error model, and also in the performance of the decoding algorithms.
As we have shown, 
the differences in the error models are somewhat unimportant.

(a)
First, faulty measurements can be included in the dynamics we consider here, and they do not change our conclusions qualitatively.
In particular, we have seen that there is a finite error threshold for the (2+1)d repetition code with faulty measurements, in Sec.~\ref{sec:faulty_meas}.

(b)
Second, while we are only measuring a fraction $p_{ZZ}^\mc{M} < 1$ of all check operators at each time step, we can still apply the MWPM to our error model, by  simply assuming that the unmeasured check operators (of fraction $1-p_{ZZ}^\mc{M}$) all have the outcome $+1$.
About half of these would be faulty, but the MWPM can nevertheless succeed in decoding for sufficiently small rate of faulty measurements $\approx \frac{1}{2}(1-p_{ZZ}^\mc{M})$.
As we have seen in Sec.~\ref{sec:MWPM_numerics}, for both the repetition code in (1+1)d and the toric code in (2+1)d,
the MWPM decoder gives a finite error threshold, as consistent with results in Refs.~\cite{DKLP2001topologicalQmemory, preskill2002MWPMnumerics}.

For the (1+1)d repetition code, the decoder in Sec.~\ref{sec:unlocated_error} is not preferable as compared to the MWPM decoder, since the former does not give a finite error threshold.\footnote{In this case, the decoder here might be useful in practice for code blocks of intermediate sizes, for its rather low time complexity of $O(TL)$, as compared to the MWPM which has a time complexity $O((TL)^3 \ln (TL))$~\cite{Kolmogorov09blossomv}.}
A generalization of this decoder -- from ``error-avoiding paths'' to ``error-avoiding membranes'' -- can be made for the (2+1)d toric code, as we explain in Appendix~\ref{app:dynamical_toric_code}.
However, we have not found a way of efficiently summing over sufficiently many such membranes that would give a finite threshold.
Thus, for the toric code in (2+1)d, the decoder in Sec.~\ref{sec:unlocated_error} is  not preferable, either.

For the (2+1)d repetition code, a natural generalization of the MWPM decoder looks for a perfect matching between line-like defects ({``Ising vortex loops" in a random bond (2+1)d Ising model}), connected by two dimensional surfaces ({Ising domain walls}), optimized so that they have minimal total area~\cite{DKLP2001topologicalQmemory, preskill2002MWPMnumerics}. However, this problem is computationally NP-hard.
Interestingly, on the other hand, our decoder in Sec.~\ref{sec:unlocated_error_2plus1d} is straightforward to implement {taking only polynomial time, linear in the circuits space-time ``volume", $O(L^d T)$}, and gives a finite threshold (see Fig.~\ref{fig:decoding_result_2+1d}).

Finally, we note two differences in the physics of the MWPM decoder and the decoder described here.

(a) 
The MWPM aims to optimize an ``energy'' by finding the ground state in the RBIM, whereas our decoder does not involve an optimization procedure or an explicit estimate of the error history, but is instead highly ``entropic'', in the sense that it receives contributions from a large number of paths, whether they are erroneous or not.
Its performance might be improved by associating weights to paths based on their geometry and prior knowledge of the error model~\cite{poulin2010RGdecoder}, which is itself an interesting future direction.

(b)
For the (2+1)d toric code, the basic objects appearing in the MWPM decoder are one-dimensional worldlines of topological defects, whereas our decoder involves summing over two-dimensional surfaces (see Appendix~\ref{app:dynamical_toric_code}).
On the other hand, for the (2+1)d repetition code, the MWPM decoder involves two-dimensional {domain walls}, while our decoder sums over one-dimensional paths.
To further explore this apparent ``duality'', a description of the underlying statistical mechanics model of our decoder will be needed.









\section{Outlook \label{sec:outlook}}

In this work, we explored two notions of phases in 
the $\mb{Z}_2$ baseline circuit, 
namely phases defined by their entanglement properties (``phases of entanglement''), such as the spin glass, the paramagnetic, and the trivial phases; and phases defined with respect to a given decoder.
Here, we comment on the relation between the two.

The spin glass phase has an immediate connection with the decoding problem: it is a necessary condition for successful decoding in the first place because of the nonzero channel capacity, and becomes a sufficient condition when the error locations are known.\footnote{Strictly speaking we have only demonstrated this for the baseline circuit, but it should also be correct for SG phases in generic $\mb{Z}_2$ circuits, e.g. the SG phases in Figs.~\ref{fig:phase_diagram_twochain}, \ref{fig:z2_circuit_plusU}, \ref{fig:bath_transition}, as long as the error-avoiding spanning paths percolate.}
On the other hand, other phases in the generic phase diagram, while rich and interesting in themselves, are unrelated to the decoding problem.
In particular, while $X$ measurements and $X$ dephasing are quite different and favor different phases (namely the paramagnetic phase and the trivial phase, respectively), they appear the same to the check operators measurements, and both become bit-flips on the repetition code.

Using the Kramers-Wannier duality, one might hope to try some kind of decoding on the paramagnetic phase by tracking the $X$ measurement outcomes, in a similar fashion as for the repetition code.
As the paramagnetic phase has a zero channel capacity (thus encodes zero logical qubits), this decoding problem is only suitable for an experimental demonstration of the PM order. 

In general, to make phases of entanglement and entanglement phase transitions in monitored dynamics experimentally relevant,
some kind of decoding seems necessary.\footnote{An exception would be monitored dynamics that is a spacetime dual of a unitary time evolution~\cite{ippoliti2020postselectionfree, grover2021rotation, ippoliti2021fractal}.}
Depending on details of the error model, the success of the decoder may or may not be directly inferred from the entanglement structure.
The experimental accessibility {of entanglement phases and transitions}, may be more severely limited by their decodability than their entanglement structure (e.g. channel capacity).
In this regard, stabilizer circuits are {appealing} not only because they are numerically simulable, but also because they might be decodable ({in polynomial time}).

Looking forward, it would be worth revisiting various measurement protected quantum phases~\cite{ippoliti2020measurementonly, sang2020protected} in monitored stabilizer circuits, and examine their properties as a code, including the corresponding decoding problem.
Entanglement can be a useful guidance in these explorations, e.g. in providing geometrical intuition.

Finally, we return to the volume law phase of the {random} hybrid Clifford circuit, which has a finite {code rate} and was shown to be a robust encoder~\cite{choi2019qec, gullans2019purification, fan2020selforganized, li2020capillary, li2021dpre}.
The final state can be decoded, provided that a classical description of all the unitaries and the measurement locations and outcomes is known to the decoder~\cite{monroe2021TrappedIonCliffordTransition}. 
It is an interesting question whether there is a \emph{robust} decoder of the volume law phase that can succeed even with errors, as in the $\mb{Z}_2$ circuit.\footnote{We note, however, that the two circuits can respond quite differently to errors.
Consider the role of local decoherence: for the $\mb{Z}_2$ circuit the symmetry-respecting decoherence acts as a mere ``dilution'' in the underlying percolation problem (see Sec.~\ref{sec:simple_z2_circuit} and Appendix~\ref{app:details_mapping_percolation}), but in random hybrid circuits (without any physical symmetry) decoherence acts as a ``magnetic field'' that breaks an emergent permutation symmetry~\cite{andreas2019hybrid, choi2019spin, li2020cft, li2021dpre, gullans2020lowdepth, ippoliti2021fractal, zhangpengfei2021magneticfield}, and will destroy the entanglement transition  when occuring at a finite density in the circuit bulk.}
An answer to this question will help us understand its relevance to practical fault tolerance applications;
we note a related discussion in Ref.~\cite{hastings2021FloquetCode}.



\section*{Acknowledgements}

We thank Ehud Altman, Yimu Bao, Michael Gullans, Tim Hsieh, David Huse, Shengqi Sang, Sagar Vijay, and Tianci Zhou for helpful discussions.
This work was supported by the Heising-Simons Foundation (Y.L. and M.P.A.F.),
and by the Simons Collaboration on Ultra-Quantum Matter, which is a grant from the Simons Foundation (651457, M.P.A.F.).
Use was made of computational facilities purchased with funds from the National Science Foundation (CNS-1725797) and administered by the Center for Scientific Computing (CSC). The CSC is supported by the California NanoSystems Institute and the Materials Research Science and Engineering Center (MRSEC; NSF DMR-1720256) at UC Santa Barbara.

\bibliography{refs}

\appendix


\section{Mapping the baseline circuit to percolation (Sec.~\ref{sec:simple_z2_circuit})
\label{app:details_mapping_percolation}}


Following the loop representation in Ref.~\cite{nahum2019majorana} in the $q=0$ limit, we map the circuit model in Fig.~\ref{fig:z2_circuit} to a bond percolation problem on the square lattice [Fig.~\ref{fig:perc_config}(a)].

Recall that in the $q=0$ limit, each $X$ measurement (occuring with probability $p_X^\mc{M} = 1-p_{ZZ}^\mc{M}$) corresponds to a \emph{broken vertical bond}, whereas each $ZZ$ measurement (occuring with probability $p_{ZZ}^\mc{M}$) corresponds to an \emph{unbroken horizontal bond}.
Thus, whether the bond is vertical or horizontal, it is unbroken with probability $p_{ZZ}^\mc{M}$.
The bond percolation transition is at $\left(p_{ZZ}^\mc{M}\right)_c = 1/2$, where each bond has an equal probability of being broken and unbroken.

With $q>0$, we represent $X$ dephasing with a new type of vertex, with red color and also a red, dashed, vertical \emph{decorated} ``incoherent'' bond across it.
We denote by $p_X^\mc{E} = q(1-p_{ZZ}^\mc{M})$ its probability of occuring.
We may also call the unbroken and undecorated bonds ``coherent''.

As we will see below, the SG phase is the percolating phase of the coherent, unbroken bonds, whereas the PM phase is the percolating phase of the broken bonds.
Here, the new ingredient is the presence of ``decorated bonds'' from dephasing channels.
Consequently, there can now be an intermediate ``trivial'' phase where neither the coherent unbroken bonds nor the broken bonds percolate.

\subsection{Decomposition of the dynamical state into quasi-GHZ states}

We now discuss the state of the circuit 
in the course of its time evolution.
When $q=0$, the state is always a product of GHZ states~\cite{sang2020protected, sang2020negativity} of the following type
\begin{align}
    \label{eq:GHZ_stab}
    \ket{\psi} = \prod_k \frac{1}{\sqrt{2}} \(\ket{0_{j_1} 0_{j_2} \ldots 0_{j_{n_k}} }
    +
    \ket{1_{j_1} 1_{j_2} \ldots 1_{j_{n_k}} }
    \),
\end{align}
where each factor is a GHZ state, and has stabilizers
\begin{align}
    \{X_{j_1} X_{j_2} \ldots X_{j_{n_k}}, Z_{j_1} Z_{j_2}, Z_{j_2} Z_{j_3}, \ldots, Z_{j_{n_k-1}} Z_{j_{n_k}} \}.
\end{align}
After mapping to bond percolation, two qubits on the upper boundary are in the same GHZ cluster \emph{if and only if} they are connected by a path of unbroken bonds in the bulk.
It thus defines an equivalence relation (denoted ``$\sim$''), and thus a partition of all the qubits.
Moreover, the bulk geometry requires that if $j_1 < j_2 < j_3 < j_4$ and $j_1 \sim j_3, j_2 \sim j_4$, we must have $j_1 \sim j_2 \sim j_3 \sim j_4$.

The effect of a $X$ dephasing on qubit $j_1$ is, in terms stabilizers in Eq.~\eqref{eq:GHZ_stab},
\begin{align}
    \label{eq:quasiGHZ_stab}
    &\quad
    \{X_{j_1} X_{j_2} \ldots X_{j_{n_k}}, Z_{j_1} Z_{j_2}, Z_{j_2} Z_{j_3}, \ldots, Z_{j_{n_k-1}} Z_{j_{n_k}} \} \nn
    \rightarrow& \quad
    \{X_{j_1} X_{j_2} \ldots X_{j_{n_k}}, 
    Z_{j_2} Z_{j_3}, \ldots, Z_{j_{n_k-1}} Z_{j_{n_k}} \},
\end{align}
where we have ``lost'' the stabilizer $Z_{j_1} Z_{j_2}$ to the dephasing channel.
Qubit $j_1$ now has classical correlations with qubits $j_{2, \ldots, n_k}$ via the string operator $X_{j_1} X_{j_2} \ldots X_{j_{n_k}}$, but no quantum entanglement.
We call the state in Eq.~\eqref{eq:quasiGHZ_stab} -- and its genrealizations when more qubits are dephased -- a ``quasi-GHZ state''.\footnote{Below, we will sometimes refer to the two types of stabilizers of a quasi-GHZ state as the ``$XXXX$'' and the ``$ZZ$'', for convenience.}

For concreteness, we take the circuit initial state a product of $\frac{1}{\sqrt{2}}\(\ket{0}_j + \ket{1}_j\)$ on each qubit, which has stabilizers $\{X_1, X_2, \ldots, X_L\}$, and itself a product of quasi-GHZ states.
At any stage of the circuit when $q>0$, the dynamical state of the system is always a product of quasi-GHZ states, as can be 
verified by induction.

As we have seen, the $X$ dephasing channel does not recluster the qubits, but only makes each cluster more incoherent.
Consequently, two qubits on the boundary are in the same quasi-GHZ cluster \emph{if and only if} they are connected by path in the bulk, where the bulk can now have both unbroken bonds and decorated incoherent bonds.

For what we discuss below, we always take periodic boundary conditions for the circuit, and focus on the steady state $T \gg L$, whence the circuit has the geometry of a very long cylinder.

\subsection{The spin glass susceptibility at SG-Trivial transition}

Recall that $\chi_{\rm SG}(L)$ is a sum of $\lv \avg{Z_i Z_j} \rv^2$.
Using the decomposition above, we see that the latter is $1$ if $Z_i Z_j$ is a stabilizer of the state, but $0$ otherwise.
It is tedious (but straightforward, again by induction) to convince oneself that this happens \emph{if and only if} qubits $i$ and $j$ (living on the upper boundary) are connected by a path containing \textit{only} coherent (black), unbroken (solid) bonds.
To $\chi_{\rm SG}$, the decorated ``incoherent'' bonds are effectively broken, as if an $X$ measurement is made.
Indeed, as one can check in Eq.~\eqref{eq:quasiGHZ_stab}, the effect of $X$ dephasing is identical to an $X$ measurement on $\chi_{\rm SG}$. 

It immediately follows that a phase transition in $\chi_{\rm SG}(L)$ occurs when the coherent unbroken bonds percolate.
From above, we see that the critical point is at $p_{ZZ}^\mc{M} = 1/2$.
Here $\lv \avg{Z_i Z_j} \rv^2 \propto |i-j|^{-2h_{\rm MI}}$ and $h_{\rm MI}$ is the scaling dimension of the boundary spin operator in critical percolation~\cite{buechler2020projectiveTFIM}.
Thus, compare Eq.~\eqref{eq:chi_SG_def}, we have
\begin{align}
    \chi_{\rm SG}(L) \propto L^{1-2h_{\rm MI}}, \quad p_{ZZ}^\mc{M} = 1/2.
\end{align}
Finite size scaling of this quantity near the critical point is shown in Fig.~\ref{fig:phase_diagram}(d), where we defined $\gamma_{\rm SG} = 1-2h_{\rm MI} = 1/3$.

\subsection{The paramagnetic susceptibility at PM-Trivial transition}

Recall that $\chi_{\rm PM}(L)$ is a sum of $\lv \avg{X_i X_{i+1} \ldots X_j} \rv^2$.
Again, this is $1$ when $X_i X_{i+1} \ldots X_j$ is a stabilizer, but $0$ otherwise.
Using the quasi-GHZ decomposition, one sees that 
this happens \emph{if and only if} no quasi-GHZ cluster has qubits inside $A = [i,j]$ and qubits in $\overline{A}$ at the same time (``no spanning quasi-GHZ'').
Moreover, focusing on the $XXXX$ string stabilizer of the quasi-GHZ cluster, one can check that this condition is the same regardless of whether the clusters are quasi-GHZ or GHZ.
Indeed, a $X$-dephasing never affects the value of $\chi_{\rm PM}(L)$. 
In comparison, an $X$ measurement breaks the $XXXX$ string apart, and can only increase $\chi_{\rm PM}(L)$. 

The ``no spanning quasi-GHZ'' condition above can be translated as follows in the percolation problem: (*) 
no qubits inside $A$ is connected to qubits inside $\overline{A}$, through paths that can now contain either \emph{coherent or incoherent} bonds.
Equivalently, all qubits in $A$ can be separated from those $\overline{A}$ by a cut that only goes through \emph{broken} bonds.
Therefore, the transition in $\chi_{\rm PM}(L)$ corresponds to one in which the broken bonds percolate.
A vertical broken bond occurs with probability $p_X^\mc{M} = (1-q)(1-p_{ZZ}^\mc{M})$, and a horizontal broken bond occurs with probability $1-p_{ZZ}^\mc{M}$.
The critical point of this anistropic percolation problem is at~\cite{SykesEssam1963}
\begin{align}
\label{eq:modelB_PM_phase_boundary}
    p_X^\mc{M} + \(1-p_{ZZ}^\mc{M}\) = 1
    \ \Leftrightarrow\ 
    p_X^\mc{M} = p_{ZZ}^\mc{M}
    \ \Leftrightarrow\ 
    q = \frac{1-2p_{ZZ}^\mc{M}}{1-p_{ZZ}^\mc{M}}.
\end{align}

Moreover, at the critical point, condition (*) implies that~\cite{buechler2020projectiveTFIM}
\begin{align}
    \lv \avg{X_i X_{i+1} \ldots X_j} \rv^2 \propto |i-j|^{-2h_{\rm MI}},
\end{align}
hence again
\begin{align}
    \chi_{\rm PM}(L) \propto L^{1-2h_{\rm MI}}, \quad 
    q = \frac{1-2p_{ZZ}^\mc{M}}{1-p_{ZZ}^\mc{M}}
\end{align}
Finite size scaling of this quantity near the critical point is shown in Fig.~\ref{fig:phase_diagram}(c), where we defined $\gamma_{\rm PM} = 1-2h_{\rm MI} = 1/3$.

At the PM critical point $p_X^\mc{M} = p_{ZZ}^\mc{M}$, the loop ensemble -- defined by cluster boundaries of \emph{broken} bonds -- becomes critical, and can be obtained from Fig.~\ref{fig:perc_config}(a) by rotating the red vertices by $90^\circ$, and treating them as if having a coherent solid bond.
The transfer matrix that generates the loop ensemble is thus~\cite{temperley_lieb_1971, blote_nienhuis_1989, jacobsen2009book}
\begin{widetext}
\begin{align}
\label{eq:TL_transfer_rescaled}
    e^{\hat{H}}
    =&
    \lz 
    \prod_{j\, {\rm even}} \bigg\{ [(1-p_X^\mc{M} - p_X^\mc{E}) + p_X^\mc{E}] \cdot \TLI\, + 
    p_X^\mc{M}
    \cdot \TLE \bigg\}_j
    \rz
    \cdot
    \lz
    \prod_{j\, {\rm odd}} \bigg\{
    [1-p_{ZZ}^\mc{M}] \cdot \TLI\, + p_{ZZ}^\mc{M} \cdot \TLE
    \bigg\}_j
    \rz \nn
    =&
    \lz
    \prod_{j\, {\rm even}} \bigg\{
    [1-p_{ZZ}^\mc{M}] \cdot \TLI\, + p_{ZZ}^\mc{M} \cdot \TLE 
    \bigg\}_j
    \rz
    \cdot
    \lz
    \prod_{j\, {\rm odd}} \bigg\{
    [1-p_{ZZ}^\mc{M}] \cdot \TLI\, + p_{ZZ}^\mc{M} \cdot \TLE 
    \bigg\}_j
    \rz.
\end{align}
\end{widetext}
Thus, it is related to the usual Temperley-Lieb repesentation of critical percolation (where $p_{ZZ}^\mc{M}=1/2$) by a rescaling of the ``temporal'' direction.
Such a rescaling
merely ``dilutes'' the loops, without affecting any of the universal properties.

\subsection{Scaling of mutual information}

\begin{figure}
    \centering
    \includegraphics[width=\columnwidth]{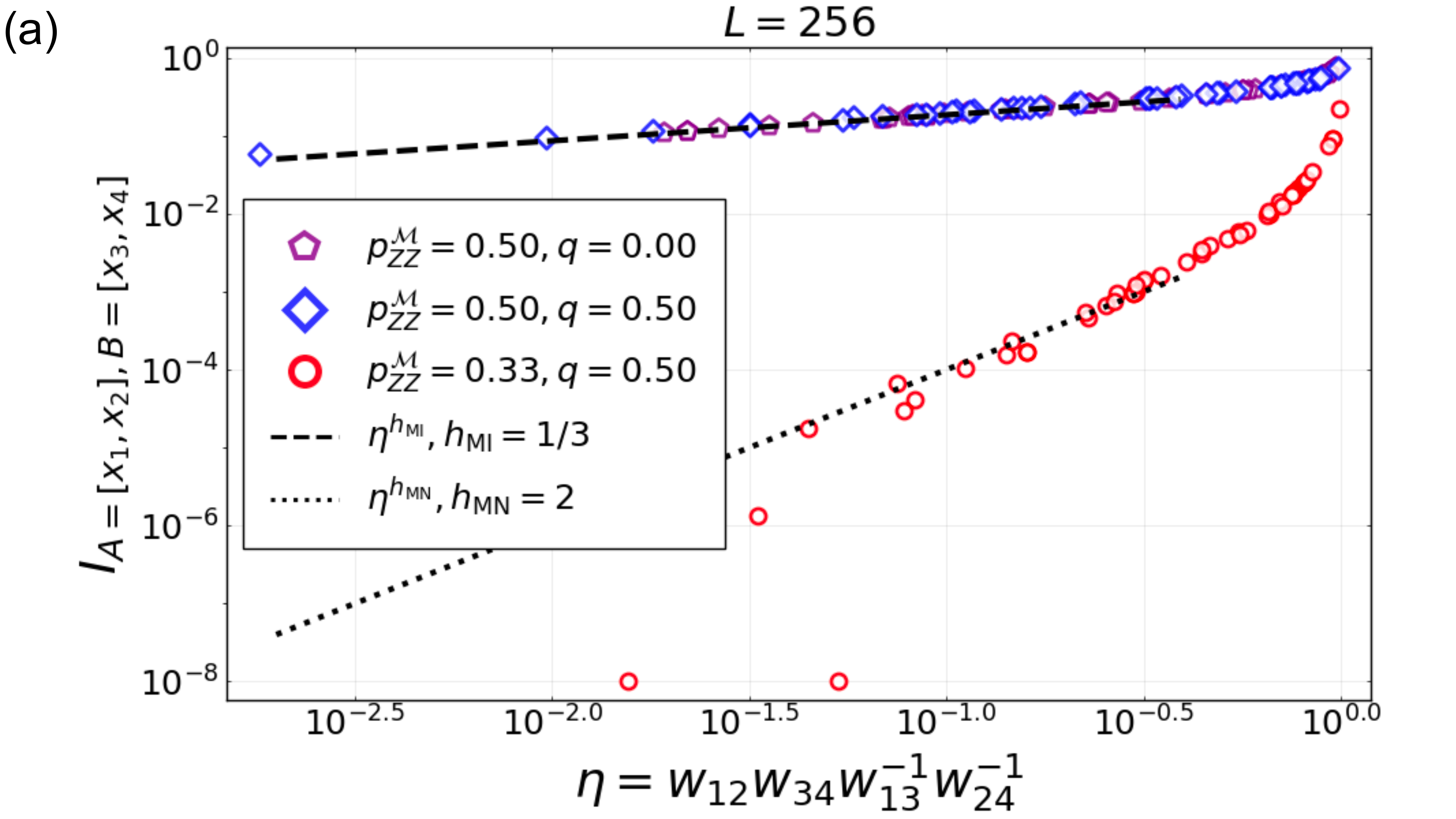}
    \includegraphics[width=\columnwidth]{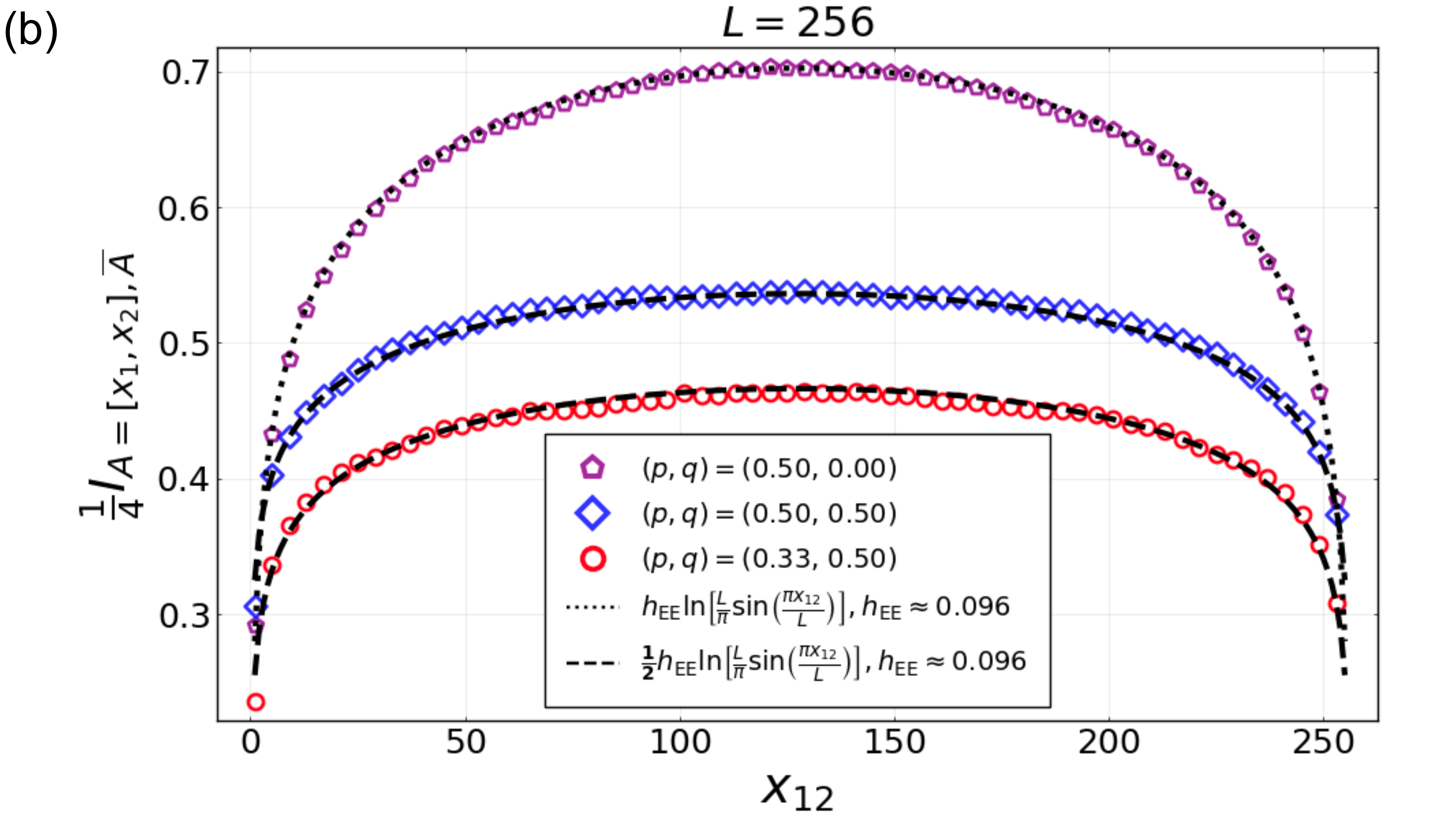}
    \caption{Mutual information $I_{A,B}$ between two disjoint regions $A = [x_1, x_2]$ and $B = [x_3, x_4]$ at the PM-Trivial transition (red circles), at the SG-Trivial transition (blue diamonds), and at the SG-PM transition (purple pentagons).
    In (a), we collapse data obtained by varying $\{x_j\}$ against the cross ratio $\eta$.
    The collapse confirms the presence of conformay invariance.
    The powerlaws of $I_{A,B}$ at small $\eta$ are universal scaling dimensions of critical percolation, which we call $h_{\rm MI}$ and $h_{\rm MN}$ following Ref.~\cite{sang2020negativity}, and are summarized in Eq.~\eqref{eq:I_AB_exp_summary} and Table~\ref{table:critical_exponents}.
    In (b), we plot the mutual information $I_{A, \ovl{A}}$ across a bipartition $(A, \ovl{A})$, versus the length of $A = [x_1, x_2]$, namely $L_A = x_{12}$.
    The results at the PM-Trivial and SG-Trivial transitions are consistent with Eq.~\eqref{eq:I_AAbar_critical_half},  with a universal coefficient that is half of that at the PM-SG transition (see Eq.~\eqref{eq:I_AAbar_critical_one})
    }
    \label{fig:baseline_MI_scaling}
\end{figure}

\subsubsection{Between two sites} 

Consider the mutual information between sites $i$ and $j$.
One can verify the following
\begin{enumerate}[(i)]
\item
When $i$ and $j$ are in different quasi-GHZ clusters, they have zero mutual information.
\item
When $i$ and $j$ are in the same quasi-GHZ cluster with $\ge 3$ qubits, they have mutual information $1$ \emph{if and only if} $Z_i Z_j$ is a stabilizer.
\item
When $i$ and $j$ are in the same quasi-GHZ cluster with exactly $2$ qubits, they always have at least mutual information $1$, but can have mutual information $2$ if $Z_i Z_j$ is also a stabilizer (in which case it becomes a GHZ cluster).
\end{enumerate}

At the SG transition $p_{ZZ}^\mc{M}=1/2$, 
\begin{enumerate}[(i)]
\item 
the second probability is $\propto |i-j|^{-2/3}$, as we have discussed above.
\item
The third probability is $\propto |i-j|^{-4}$ and is subleading.
Previously, this probability is related to the ``mutual negativity''~\cite{sang2020negativity} (see also Ref.~\cite{shi2020negativity}) in critical percolation.
\end{enumerate}

At the PM transition $p_X^\mc{M} = p_{ZZ}^\mc{M}$,
\begin{enumerate}[(i)]
\item 
The second probability decays exponentially in $|i-j|$ as $\lv \avg{Z_i Z_j} \rv^2$, whence the SG order parameter vanishes.
\item
For the third probability, its leading contribution is given by the correlation function $\avg{\TLE_i \TLE_j}$ in the loop ensemble generated by Eq.~\eqref{eq:TL_transfer_rescaled}, which is a boundary correlation function of stress tensors~\cite{sang2020negativity}.
Consequently, this
probability is still $\propto |i-j|^{-4}$.
\end{enumerate}

These results when applied to
two 
antipodal sites $A = [0, \epsilon]$ and $B = [L/2, L/2+\epsilon]$ become
\begin{align}
\label{eq:I_AB_exp_summary}
 I_{A,B} 
 \propto
 \begin{cases}
        L^{-2/3}, & \text{SG-Trivial transition $(p_{ZZ}^\mc{M} = 1/2)$} \\
        L^{-4}, & \text{PM-Trivial transition $(p_{ZZ}^\mc{M} = p_X^\mc{M})$} \\
        1, & \text{Inside SG phase} \\
        0, & \text{Inside PM and Trivial phases} \\
        L^{-2/3}, & \text{SG-PM transition $(p_{ZZ}^\mc{M} = 1/2, q = 0)$}
    \end{cases}.
\end{align}
The last line is a known result from Ref.~\cite{buechler2020projectiveTFIM}.

At the critical point, we consider general disjoint subregions $A = [x_1, x_2]$ and $B = [x_3, x_4]$ with arbitrary endpoints, where the mutual information becomes a function 
of the cross ratio $\eta$, due to the emergent conformal symmetry.
With periodic boundary conditions, we take $\eta = \frac{w_{12} w_{34}}{w_{13} w_{24}}$, where $w_{ij} = \sin\(\frac{\pi}{L} x_{ij}\)$ is the chord distance.\footnote{The chord distance follows from a conformal mapping of the semi-infinite cylinder to the upper half plane. When $A = [0, \epsilon]$ and $B = [L/2, L/2+\epsilon]$, the cross ratio scale as $\eta \propto L^{-2}$.}
The numerical results for $I_{A,B}$ with varying endpoints $\{x_j\}$
are shown in Fig.~\ref{fig:baseline_MI_scaling}(a).
The data collapse is consistent with conformal invariance, and the powerlaws of $I_{A,B}$ at small $\eta$ are consistent with Eq.~\eqref{eq:I_AB_exp_summary}.

\subsubsection{Between a bipartition $(A, \ovl{A})$}

Recall that for a stabilizer state \emph{in the clipped gauge}~\cite{nahum2017KPZ, li2019hybrid}, the bipartite mutual information $I_{A,\overline{A}}$ is the number of stabilizers that cross the cut between $A$ and $\overline{A}$.
For example, for a coherent GHZ state, a bipartition would have mutual information $2$, one from the ``$XXXX$'' string, and one from an ``$ZZ$'' operator, both crossing the cut.
For our decomposition of the dynamical state into a product of quasi-GHZ states, the stabilizers are already in the clipped gauge.

At the coherent critical point $(p_{ZZ}^\mc{M},q) = (1/2, 0)$, the length distribution of stabilizers -- counting both the ``$XXXX$'' and the ``$ZZ$'' Pauli strings -- follows an inverse square law with a universal coefficient~\cite{nahum2019majorana, sang2020negativity},
\begin{align}
    P(\ell) \approx \frac{\sqrt{3}}{2\pi} \ell^{-2},
\end{align}
which leads to the following expression of $I_{A,\overline{A}}$ in a periodic system, when $A = [x_1, x_2]$ has two endpoints:
\begin{align}
\label{eq:I_AAbar_critical_one}
    \frac{1}{4} I_{A, \overline{A}} =&\ \frac{1}{4} \times 2 \times \frac{\sqrt{3}}{2\pi} \times \ln 2 \times
    \ln \( \frac{L}{\pi} \sin \frac{\pi x_{12}}{L} \) \nn
    =&\ h_{\rm EE} \ln \( \frac{L}{\pi} \sin \frac{\pi x_{12}}{L} \).
\end{align}
Here, $h_{\rm EE} = \frac{\sqrt{3}}{4\pi} \ln 2 \approx 0.096$.
If we count the $XXXX$ and the $ZZ$ stabilizers separately, they will each have an inverse square law ``critical'' distribution, with the same coefficient $\frac{1}{2} \frac{\sqrt{3}}{2\pi}$, due to Kramers-Wannier duality.

With decoherence, the picture is slightly different, since the $XXXX$ and the $ZZ$ stabilizers are not always critical at the same time.
Again we consider the state decomposition into quasi-GHZ clusters, and calculate the mutual information for each cluster.
\begin{enumerate}[(i)]
\item
At the SG-Trivial transition, the unbroken coherent bonds are critical, whereas the unbroken bonds \textit{plus} the incoherent bonds are ``supercritical''.
As a consequence, there will be a few (order one) extensive quasi-GHZ clusters, each having a very long $XXXX$ string.
Thus, there is at most $O(1)$ contribution from $XXXX$ strings to any $I_{A, \overline{A}}$.

On the other hand, since the incoherent bonds are no different from broken bonds to the $ZZ$ operators, the length distribution of the $ZZ$s will still be critical, having an inverse square law with coefficient $\frac{1}{2}\frac{\sqrt{3}}{2\pi}$.

Thus, as compared with the coherent percolation at $p_{ZZ}^\mc{M} = 1/2, q = 0$, only half the stabilizers -- namely the $ZZ$s -- contribute to the mutual information.

\item
At the PM-Trivial transition, the broken bonds \textit{plus} the incoherent bonds are now critical, whereas the unbroken bonds alone are ``subcritical''.
Therefore, now the $XXXX$ strings will have a critical inverse square law length distribution with coefficient $\frac{1}{2} \frac{\sqrt{3}}{2\pi}$,\footnote{Note that 
this coefficient is universal number of the percolation universality class.
It characterizes the critical loop ensemble, and remains unchanged under rescaling of the temporal direction as in Eq.~\eqref{eq:TL_transfer_rescaled}.
Thus, we have the same coefficient despite at a value of $p_{ZZ}^\mc{M} = p_X^\mc{M}$ (the PM critical point) different than $1/2$.
} whereas most of the $ZZ$ operators are short, of length $O(1)$.

Thus, as compared with the coherent percolation at $p_{ZZ}^\mc{M} = 1/2, q = 0$, now the other half of the stabilizers -- namely the $XXXX$ strings -- contribute to the mutual information.
\end{enumerate}

These arguments 
imply that the coefficient of the mutual information at these two transitions are both half of the SG-PM percolation transition at $p_{ZZ}^\mc{M}=1/2, q=0$, namely
\begin{align}
\label{eq:I_AAbar_critical_half}
    \frac{1}{4} I_{A, \overline{A}} = \frac{1}{2} h_{\rm EE} \ln \( \frac{L}{\pi} \sin \frac{\pi x_{12}}{L} \).
\end{align}

We compute $I_{A,\ovl{A}}$ at the three transitions, and plot the results in Fig.~\ref{fig:baseline_MI_scaling}(b).
They are fully consistent with Eqs.~(\ref{eq:I_AAbar_critical_one}, \ref{eq:I_AAbar_critical_half}).

Inside the three phases, the bipartite mutual information $I_{A, \ovl{A}}$ always obeys an area law (data not displayed).


\subsection{Inclusion of $ZZ$ decoherence \label{sec:ZZ_dep_channel}}

\begin{figure}[t]
    \centering
     \includegraphics[width=.45\textwidth]{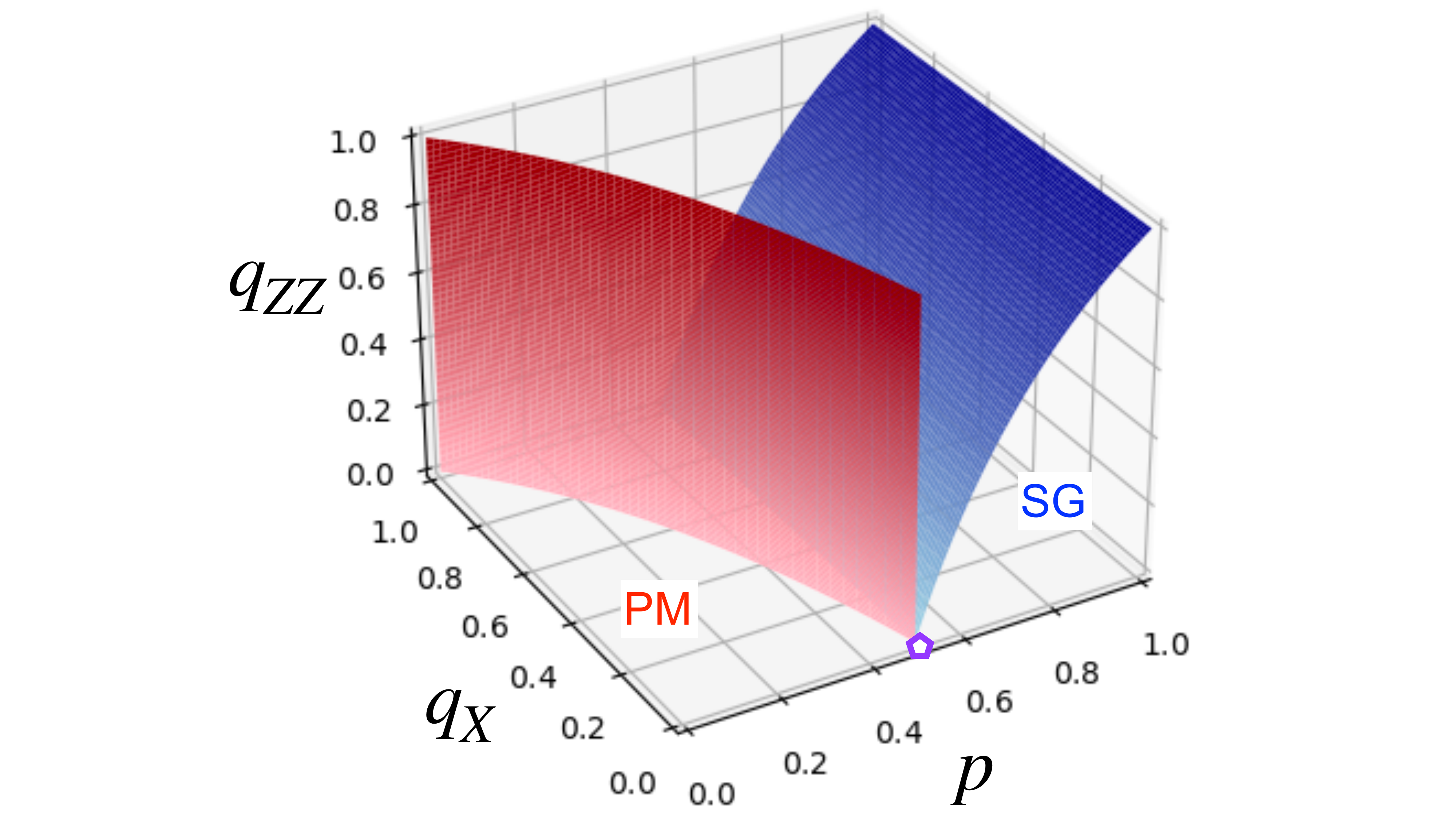}
\caption{The phase diagram of the model with both decoherences in $X$ and $ZZ$.
Here, the PM phase boundary is given by $q_X = \frac{1-2p}{1-p}$, whereas the SG phase boundary is $q_{ZZ} = \frac{2p-1}{p}$.
The phase transitions are still in the same universality class as when $q_{ZZ} = 0$.
}
    \label{fig:phase_diagram_3D}
\end{figure}

\begin{figure*}[t]
    \centering
    \includegraphics[width=\columnwidth]{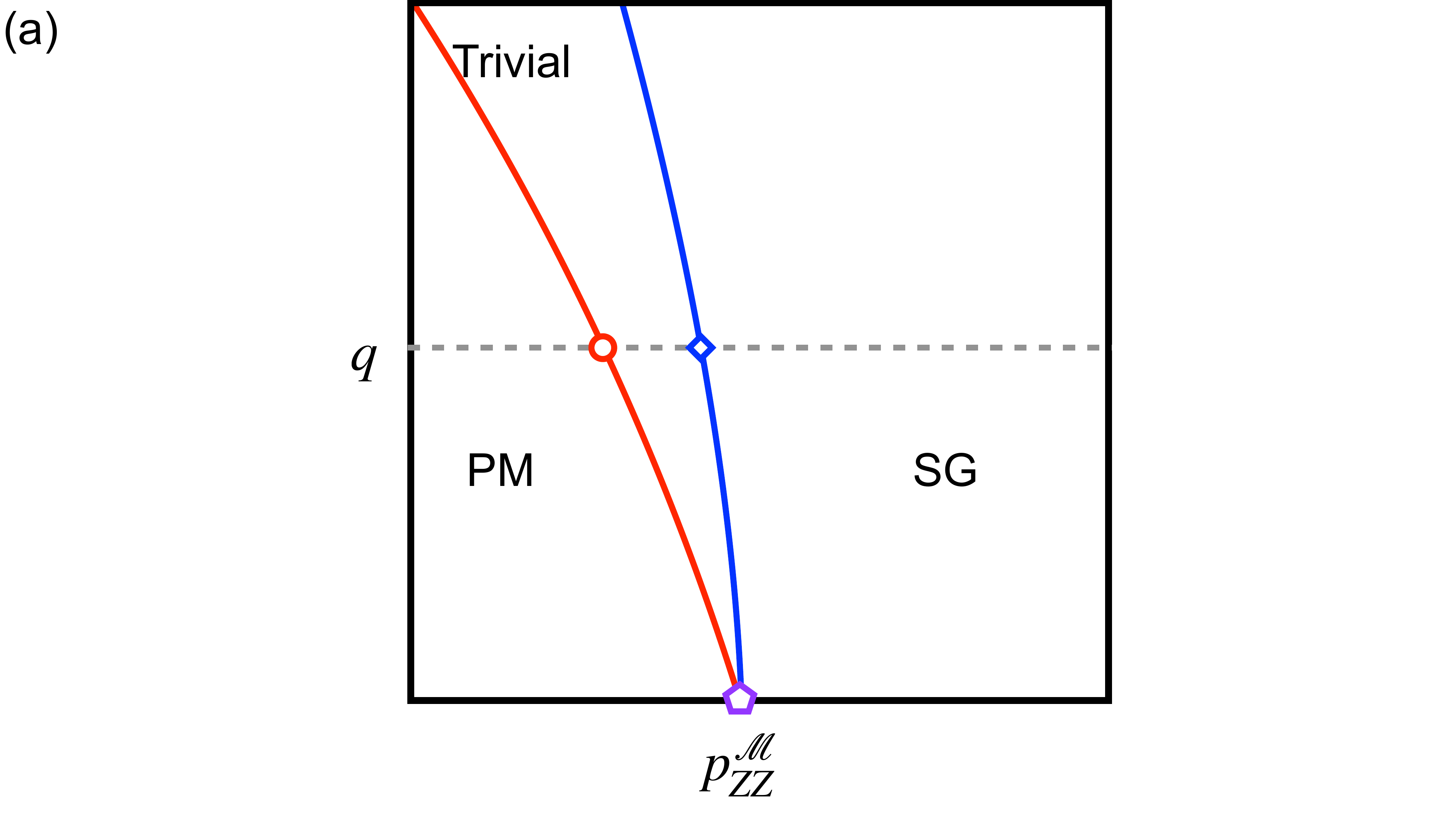}
    \includegraphics[width=\columnwidth]{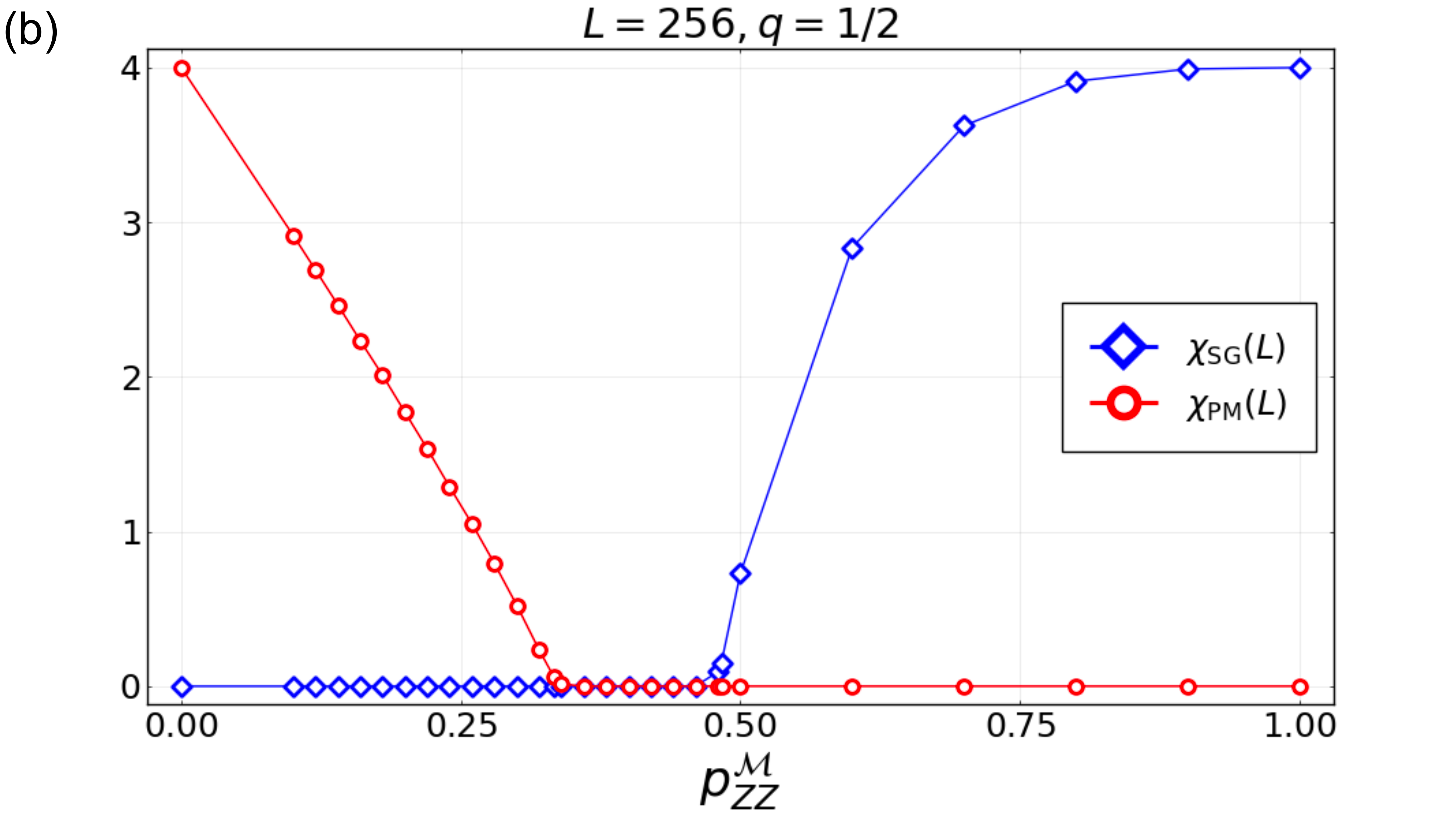}
    \includegraphics[width=\columnwidth]{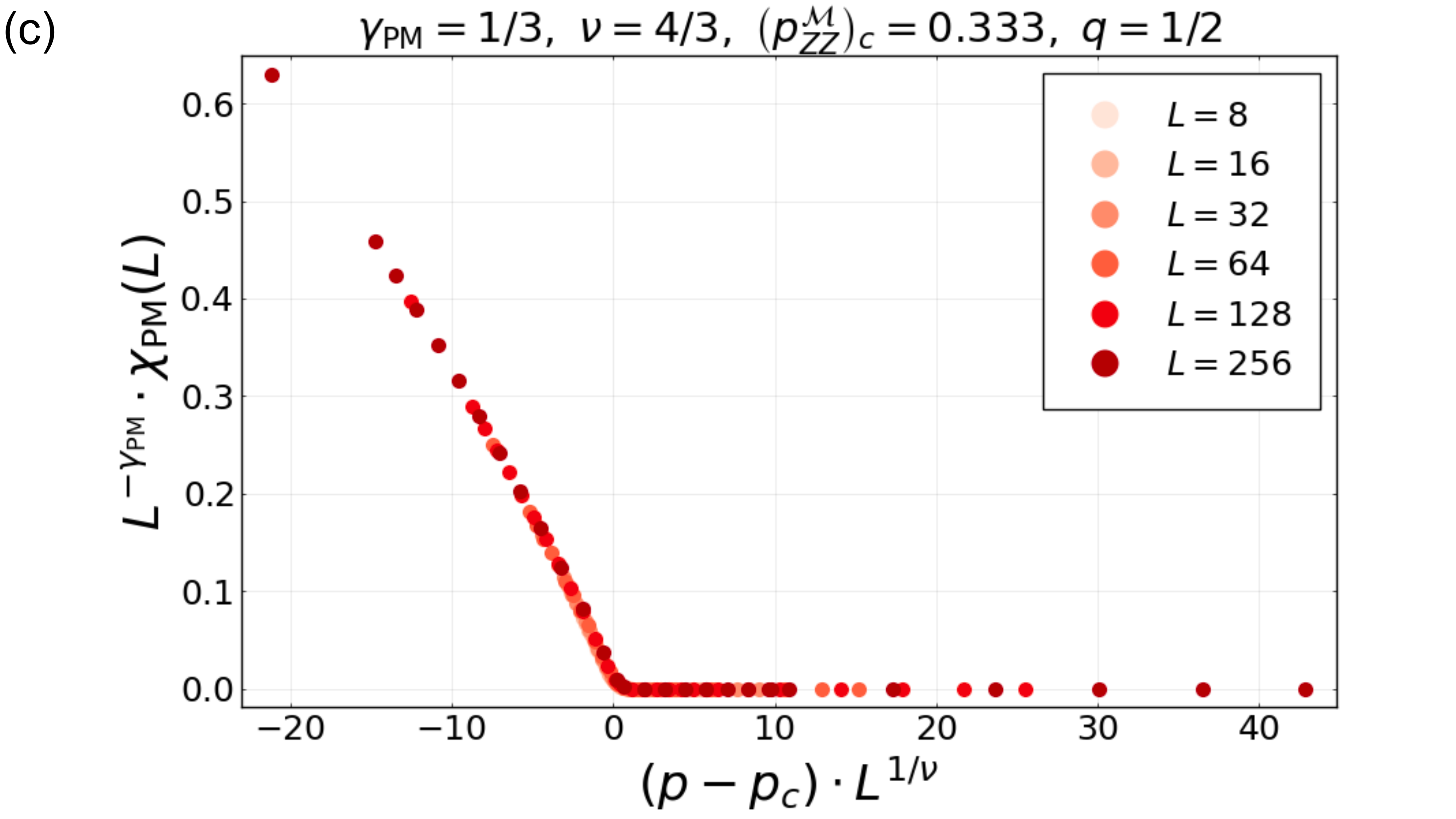}
    \includegraphics[width=\columnwidth]{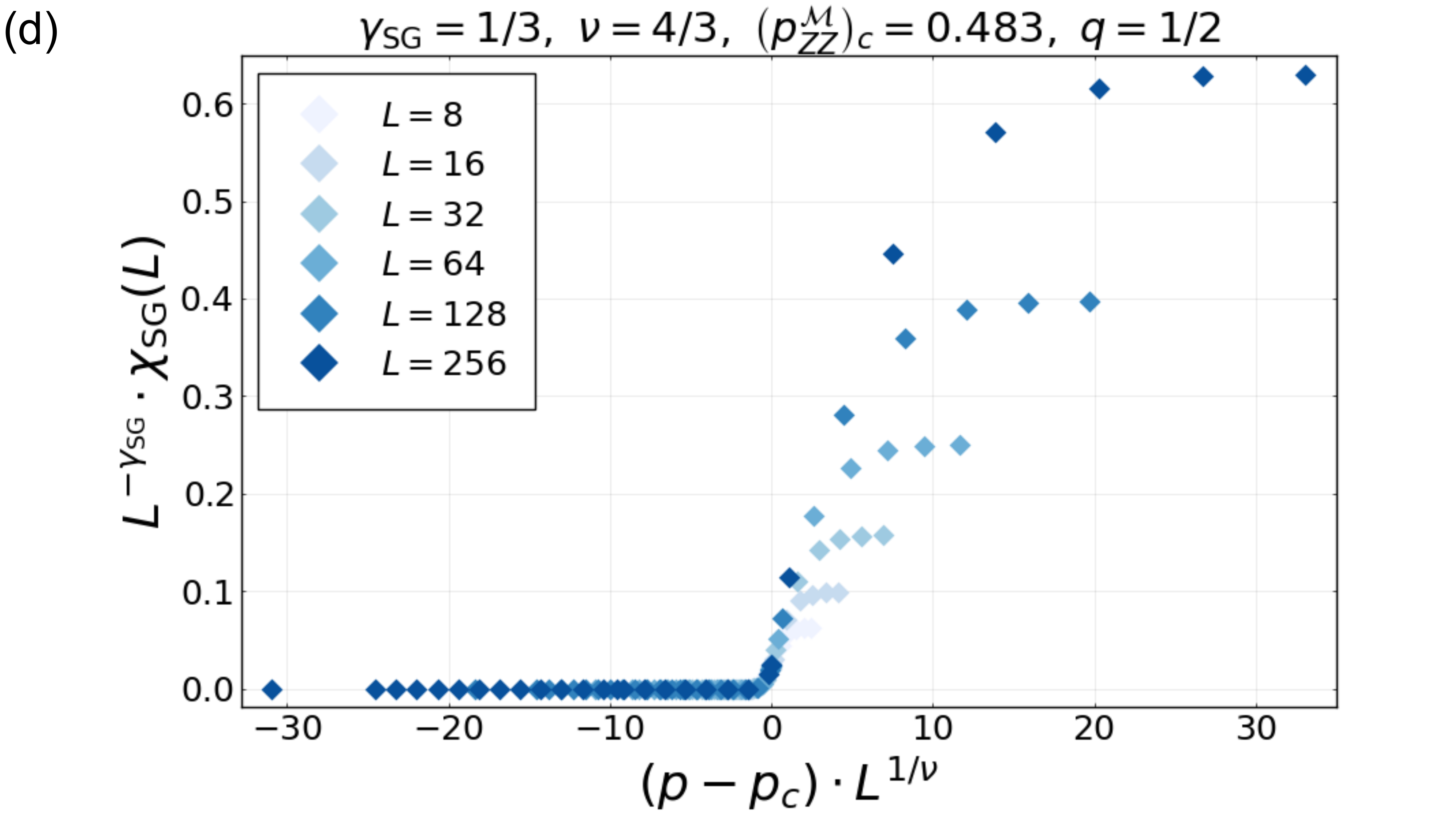}
    \includegraphics[width=\columnwidth]{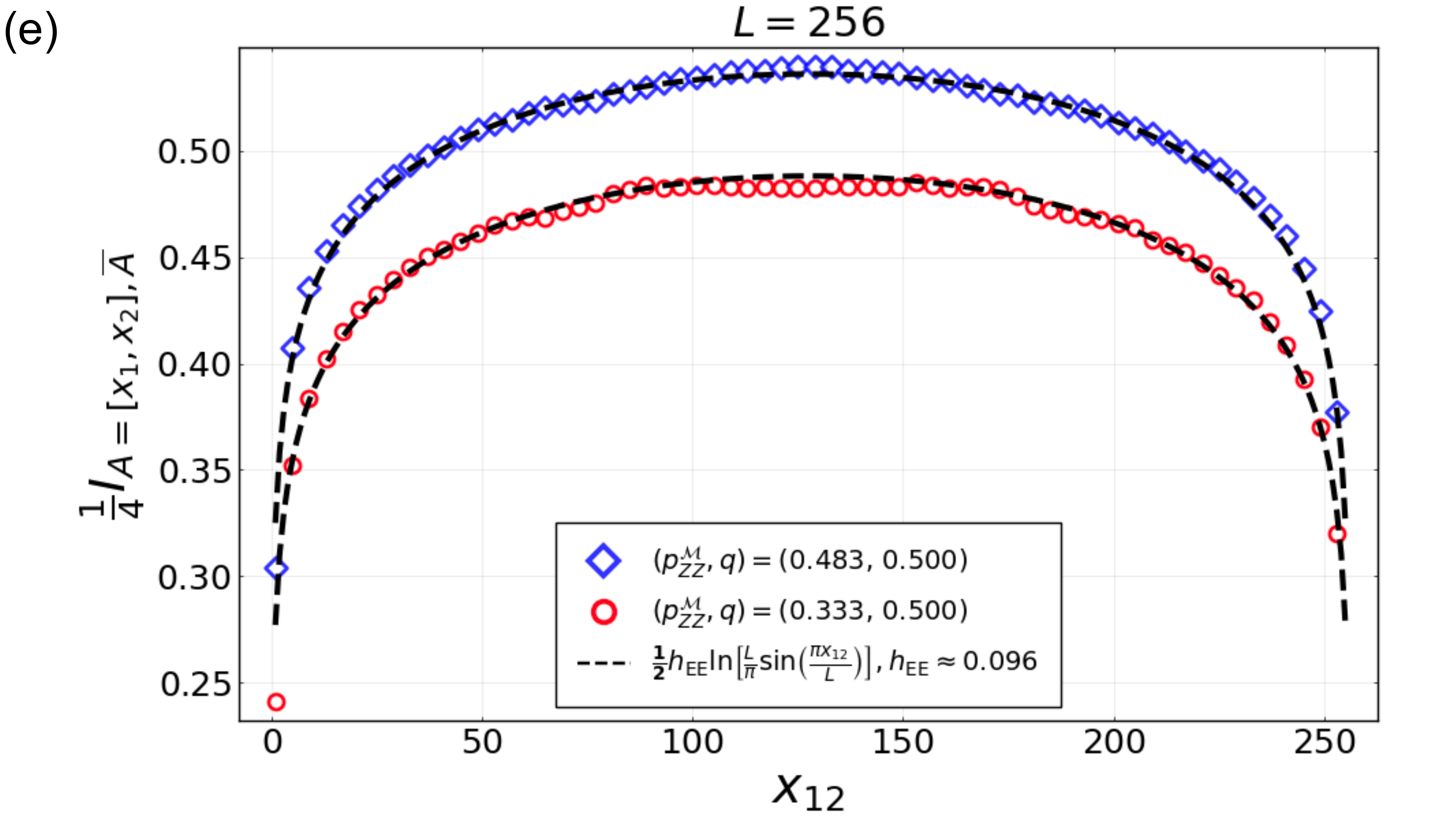}
    \includegraphics[width=\columnwidth]{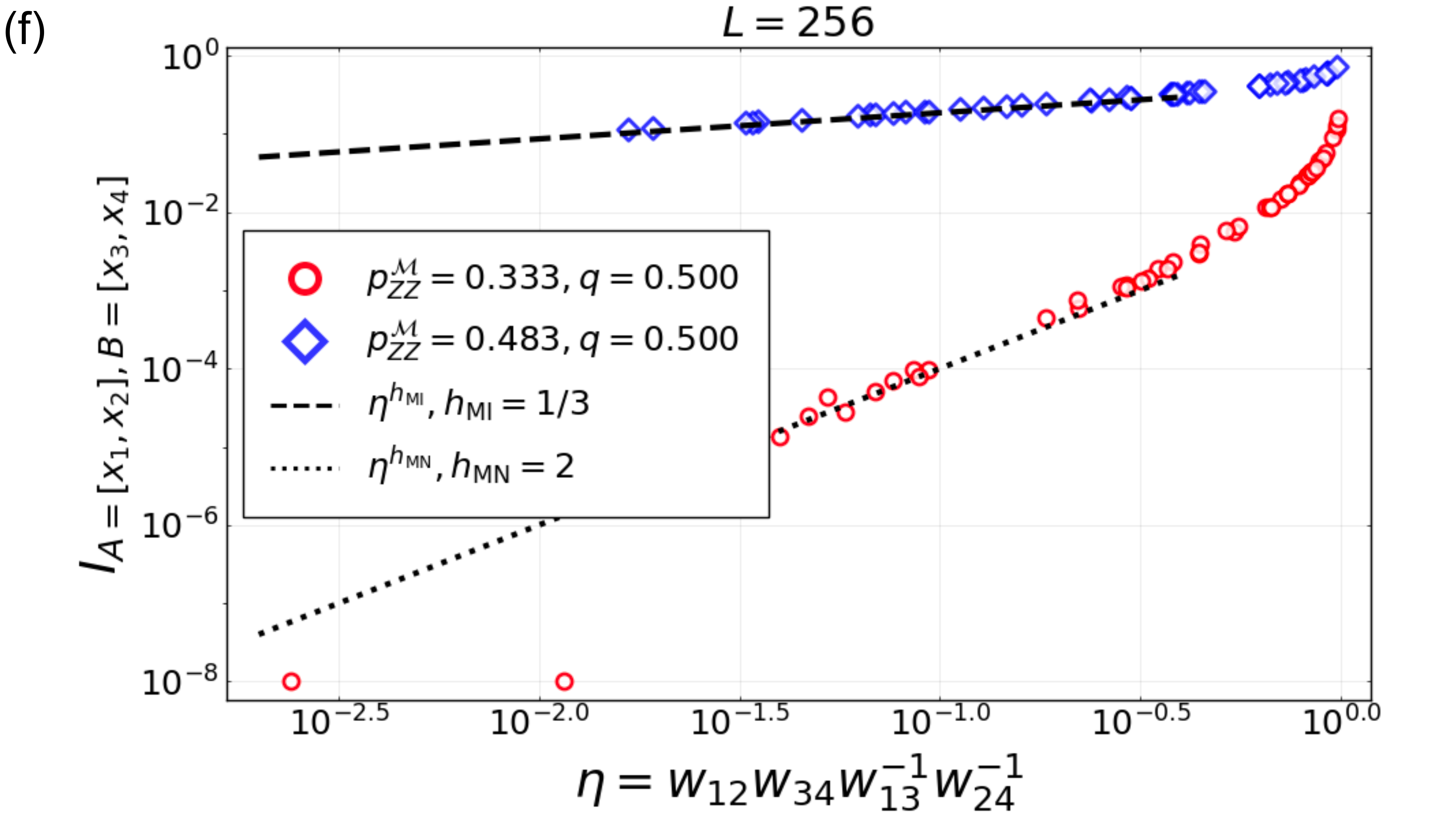}
    \caption{Numerical results for the ``two-leg ladder'' circuit model [Fig.~\ref{fig:phase_diagram_twochain}(a)]~\cite{bao2021enriched}.
    We confirm the phase diagram in Fig.~\ref{fig:phase_diagram_twochain}(b), and find the critical points are in the universality class of critical percolation.
    In particular, (a,b,c,d) are consistent with Fig.~\ref{fig:phase_diagram}, and (e,f) are consistent with Fig.~\ref{fig:baseline_MI_scaling}.
    }
    \label{fig:twochain_numerics}
\end{figure*}

Here we briefly discuss a slight generalization of the baseline model, where we replace a fraction of the $ZZ$ measurements by $ZZ$ dephasings,
\begin{align}
    \label{eq:ZZ_dep_channel}
    \rho \to \mathcal{E}_{Z_j Z_{j+1}}
    (\rho) = \frac{1}{2} \(\rho + (Z_j Z_{j+1}) \cdot \rho \cdot (Z_j Z_{j+1}) \).
\end{align}
The parameters are now
\begin{align}
    p_{ZZ}^\mc{M} =&\ (1-q_{ZZ}) p, \\
    p_{ZZ}^\mc{E} =&\ q_{ZZ} p, \\
    p_{X}^\mc{M} =&\ (1-q_{X})(1-p), \\
    p_{X}^\mc{E} =&\ q_{X}(1-p).
\end{align}
After mapping to bond percolation, we now have both horizontal and vertical incoherent bonds. 

The SG transition is now when coherent, unbroken bonds percolate, and the PM transition is still when broken bonds percolate.
We have at the SG transition
\begin{align}
    \label{eq:modelC_SG_boundary}
    &\ p_{ZZ}^\mc{M} + (1 - p_X^\mc{M} - p_X^\mc{E}) = 1 \nn
    \Rightarrow&\ q_{ZZ} = \frac{2p-1}{p},
\end{align}
and at the PM transition
\begin{align}
    \label{eq:modelC_PM_boundary}
    &\ p_{X}^\mc{M} + (1 - p_{ZZ}^\mc{M} - p_{ZZ}^\mc{E})= 1 \nn
    \Rightarrow&\ q_{X} = \frac{1-2p}{1-p}.
\end{align}
Notice that the critical value of $q_X$ in Eq.~\eqref{eq:modelC_PM_boundary} is the same as when $q_{ZZ} = 0$ (see Eq.~\eqref{eq:modelB_PM_phase_boundary}).

The three-dimensional phase diagram is shown in Fig.~\ref{fig:phase_diagram_3D}.
There is perhaps no surprise that the two phase boundaries have same shape, as one would expect from Kramers-Wannier duality.
The universality classes of these transitions are unaffected by the nonzero $q_{ZZ}$.

\subsection{Equivalence between decoherence and bath-system coupling (Sec.~\ref{sec:intro_bath})
\label{app:twochain_numerics}
}


Here, we provide numerical evidences for the phase diagram in Fig.~\ref{fig:phase_diagram_twochain}(b) of the ``two-leg ladder'' circuit model [Fig.~\ref{fig:phase_diagram_twochain}(a)]~\cite{bao2021enriched}.
The results are shown in 
Fig.~\ref{fig:twochain_numerics}.

In particular, Fig.~\ref{fig:twochain_numerics}(a,b,c,d) should be compared with Fig.~\ref{fig:phase_diagram}, and Fig.~\ref{fig:twochain_numerics}(e,f) should be compared with Fig.~\ref{fig:baseline_MI_scaling}, where we find consistency throughout.


\section{Correctness of the decoding algorithm with located errors (Sec.~\ref{sec:located_error}) \label{app:decoding_alg_proof}}

In section, we prove the correctness of the decoding algorithm in Sec.~\ref{sec:located_error}, for the dynamical repetition code with \emph{located} errors in the spin glass phase.
To do that, we first outline a general procedure of verification for all stabilizer codes, before focusing on this particular case.
Our approach here is perhaps not the simplest for the repetition code, but can be easily generalized to slightly more complicated cases, such as the toric code (see Appendix~\ref{app:dynamical_toric_code}).

Let the stabilizer group associated with the code be denoted $\mc{S}$.
Recall that $\mc{S}$ is an abelian group of mutually-commuting Pauli string operators on $L$ qubits.
The projection operator on the code space is 
\env{align}{
    \mc{P}_{\mc{S}} = 
    2^{-L}
    \sum_{g \in \mc{S}} g.
}
We have $\(\mc{P}_{\mc{S}}\)^2 = \mc{P}_{\mc{S}}$.
Let us assume the stabilizer code has only one logical qubit (so that $\log_2 |\mc{S}| = L-1$), and denote the logical $X$ and $Z$ operators by 
$\mathbf{X}$ and $\mathbf{Z}$.
These two operators anticommute with each other, but commute with all stabilizers.
We also define the logical $Y$  operator as $\mathbf{Y} \coloneqq - i \mathbf{Z} \mathbf{X}$.

A general density matrix inside the two-dimensional code space may be written as a state inside the ``logical Bloch sphere'',
\env{align}{
\label{eq:general_state_code_space}
    \rho_0 = \frac{1}{2} \mc{P}_{\mc{S}} \(\mathbb{1} + x \mathbf{X} + y \mathbf{Y} + z \mathbf{Z} \) \mc{P}_{\mc{S}}.
}
Here, $x, y, z$ are real numbers, satisfying $x^2 + y^2 + z^2 \le 1$.
We will always take the initial state of the circuit in Fig.~\ref{fig:z2_circuit} to be of this type.
Notice that we are generalizing the case in Sec.~\ref{sec:located_error} by also including possibly mixed initial states.

The procedure of error and error correction, as outlined in Sec.~\ref{sec:located_error}, are as follows.
\begin{enumerate}
\item
We send the initial state through a circuit comprised of Pauli measurements and Pauli decoherences.
Among the Pauli measurements, some are of stabilizers in $\mc{S}$.
We denote the quantum operation of the circuit by $\mc{E}$.
\item
After the circuit terminates, we measure all the stabilizers.
This quantum operation is denoted $\mc{M}$.
\item
Based on the stabilizer measurement outcomes at the final state, as well as the stabilizer measurements throughout the circuit bulk, we apply a unitary gate, denoted $\mc{U}$.
\end{enumerate}
The resultant state should now be equal to the initial state $\rho_0$,
\env{align}{
    \label{eq:recovery_map_condition}
    \rho_0 = (\mc{U} \circ \mc{M} \circ \mc{E}) (\rho_0).
}

The state evolution is in general a bit complicated to track, although possible in special cases~\cite{buechler2020projectiveTFIM}.
To simplify the proof, 
we rewrite $\rho_0$ as a linear combination of stabilizer density matrices,
\env{align}{
    \label{eq:decomp_rho0_stab}
    &  \rho_0 \nn 
    =&\, (1-x-y-z) \rho_{\mathbb{1}} + x \rho_\mathbf{X} + y \rho_\mathbf{Y} + z \rho_\mathbf{Z} \nn
    \coloneqq&\,
    \alpha_\mathbb{1} \rho_{\mathbb{1}}
    + \alpha_\mathbb{\mathbf{X}} \rho_\mathbf{X}
    + \alpha_\mathbb{\mathbf{Y}} \rho_\mathbf{Y}
    + \alpha_\mathbb{\mathbf{Z}} \rho_\mathbf{Z},
}
where
\env{align}{
    \rho_{\mathbb{1}} =&\ \frac{1}{2} \mc{P}_{\mc{S}}
    \cdot \mathbb{1} \cdot
    \mc{P}_{\mc{S}} = \frac{1}{2} \mc{P}_\mc{S}, \\
    \rho_{\mathbf{X}} =&\ \mc{P}_{\mc{S}} \cdot \frac{\mathbb{1} + \mathbf{X}}{2} \cdot
    \mc{P}_{\mc{S}}, \\
    \rho_{\mathbf{Y}} =&\ \mc{P}_{\mc{S}} \cdot \frac{\mathbb{1} + \mathbf{Y}}{2} \cdot
    \mc{P}_{\mc{S}}, \\
    \rho_{\mathbf{Z}} =&\ \mc{P}_{\mc{S}} \cdot \frac{\mathbb{1} + \mathbf{Z}}{2} \cdot
    \mc{P}_{\mc{S}}.
}
Each $\rho_g$ ($g \in \{ \mathbb{1}, \mathbf{X}, \mathbf{Y}, \mathbf{Z} \}$) is a normalized stabilizer state, whose stabilizer group is $\mc{S}_g \coloneqq \mc{S} \cup g \mc{S}$.
Our strategy is then to work with each summand of Eq.~\eqref{eq:decomp_rho0_stab} separately.

This is not entirely trivial, since processes $\mc{M}$ and $\mc{E}$ involve projective measurements on the state whose results are subsequently recorded, and are threrefore nonlinear, due to the normalization of the state with the Born probabilities.
To treat this,
we represent the evolution using $\widetilde{\mc{M}}$ and $\widetilde{\mc{E}}$, the \emph{linear} counterparts of $\mc{M}$ and $\mc{E}$ \emph{without} normalization, and ``delay'' the normalization until the very end.
The condition Eq.~\eqref{eq:recovery_map_condition} becomes
\env{align}{
    \label{eq:recovery_map_condition_un_normalized}
    \rho_0 = \mc{U} \lz 
    \frac{(\widetilde{\mc{M}} \circ \widetilde{\mc{E}}) (\rho_0) }
    { \mathrm{Tr}(\widetilde{\mc{M}} \circ \widetilde{\mc{E}}) (\rho_0)} 
    \rz.
}
Using the decomposition in Eq.~\eqref{eq:decomp_rho0_stab}, this is equivalent to
\env{align}{
    & \rho_0 \nn
    =&\
    \mc{U} \lz 
    \frac{(\widetilde{\mc{M}} \circ \widetilde{\mc{E}}) (\sum_{g} \alpha_g \rho_g) }
    { \mathrm{Tr}(\widetilde{\mc{M}} \circ \widetilde{\mc{E}}) (\rho_0)} 
    \rz \nn
    =&\
    \mc{U} \lz 
    \sum_{g} \alpha_g
    \frac
    { \mathrm{Tr}(\widetilde{\mc{M}} \circ \widetilde{\mc{E}}) (\rho_g)}
    { \mathrm{Tr}(\widetilde{\mc{M}} \circ \widetilde{\mc{E}}) (\rho_0)}
    \frac{(\widetilde{\mc{M}} \circ \widetilde{\mc{E}}) ( \rho_g) }
    { \mathrm{Tr}(\widetilde{\mc{M}} \circ \widetilde{\mc{E}}) (\rho_g)} 
    \rz \nn
    =&\
    \sum_{g} \alpha_g
    \frac
    { \mathrm{Tr}(\widetilde{\mc{M}} \circ \widetilde{\mc{E}}) (\rho_g) }
    { \mathrm{Tr}(\widetilde{\mc{M}} \circ \widetilde{\mc{E}}) (\rho_0)}
    \cdot
    (\mc{U} \circ \mc{M} \circ \mc{E}) ( \rho_g).
}
Here, $g \in \{ \mathbb{1}, \mathbf{X}, \mathbf{Y}, \mathbf{Z} \}$.
The RHS will be equal to $\rho_0$, if both conditions below are met for each $g$,
\begin{align}
\label{eq:UME_equal}
\rho_g =&\ (\mc{U} \circ \mc{M} \circ \mc{E}) ( \rho_g), \\
\mathrm{Tr}(\widetilde{\mc{M}} \circ \widetilde{\mc{E}}) (\rho_g) =&\ \mathrm{Tr}(\widetilde{\mc{M}} \circ \widetilde{\mc{E}}) (\rho_0).
\end{align}
Both conditions can be verified by tracking the evolution of the stabilizer states $\rho_g$ under $\mc{M} \circ \mc{E}$.\footnote{If not analytically, at least numerically.}
The second condition is equivalent to the following condition which is easier to verify,\footnote{
Assuming that Eq.~\eqref{eq:trace_ME_equal} holds, we have
\begin{align}
    &\mathrm{Tr}(\widetilde{\mc{M}} \circ \widetilde{\mc{E}}) (\rho_0) \nn
    =&\
    \sum_g \alpha_g \mathrm{Tr}(\widetilde{\mc{M}} \circ \widetilde{\mc{E}}) (\rho_g)\nn
    =&\
    \sum_g \alpha_g \mathrm{Tr}(\widetilde{\mc{M}} \circ \widetilde{\mc{E}}) (\rho_\mathbb{1}) \nn
    =&\
    \mathrm{Tr}(\widetilde{\mc{M}} \circ \widetilde{\mc{E}}) (\rho_\mathbb{1}), \nonumber
\end{align}
where we used that $\sum_g \alpha_g = 1$.
}
\begin{align}
\label{eq:trace_ME_equal}
    &\mathrm{Tr}(\widetilde{\mc{M}} \circ \widetilde{\mc{E}}) (\rho_\mathbb{1})
    = \mathrm{Tr}(\widetilde{\mc{M}} \circ \widetilde{\mc{E}}) (\rho_\mathbf{X})\nn
    &\quad = \mathrm{Tr}(\widetilde{\mc{M}} \circ \widetilde{\mc{E}}) (\rho_\mathbf{Y})
    = \mathrm{Tr}(\widetilde{\mc{M}} \circ \widetilde{\mc{E}}) (\rho_\mathbf{Z}).
\end{align}

Intuitively, the decoding algorithm is successful if it works on each of branch of stabilizer states, \emph{and} for a given history of measurement outcomes the trajactory occurs with the same Born probability for each of the branch, so that the amplitude coherence between different branches are preserved.


\subsection{Verification for the dynamic repetition code with error-avoiding spanning paths}

Recall that the repetition code has stabilizers $Z_j Z_{j+1}$ for $1 \le j < L$, and the logical operators can be chosen to be $\mathbf{X} = X_1 \ldots X_L$, $\mathbf{Z} = Z_j$ for any $1 \le j \le L$.
Below we try to describe $(\mc{M} \circ \mc{E}) ( \rho_g)$ and $\mathrm{Tr}(\widetilde{\mc{M}} \circ \widetilde{\mc{E}}) (\rho_g)$ for $g \in \{ \mathbb{1}, \mathbf{X}, \mathbf{Y}, \mathbf{Z} \}$, and show that they satisfy the conditions above, \emph{provided} that an error-avoiding spanning path exists.

Let us first list a convenient basis of the stabilizer group for each of $\rho_g$; we denote the corresponding basis $\mc{G}(\rho_g)$.
Our choices are
\env{align}{
\label{eq:stab_initial_1}
    \mc{G}(\rho_{\mathbb{1}}) =&\ \{+ Z_j Z_{j+1} : 1 \le j < L \}, \\
\label{eq:stab_initial_X}
    \mc{G}(\rho_{\mathbf{X}}) =&\ \mc{G}(\rho_{\mathbb{1}}) \cup \{\mathbf{X}\}, \\
\label{eq:stab_initial_Z}
    \mc{G}(\rho_{\mathbf{Z}}) =&\ \mc{G}(\rho_{\mathbb{1}}) \cup \{\mathbf{Z} \}, \\
\label{eq:stab_initial_Y}
    \mc{G}(\rho_{\mathbf{Y}}) =&\ \mc{G}(\rho_{\mathbb{1}}) \cup \{ - i  \mathbf{Z} \mathbf{X} \}.
}
Below, we show by induction that at all circuit time steps $t$, stabilizers of the corresponding dynamical states $\mc{E}_t(\rho_g)$ can be chosen so that they satisfy the following conditions:
\env{align}{
    \label{eq:cond_detail_X}
    \mc{G}(\mc{E}_t(\rho_\mathbf{X})) =&\ \mc{G}(\mc{E}_t(\rho_\mathbb{1})) \cup \{\mathbf{X}\}, \\
    \label{eq:cond_detail_Z}
    \mc{G}(\mc{E}_t(\rho_\mathbf{Z})) =&\ \mc{G}(\mc{E}_t(\rho_\mathbb{1})) \cup \{ Z_{\rm cum}(\pi) \cdot Z_v \}, \\
    \label{eq:cond_detail_Y}
    \mc{G}(\mc{E}_t(\rho_\mathbf{Y})) =&\ \mc{G}(\mc{E}_t(\rho_\mathbb{1})) \cup \{ -i Z_{\rm cum}(\pi) \cdot Z_v \cdot \mathbf{X} \}, \\
    \label{eq:cond_detail_norm}
    \mathrm{Tr}[\widetilde{\mc{E}_t} (\rho_\mathbb{1})]
    =&\ \mathrm{Tr}[\widetilde{\mc{E}_t} (\rho_\mathbf{X})] = \mathrm{Tr} [ \widetilde{\mc{E}_t} (\rho_\mathbf{Y})]
    = \mathrm{Tr}[\widetilde{\mc{E}_t} (\rho_\mathbf{Z})].
}
Here, by assumption, there exists an error-avoiding spanning path (as defined in Sec.~\ref{sec:located_error}) $(u, 0) \xrightarrow[]{\pi}(v, t)$, and $Z_{\rm cum}$ is the product of all $Z_j Z_{j+1}$ measurement outcomes along $\pi$ [Eq.~\eqref{eq:Z_cum_def}].

Conditions in Eq.~(\ref{eq:cond_detail_X},
\ref{eq:cond_detail_Z}, \ref{eq:cond_detail_Y}, \ref{eq:cond_detail_norm}) obviously hold for the initial state.
Assume they are correct at all times before $t$.
\begin{enumerate}
\item
At odd $t$, the circuit involves measurements of $Z_j Z_{j+1}$.
Let us consider these one by one, and focus on, say, $g = Z_1 Z_2$.

Consider the evolution of $\mc{E}_t(\rho_\mathbb{1})$.
There are two possibilities,
\begin{enumerate}
\item
    $g$ commutes with all elements of $\mc{G}(\mc{E}_t(\rho_\mathbb{1}))$.
    There are two cases:
    \begin{enumerate}
        \item
        $g$ is already be an element of the group generated by $\mc{G}(\mc{E}_t(\rho_\mathbb{1}))$,
        \begin{align}
            g \in \avg{\mc{G}(\mc{E}_t(\rho_\mathbb{1}))}.
        \end{align}
        The measurement outcome will then be deterministic, fixed by $\mc{G}(\mc{E}_t(\rho_\mathbb{1}))$.
        In this case, $\mc{G}(\mc{E}_t(\rho_\mathbb{1}))$ remains unchanged.
        \item
        $g$ does not belong to the following larger group (which contains  $\avg{\mc{G}(\mc{E}_t(\rho_\mathbb{1}))}$)
        \begin{align}
            g \notin \avg{\mc{G}(\mc{E}_t(\rho_\mathbb{1})) \cup \{ \mathbf{X}, Z_{\rm cum}(\pi) \cdot Z_v, -i Z_{\rm cum}(\pi) \cdot Z_v \cdot \mathbf{X}\} },
        \end{align}
        In this case, $\mc{G}(\mc{E}_t(\rho_\mathbb{1}))$ becomes $\mc{G}(\mc{E}_t(\rho_\mathbb{1})) \cup \{\pm g\}$, corresponding to measurement outcome of $g$ being $\pm 1$, with equal probabilities $1/2$.
    \end{enumerate}
    Notice that it is not possible for 
    \begin{align}
        g \notin \avg{\mc{G}(\mc{E}_t(\rho_\mathbb{1}))} 
    \end{align}
    but at the meantime
    \begin{align}
        g \in \avg{\mc{G}(\mc{E}_t(\rho_\mathbb{1})) \cup \{ \mathbf{X}, Z_{\rm cum}(\pi) \cdot Z_v, -i Z_{\rm cum}(\pi) \cdot Z_v \cdot \mathbf{X}\} },
    \end{align}
    for then $g$ must anticommute with at least one of $\{ \mathbf{X}, Z_{\rm cum}(\pi) \cdot Z_v, -i Z_{\rm cum}(\pi) \cdot Z_v \cdot \mathbf{X}\}$, which is
    not the case (with $g = Z_j Z_{j+1}$).

\item
    $g$ anticommutes at least one element of $\mc{G}(\mc{E}_t(\rho_\mathbb{1}))$.
    Without loss of generality, assume there is exaclty one such element, denoted $h$.
    The evolution of $\mc{G}(\mc{E}_t(\rho_\mathbb{1}))$ would be to replace $h$ with $\pm g$, with equal probabilities $1/2$.
\end{enumerate}

In all cases above, the update of $\mc{G}(\mc{E}_t(\rho_\mathbb{1}))$ when combined with any of $\{ \mathbf{X}, Z_{\rm cum}(\pi) \cdot Z_v, -i Z_{\rm cum}(\pi) \cdot Z_v \cdot \mathbf{X}\}$, are legal updates of $\mc{G}(\mc{E}_t(\rho_{\mathbf{X}, \mathbf{Y}, \mathbf{Z}}))$ as well.
Here, we use again the fact that $g = Z_j Z_{j+1}$ commutes with each of $\{ \mathbf{X}, Z_{\rm cum}(\pi) \cdot Z_v, -i Z_{\rm cum}(\pi) \cdot Z_v \cdot \mathbf{X}\}$

Moreover, the measurement outcome of $g$ obeys the same probability distribution for all four evolutions.

We have shown that conditions Eq.~(\ref{eq:cond_detail_X},
\ref{eq:cond_detail_Z},
\ref{eq:cond_detail_Y},  \ref{eq:cond_detail_norm}) are preserved under one $Z_j Z_{j+1}$ measurement -- hence all of them -- at odd $t$.

\item
After the $Z_j Z_{j+1}$ are measured at an odd time step $t$, we have the freedom of moving the upper endpoint of the error-avoding spanning path $\pi$.

In particular, we can choose a different path $\pi^\p$ ending at another site $(v^\p, t)$.
In other words, we can make the following replacement of the generating set of $\mc{E}_t(\rho_\mathbf{Z})$,
\begin{align}
    & \mc{G}(\mc{E}_t(\rho_\mathbf{Z})) = 
    \mc{G}(\mc{E}_t(\rho_\mathbb{1})) \cup \{ Z_{\rm cum}(\pi) \cdot Z_v \} \nn 
    \to \quad &
    \mc{G}(\mc{E}_t(\rho_\mathbf{Z})) = 
    \mc{G}(\mc{E}_t(\rho_\mathbb{1})) \cup \{ Z_{\rm cum}(\pi^\p) \cdot Z_{v^\p} \}.
\end{align}
The two sets generate the same stabilizer group, since $v$ and $v^\p$ are connected by a path of $ZZ$ stabilizer measurements in the bulk (e.g. by joining $\pi$ and $\pi^p$), so that $ Z_{\rm cum}(\pi) Z_{\rm cum}(\pi^\p) \cdot Z_v Z_{v^\p}$ is also a stabilizer of $\mc{E}_t(\rho_\mathbb{1})$ (see Appendix~\ref{app:details_mapping_percolation}).

Thus, we have the freedom in choosing the site at which the path $\pi$ ends, and we have shown that different choices are consistent.
In particular, under the assumption that an error-avoding spanning path $\pi$ exists, there always exists a choice where the endpoint does not experience an error at the current timestep.
We can perform this ``error-avoiding gauge transformation'' at each timestep throughout the time evolution whenever needed.

\item
At even $t$, consider the measurement of $X_j$.
By the asssumption of an error-avoiding spanning path, we may assume that the path ends at a site $v \neq j$.
In this case, $X_j$ also commutes with all of $\{ \mathbf{X}, Z_{\rm cum}(\pi) \cdot Z_v, -i Z_{\rm cum}(\pi) \cdot Z_v \cdot \mathbf{X}\}$.
The reasoning above for $g = Z_j Z_{j+1}$ applies equally well here for $g = X_j$, verbatim.

Thus, 
conditions Eq.~(\ref{eq:cond_detail_X},
\ref{eq:cond_detail_Z},
\ref{eq:cond_detail_Y},  \ref{eq:cond_detail_norm}) are preserved under $X_j$ measurements at even $t$.

\item
At even $t$, consider $g = X_j$ decoherence.
We similarly assume $v \neq j$.

There are two possibilities for the evolution of $\mc{E}_t(\rho_\mathbb{1})$.
\begin{enumerate}
\item
    $g$ commutes with all elements of $\mc{G}(\mc{E}_t(\rho_\mathbb{1}))$.
    In this case, $\mc{G}(\mc{E}_t(\rho_\mathbb{1}))$ remains unchanged.
    The other three $\mc{G}(\mc{E}_t(\rho_{\mathbf{X}, \mathbf{Y}, \mathbf{Z}}))$ also remain unchanged, since $g$ commutes with all of $\{ \mathbf{X}, Z_{\rm cum}(\pi) \cdot Z_v, -i Z_{\rm cum}(\pi) \cdot Z_v \cdot \mathbf{X}\}$ as well.
\item
    $g$ anticommutes at least one element of $\mc{G}(\mc{E}_t(\rho_\mathbb{1}))$.
    Again assume there is exaclty one such element, denoted $h$.
    The evolution of $\mc{G}(\mc{E}_t(\rho_\mathbb{1}))$ would be to simply drop $h$.
    The same applies to $\mc{G}(\mc{E}_t(\rho_{\mathbf{X}, \mathbf{Y}, \mathbf{Z}}))$
\end{enumerate}
Thus, conditions Eq.~(\ref{eq:cond_detail_X}, \ref{eq:cond_detail_Z},
\ref{eq:cond_detail_Y}) are preseved under $X_j$ decoherence.

Since $X_j$ decoherence is a quantum channel that preserves the norm of the density matrix, and does not involve measurements, Eq.~(\ref{eq:cond_detail_norm}) is also preserved.

\end{enumerate}

The induction above shows that at $t = T$, $\mc{E} = \mc{E}_T$, conditions Eq.~(\ref{eq:cond_detail_X},
\ref{eq:cond_detail_Z},
\ref{eq:cond_detail_Y},  \ref{eq:cond_detail_norm}) hold.
Since $\mc{M}$ is just another layer of $Z_j Z_{j+1}$ measurements, these conditions also hold upon replacing $\mc{E}$ by $\mc{M} \circ \mc{E}$.
Here, Eq.~\eqref{eq:cond_detail_norm} becomes Eq.~\eqref{eq:trace_ME_equal}.

Lastly, we verify Eq.~\eqref{eq:UME_equal} for a particular $\mc{U}$, constructed using previous measurement outcomes.
For the states $(\mc{M} \circ \mc{E})(\rho_{\mathbb{1}, \mathbf{X}, \mathbf{Y}, \mathbf{Z}})$, we have
\begin{align}
    \mc{G}\( (\mc{M} \circ \mc{E})(\rho_{\mathbb{1}}) \) =&\ \{\pm  Z_j Z_{j+1} \}, \\
    \mc{G}\( (\mc{M} \circ \mc{E})(\rho_{\mathbf{X}}) \) =&\ \mc{G}\( (\mc{M} \circ \mc{E})(\rho_{\mathbb{1}}) \) \cup \{\mathbf{X}\}, \\
    \mc{G}\( (\mc{M} \circ \mc{E})(\rho_{\mathbf{Y}}) \) =&\ \mc{G}\( (\mc{M} \circ \mc{E})(\rho_{\mathbb{1}}) \) \cup \{ -i Z_{\rm cum}(\pi) \cdot Z_v \cdot \mathbf{X} \}, \\
    \mc{G}\( (\mc{M} \circ \mc{E})(\rho_{\mathbf{Z}}) \) =&\ \mc{G}\( (\mc{M} \circ \mc{E})(\rho_{\mathbb{1}}) \) \cup \{ Z_{\rm cum}(\pi) \cdot Z_v \}.
\end{align}
The ``error-correcting unitary'' $\mc{U}$ will first need to correct the signs for all stabilizers $Z_j Z_{j+1} = -1$, while commuting with $\{ \mathbf{X}, Z_{\rm cum}(\pi) \cdot Z_v, -i Z_{\rm cum}(\pi) \cdot Z_v \cdot \mathbf{X}\}$.
This may be achieved by apply the so-called ``destabilizers'', or in this special case of the repetition code, by applying $X_j$ (``spin flip'') for all $j\neq v$ where $Z_j Z_v = -1$ (see Sec.~\ref{sec:located_error}).
Next, we apply $\mathbf{X}$ to the system if $Z_{\rm cum}(\pi) = -1$, again  described in Sec.~\ref{sec:located_error}.
After these steps, we have
\begin{align}
    \mc{G}\( (\mc{U} \circ \mc{M} \circ \mc{E})(\rho_{\mathbb{1}}) \) =&\ \{+  Z_j Z_{j+1} \}, \\
    \mc{G}\( (\mc{U} \circ \mc{M} \circ \mc{E})(\rho_{\mathbf{X}}) \) =&\ \{+  Z_j Z_{j+1} \} \cup \{\mathbf{X}\}, \\
    \mc{G}\( (\mc{U} \circ \mc{M} \circ \mc{E})(\rho_{\mathbf{Y}}) \) =&\ \{+  Z_j Z_{j+1} \} \cup \{ -i Z_v \cdot \mathbf{X} \}, \\
    \mc{G}\( (\mc{U} \circ \mc{M} \circ \mc{E})(\rho_{\mathbf{Z}}) \) =&\ \{+  Z_j Z_{j+1} \} \cup \{ + Z_v \}.
\end{align}
Comparing with Eq.~(\ref{eq:stab_initial_1}, \ref{eq:stab_initial_X}, \ref{eq:stab_initial_Z}, \ref{eq:stab_initial_Y}), we have
\begin{align}
    &\mc{G}\( (\mc{U} \circ \mc{M} \circ \mc{E})(\rho_g)\) = \mc{G} (\rho_g)
    \nn
    \quad \Leftrightarrow \quad&
    (\mc{U} \circ \mc{M} \circ \mc{E})(\rho_g) = \rho_g.
\end{align}
Thus, we have verified Eq.~\eqref{eq:UME_equal}.
Combined with Eq.~\eqref{eq:trace_ME_equal}, we verified Eq.~\eqref{eq:recovery_map_condition}, the correctness of the decoding algorithm for a general initial state $\rho_0$ [Eq.~\eqref{eq:general_state_code_space}] in the code space.

To summarize, we have verified the correctness of the decoding algorithm described in Sec.~\ref{sec:located_error}, for the quantum repetition code with located errors.
Crucial to its correctness is the existence of an error-avoiding spanning path, as we have emphasized throughout.
From our arguments, we see how the spanning path ensures the encoded information is never directly measured or decohered, therefore preserved and recoverable in the final state.

\subsection{Decoding the dynamic toric code with error-avoiding spanning membranes \label{app:dynamical_toric_code}}

\begin{figure}[b]
    \centering
    \includegraphics[width=.5\textwidth]{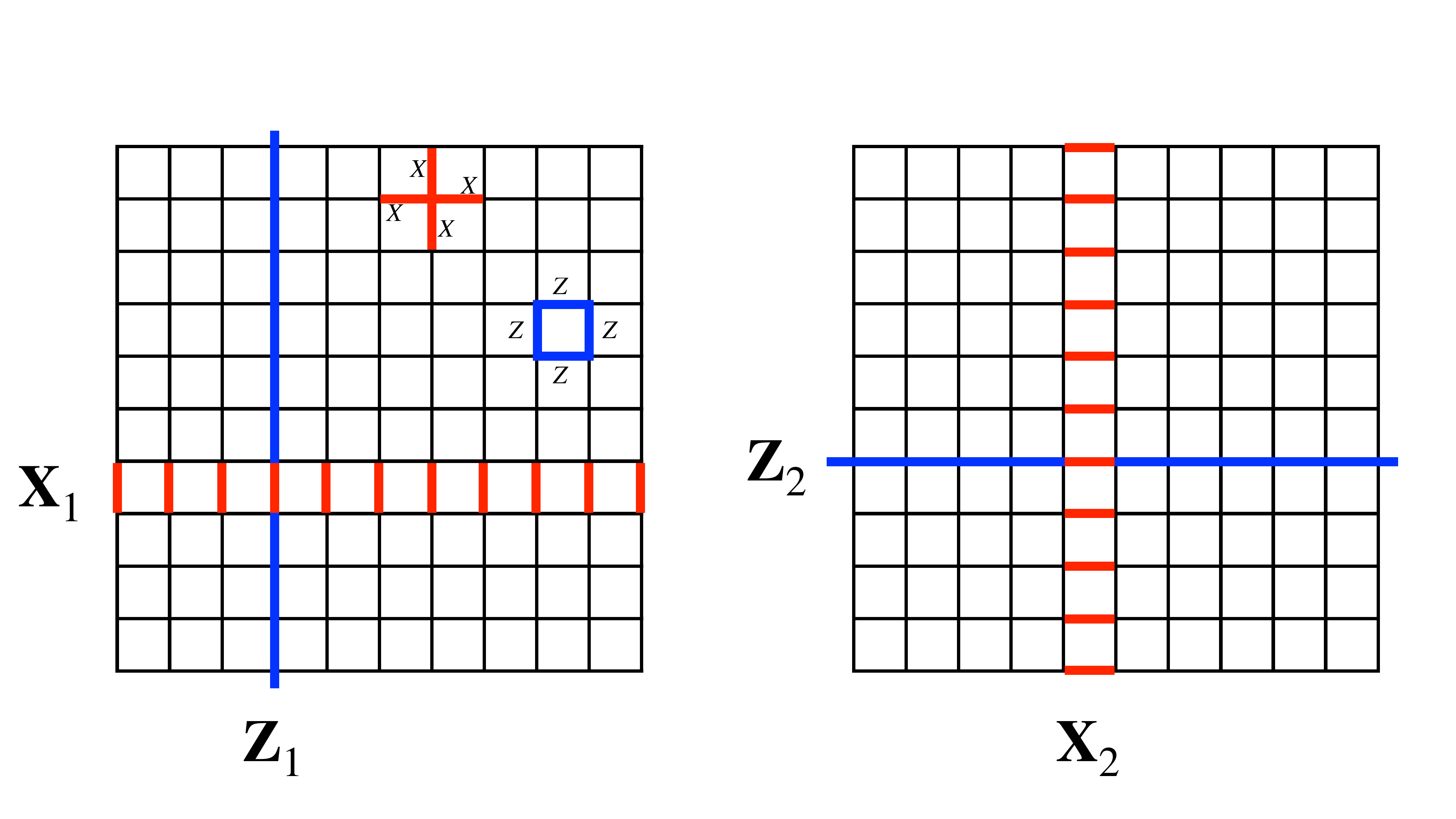}
    \caption{The toric code on a two-dimensional torus.
    The stabilizers of the toric code are the plaquette $Z$ operators (denoted $\Box_Z$) and the star $X$ operators (denoted $+_X$), shown on the left.
    The code has two logical qubits, and the logical operators (denoted $\mathbf{Z}_{1,2}$, $\mathbf{X}_{1,2}$) are strings of $X$s or $Z$s on noncontractible loops on the torus, as illustrated.
    }
    \label{fig:toric_code}
\end{figure}

\begin{figure}[t]
    \centering
    \includegraphics[width=.5\textwidth]{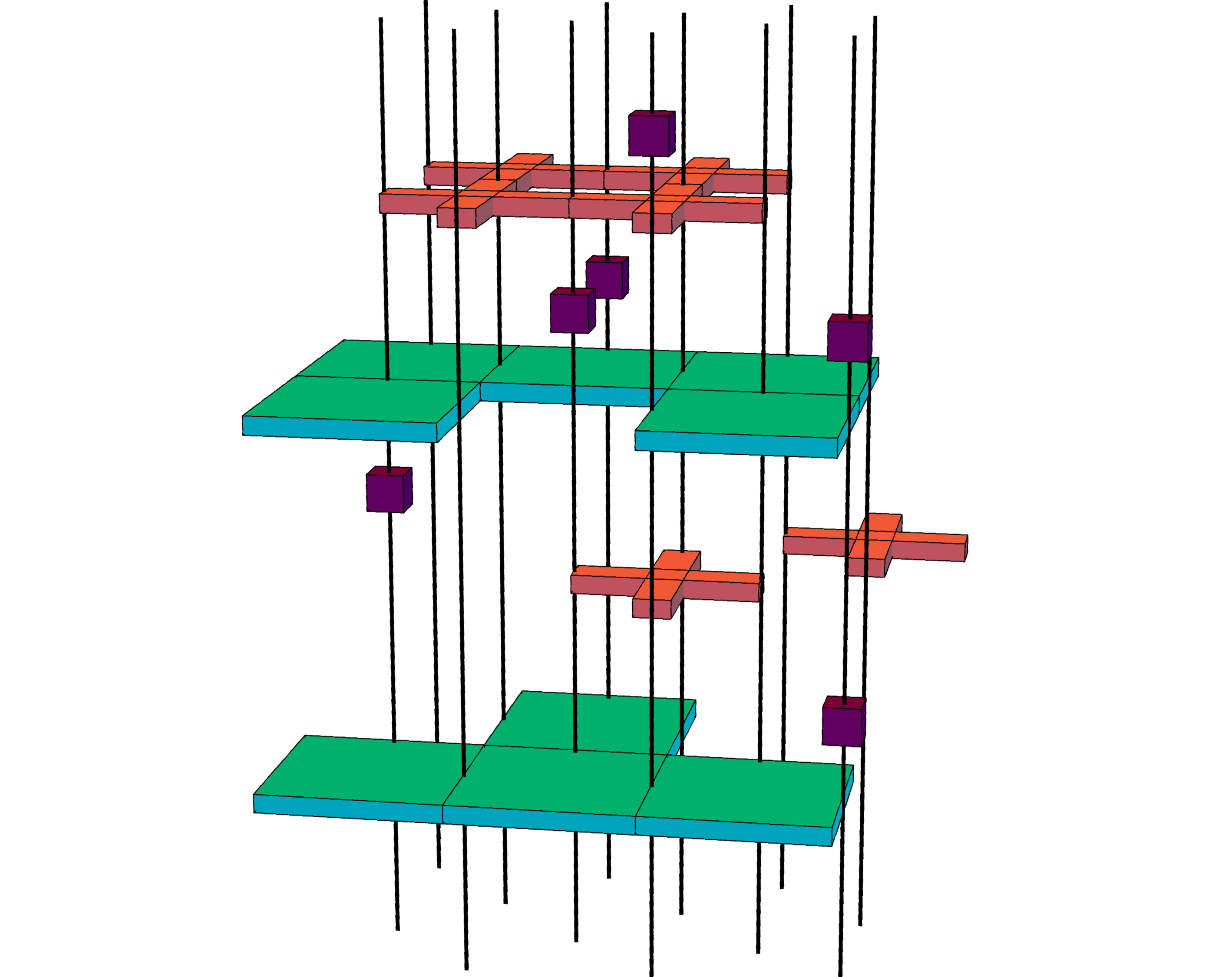}
    \caption{The circuit dynamics of the toric code in (2+1) spacetime dimensions.
    We measure the $\Box_Z$ and $+_X$ operators at alternating timesteps, with rates $p_{\Box}^{\mc{M}}$ and $p_{+}^{\mc{M}}$, respectively.
    In between the stabilizer measurements, single-qubit errors can occur at a finite rate $p^{\rm  err}$.
    }
    \label{fig:toric_code_circuit}
\end{figure}

\begin{figure}[t]
    \centering
    \includegraphics[width=.5\textwidth]{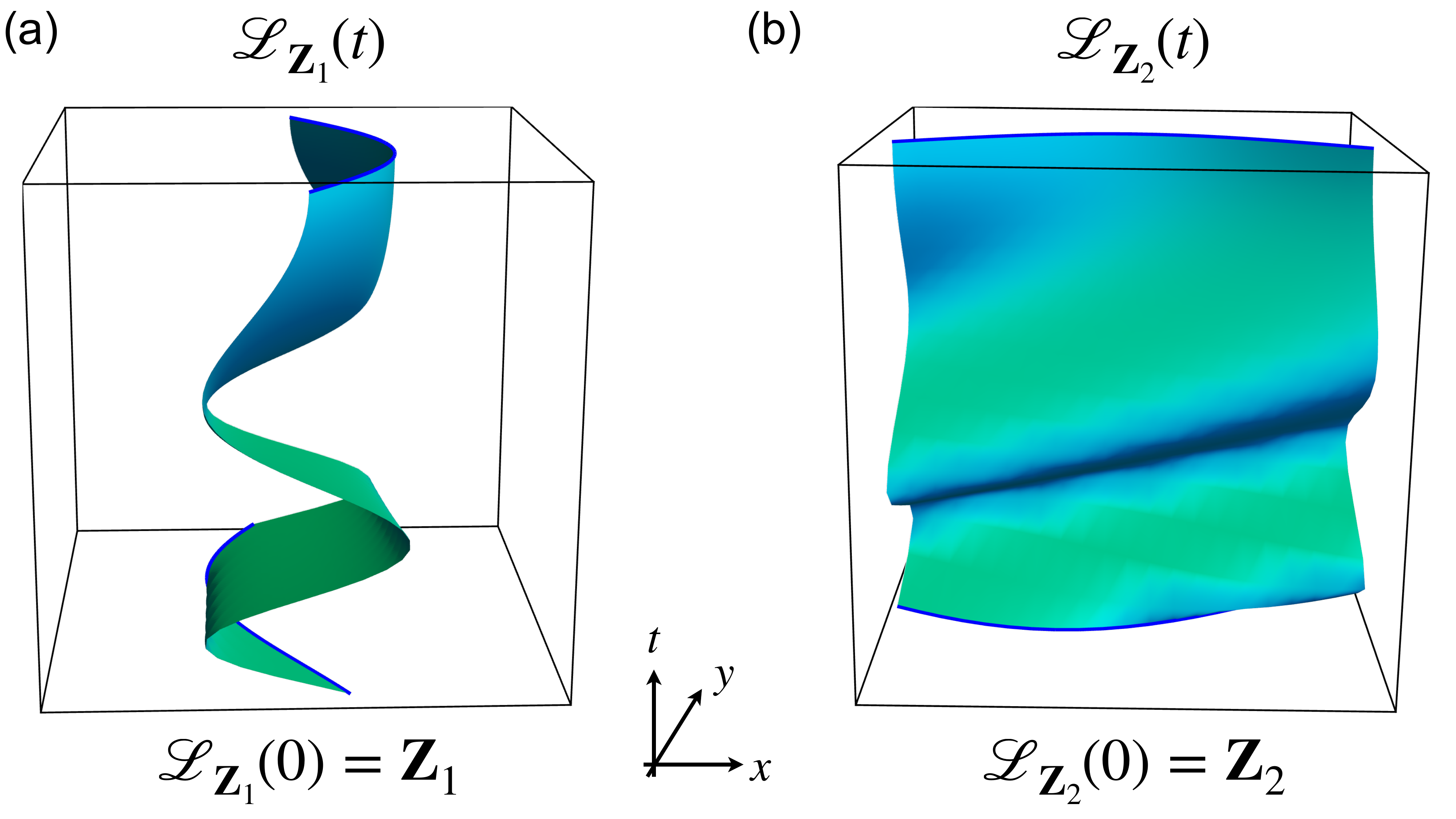}
    \includegraphics[width=.5\textwidth]{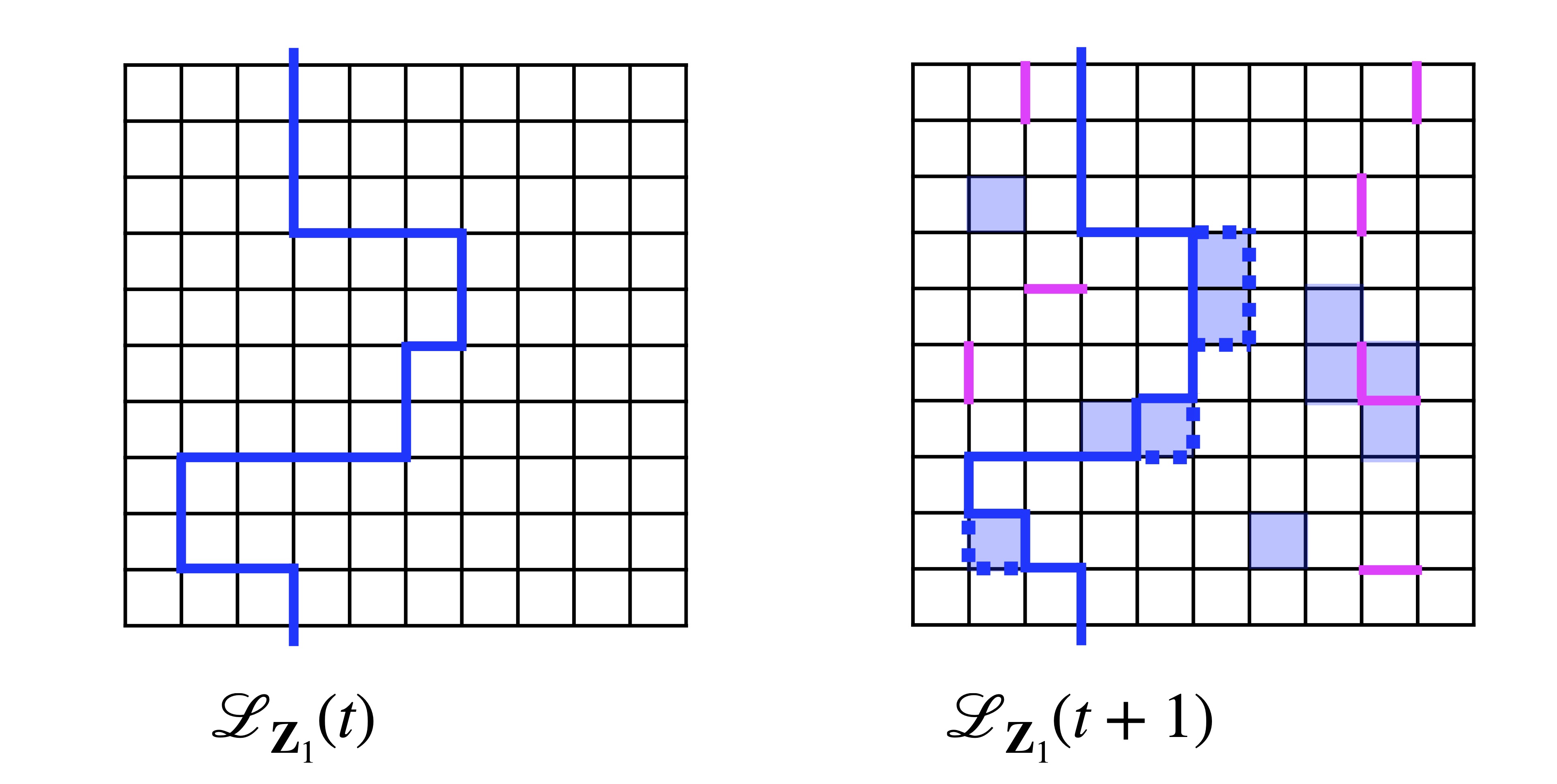}
    \caption{
    (a,b)
    dynamic logical operators defined on the error avoiding membranes.
    Faces of the membrane $\mu$ parallel to $xt$- or $yt$- planes do not contain any errors, and faces parallel the $xy$- plane must each contain a plaquette stabilizer measurement.
    In (a), the membrane $\mu$ intersects the initial and final temporal planes at two noncontractible loops along the $y$ direction (highlighted with blue color), which we idenfity as $L_{\mathbf{Z}_1}(0)$ and $L_{\mathbf{Z}_1}(t)$.
    In (b), the membrane $\mu$ is now along the $x$ direction, and defines $L_{\mathbf{Z}_2}(0)$ and $L_{\mathbf{Z}_2}(t)$.
    (c,d)
    Comparing the cross section of $\mu$ at two consecutive time steps, $t$ and $t+1$.
    Single qubit errors that occur in between are denoted as pink links.
    The dynamic logical operators can be ``wiggled'' by measurements of plaquette operators (shaded) at time $t+1$, corresponding to a ``error-avoiding gauge transformation'' of the dynamic logical operator.
    }
    \label{fig:toric_code_membrane}
\end{figure}

Using the formalism developed above, we consider a different example \emph{without} $\mb{Z}_2$ symmetry, namely a toric code~\cite{kitaev1997, DKLP2001topologicalQmemory, fowler2012surfacecode} undergoing Pauli measurements and Pauli decoherence in (2+1) spacetime dimensions.
Recall that the toric code has two types of check operators, namey the product of Pauli $Z$s on each plaquette (denoted $\Box_Z$), and the product on $X$s on each star (denoted $+_X$), see Fig.~\ref{fig:toric_code}.
On a two dimensional torus, the toric code supports two logical qubits, and the logical operators are Pauli string operators on non-contractible loops, as illustrated in Fig.~\ref{fig:toric_code}.


The dynamics we consider is shown Fig.~\ref{fig:toric_code_circuit}.
Here, time runs upwards.
Each square represents a measurement of the plaquette check operator $\Box_Z$, and each star represents a measurement of the star check operator $+_X$.
We denote the measurement rates of these $p_{\Box}^\mc{M}$ and $p_{+}^\mc{M}$.
In between the check operator measurements are the single qubit errors, which can be taken to be arbitrary, without any symmetry constraints.
However, we will focus on the case of bit-flip errors only, so that we can simplfy the problem by considering only the $\Box_Z$ measurements.
Phase errors can be tracked by the $+_X$ measurements, and can be decoded independently.
We note that a similar model was previously considered in Ref.~\cite{barkeshli2020topological}.

The analysis of the dynamics follows from what was described above for the repetition code, with slight modifications.
Here we have two logical qubits, and a four-dimensional code space.
A decomposition of a general density matrix in the code space, similar to Eq.~\eqref{eq:decomp_rho0_stab}, would read
\begin{align}
    \rho_0 = \sum_g \alpha_g \rho_g,
\end{align}
where $g$ now can take all $16$ logical Pauli matrices on the two logical qubits, including the identity.
Again, as in Eq.~(\ref{eq:cond_detail_X},
\ref{eq:cond_detail_Z}, \ref{eq:cond_detail_Y}, \ref{eq:cond_detail_norm}), on each trajectory $\mc{E}$, we can show by induction that 
\begin{align}
    \mc{G}(\mc{E}_t(\rho_{g \neq \mathbb{1}})) = 
    \mc{G}(\mc{E}_t(\rho_{\mathbb{1}})) \cup \{\mc{L}_g(t)\},
\end{align}
where $\mc{L}_g$ is a ``dynamic logical operator'' (to be defined below) for the operator $g$, and these satisfy
\begin{align}
    \mc{L}_{g_1}(t) \mc{L}_{g_2}(t) = \mc{L}_{g_1 \cdot g_2}(t).
\end{align}
In the previous example of the repetition code, $\mc{L}_{\mathbf{X}}(t) = \mathbf{X}$, and $\mc{L}_{\mathbf{Z}}(t) = Z_{\rm cum}(\pi) Z_v$.
Here for the toric code, instead of an error-avoiding spanning path, $\mc{L}_g$'s are defined on ``error-avoiding spanning membranes''.

As an example, for $g = \mathbf{Z}_1 \mathbb{1}_2$, such a membrane $\mu$ intersects the lower boundary of the circuit at a topologically nontrivial loop, and also does so at the upper boundary [Fig.~\ref{fig:toric_code_membrane}(a)].
Importantly, $\mu$ contains no errors on its vertical faces (parallel to the $xt$- and $yt$- planes), and on each of its faces parallel to the $xy$-plane at time $t$ a corresponding measurement of the plaquette stabilizer $\Box_Z$ was made.
The product of all Pauli $Z$'s on the loop at the lower boundary of $\mu$ is the logical operator $\mathbf{Z}_1 \mathbb{1}_2$ for the toric code, and
the operator $\mc{L}_{\mathbf{Z}_1 \mathbb{1}_2}(t)$ is defined to be the product of Pauli $Z$'s on the loop at the upper boundary of $\mu$, whose sign (denoted $\Box_{\rm cum}(\mu)$) is determined by the product of all plaquette stabilizers on $\mu$.
One can similarly find $\mc{L}_{\mathbb{1}_1 \mathbf{Z}_2}(t)$ [Fig.~\ref{fig:toric_code_membrane}(b)], $\mc{L}_{\mathbf{X}_1 \mathbb{1}_2}(t)$, and $\mc{L}_{\mathbb{1}_1 \mathbf{X}_2}(t)$ (hence  $\mc{L}_g(t)$ for all $g$) from their corresponding membranes, when they exist.

In [Fig.~\ref{fig:toric_code_membrane}(c,d)], we further illustrate the error-avoiding nature of these membranes.
From one timestep to the next, we can change the dynamic logical operator $L_{\mathbf{Z}_1}(t)$, by performing the error-avoiding gauge transformation, allowed by stabilizer measurements.

The decoding starts by measuring all the stabilizers, and correcting these signs using unitaries that commute with all $\mc{L}_g(t)$. 
The signs of $\mc{L}_g(t)$s can be subsequently corrected by applying logical operators, based on the sign calculated from error-avoiding spanning membranes, in a fashion similar to the repetition code.
For example, if $\mc{L}_{\mathbf{Z}_1 \mathbb{1}_2}$ is deemed to have a minus sign, we apply the logical operator $\mc{L}_{\mathbf{X}_1 \mathbb{1}_2}$ to the code block.

Given the stabilizer measurements in the bulk and locations of the errors, error-avoiding spanning membranes can be found in polynomial time.
Such two-dimensional membranes are present with probability $1$ in the ``percolating'' phase of the membranes, which is equivalent to the ``non-percolating'' phase of one-dimensional paths of errors that intersect the membrane.
Thus, the 
decodability
of the encoded information with located errors for the dynamical toric code model 
is again related to the physics of percolation, this time in three spatial dimensions.

More intriguing is the 
decodability
of the quantum information with \emph{unlocated errors}.
For concreteness, we consider the membrane whose boundaries are $\mc{L}_{\mathbf{Z}_1 \mathbb{1}_2}(0)$ and $\mc{L}_{\mathbf{Z}_1 \mathbb{1}_2}(t)$.
Naively, one can generalize the sum over paths to a sum over membranes 
\begin{align}
    f(\gamma_T) = 
    \sum_{\gamma_0}
    \sum_{\mu(t=0) = \gamma_0 \text{ and } \mu(t=T) = \gamma_T}
    \Box_{\rm cum}(\mu).
\end{align}
Here, $\gamma_{0,T}$ specify the noncontratible loops corresponding to $\mc{L}_{\mathbf{Z}_1 \mathbb{1}_2}(0)$ and $\mc{L}_{\mathbf{Z}_1 \mathbb{1}_2}(t)$, and the summation is over all membranes with these boundary conditions (which might or might not contain errors).
However, the number of possible choices of $\gamma_{0,T}$ and of $\mu$ are exponential in $L, T$, and it is not clear how to sum over all these paths.

We may consider restricting the summation to membranes that can be parametrized as a height function $x(y,t)$.
These membranes can be thought of as a generalization of the directed paths in Eq.~\eqref{eq:path_sum} to ``directed surfaces'', effectively neglecting all surfaces with overhangs.
This sum can be evaluated using a similar recursion in Eq.~\eqref{eq:f_i_t_dp}.
However, upon carrying out such a summation, we do not find a finite error threshold.
Evidently, the directed membranes are not good enough at avoiding errors.
It would be interesting if there exists an efficient summation that can decode with probability $1$ below a finite error threshold.



\section{Phases of generic $\mb{Z}_2$ circuits \label{sec:generic_phases}}

In this Appendix we introduce a few variations of the baseline circuit model [Fig.~\ref{fig:z2_circuit}], and focus on the physics of the phases these models can host, without worrying too much whether these phases can be decoded.

In Appendix~\ref{sec:perturb_unitary}, we consider perturbing the circuit with random local (Clifford) unitaries of $\mb{Z}_2$ symmetry, and find that the response of the system to the unitaries depends on whether or not the decoherences are present.

In Appendix~\ref{sec:bath_transition}, 
we introduce projective measurements in the bath, on top of the two-leg ladder circuit model (Sec.~\ref{sec:intro_bath}).
As the meausurements drive the bath into a nonthermalizing phase, the strength of decoherence diminishes, and the effects of the bath-system coupling is comparable to that of local $\mb{Z}_2$ unitaries within the system, as in  Appendix~\ref{sec:perturb_unitary}.


\subsection{Introduction of $\mb{Z}_2$ symmetric unitaries \label{sec:perturb_unitary}}


\begin{figure}[b]
    \centering
    \includegraphics[width=.48\textwidth]{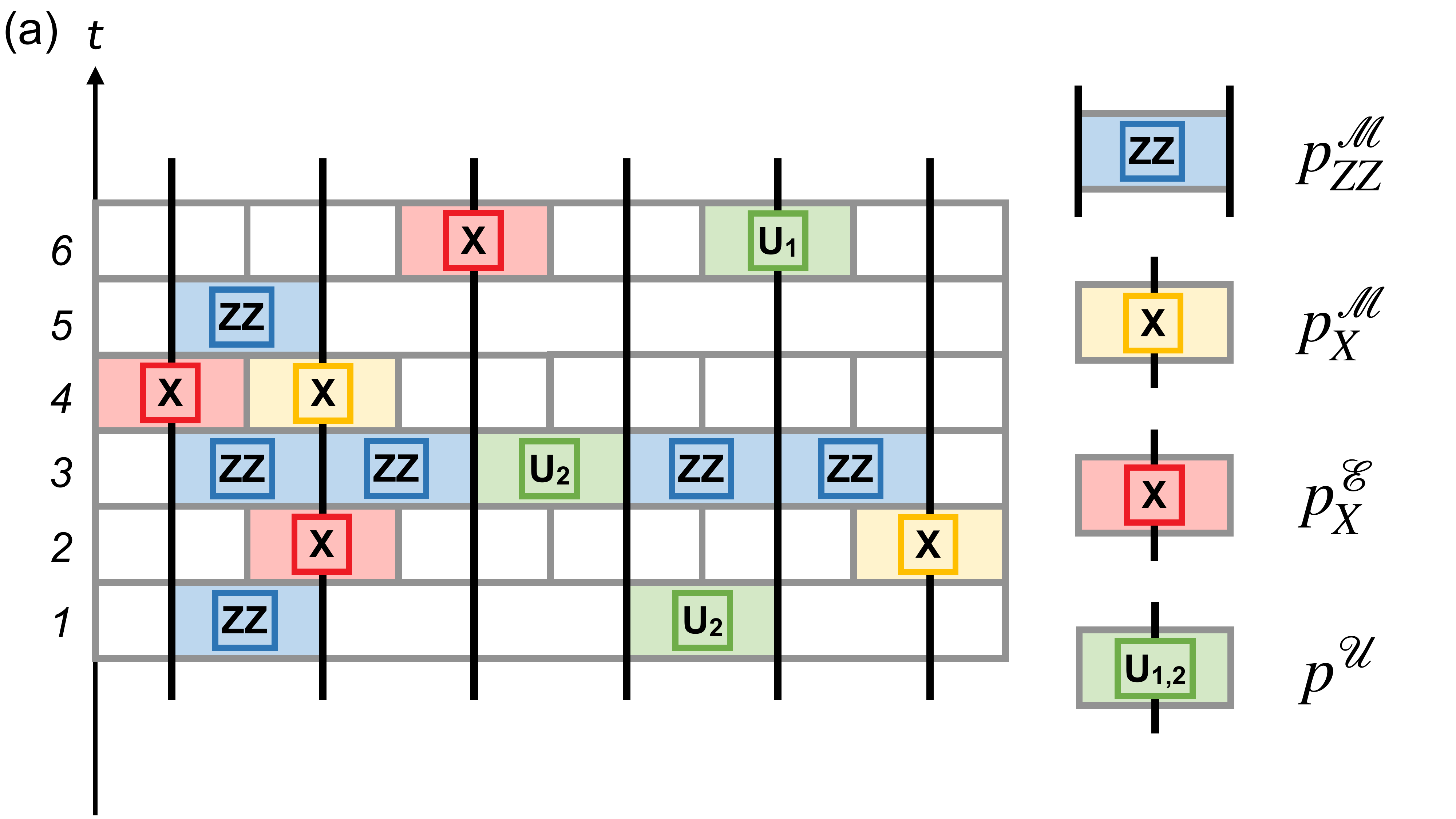}
    \includegraphics[width=.48\textwidth]{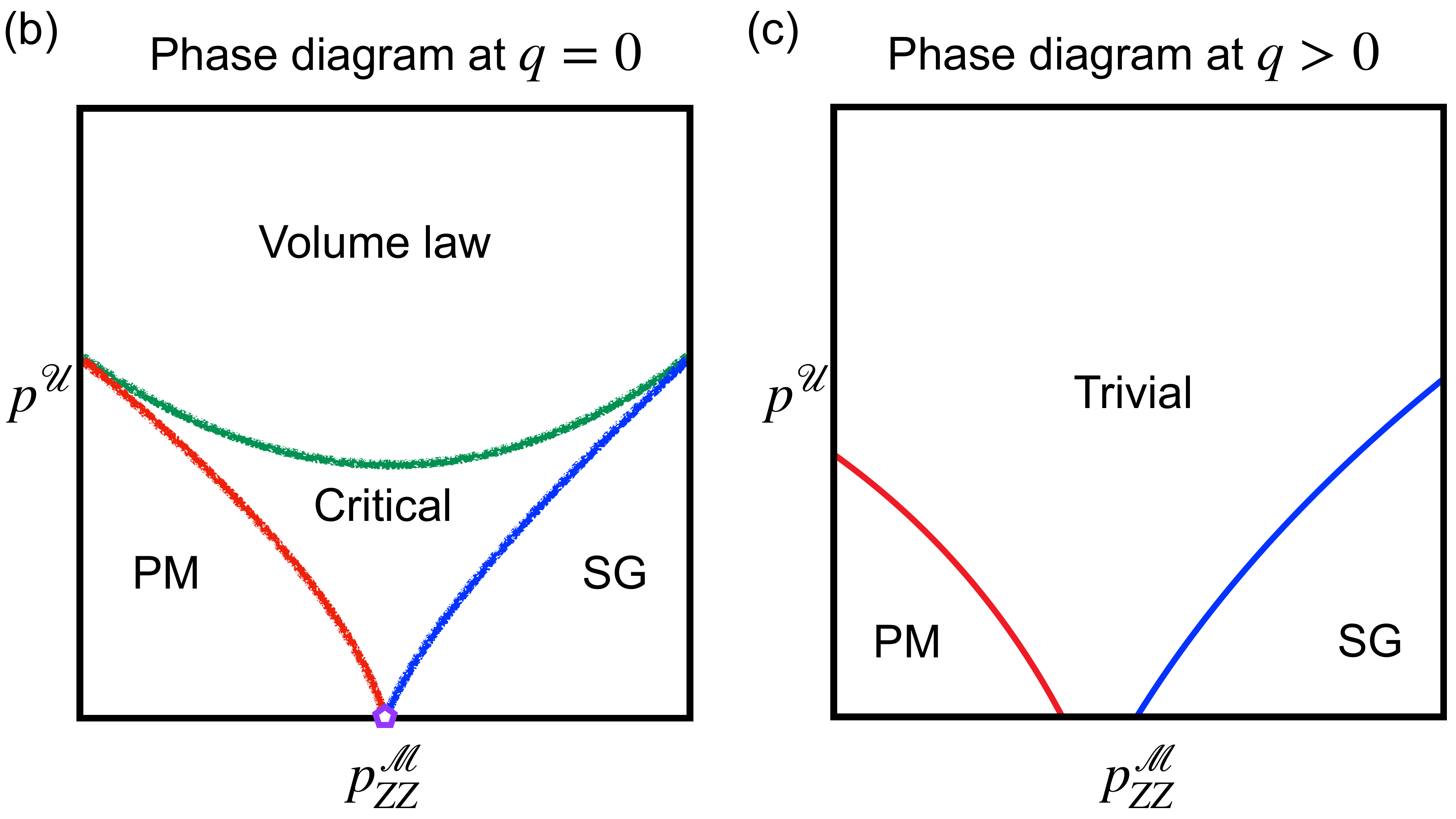}
    \caption{
    (a)
    The circuit model obtained by
    introducing $\mb{Z}_2$ symmetric unitary gates into the baseline cicuit [Fig.~\ref{fig:z2_circuit}].
    (b)
    The phase diagram in the absence of $X$ decoherence~\cite{sang2020protected}.
    We see a ``critical'' phase and a volume-law entangled phase with increasing $p^{\mc{U}}$.
    (c)
    The phase diagram in the presence of $X$ decoherence, which is much simpler than (b).
    The three phases and critical properties of phase transitions are the same as in the baseline circuit (see Fig.~\ref{fig:phase_diagram}).
    }
    \label{fig:z2_circuit_plusU}
\end{figure}

We consider the following variation [Fig.~\ref{fig:z2_circuit_plusU}(a)] of the baseline circuit.
At each spacetime location of the circuit, we apply with probability $p^\mc{U}$ a random 1-qubit or 2-qubit Clifford unitary that commute with $\mathbf{X}$ (i.e. $\mb{Z}_2$ symmetric), depending on the parity of $t$.
We only apply the other three types of gates ($\mb{Z}_2$ symmetric measurements and decoherences) when such a unitary gate is \emph{not} applied.
Using parameters in Sec.~\ref{sec:simple_z2_circuit}, the current model can be parametrized as follows,
\begin{align}
    p^{\mc{U}_1} = p^{\mc{U}_2} =&\ p^\mc{U}, \\
    p_{ZZ}^{\mc{M}} =&\ (1-p^\mc{U}) p, \\
    p_{X}^{\mc{M}}  =&\ (1-p^\mc{U}) (1-q)(1-p), \\
    p_{X}^{\mc{E}}  =&\ (1-p^\mc{U}) q (1-p).
\end{align}

The case where $q = p_X^{\mc{E}} = 0$ was previously studied in Refs.~\cite{sang2020protected, sang2020negativity}.
We reproduce the two-dimensional phase diagram (in the $p_{ZZ}^{\mc{M}}-p^{\mc{U}}$ plane) in  Fig.~\ref{fig:z2_circuit_plusU}(b).
The SG and PM phases are still present for small $p^{\mc{U}}$.
A volume law entangled phase appears when $p^{\mc{U}}$ is large, regardless of the value of $p_{ZZ}^{\mc{M}}$.
In between the SG, PM and the volume law phase is a ``critical'' phase which exhibits logarithmic scaling of entanglement entropy~\cite{sang2020protected}.
Except at the solvable point $(p_{ZZ}^{\mc{M}}, p^{\mc{U}}) = (1/2, 0)$, the phase transitions appear to be in universality classes different from 2d critical percolation, and their exact nature remains unclear.

The phase diagram is suprisingly simple when {decoherence is present}, $q > 0$ (hence $p_X^{\mc{E}} > 0$), as we depict in Fig.~\ref{fig:z2_circuit_plusU}(c).
Here, the trivial phase between PM and SG at $p^{\mc{U}} = 0$ [Fig.~\ref{fig:phase_diagram}] is \emph{enabled} by unitaries, which eventually takes over at large $p^{\mc{U}}$.
The transitions between this phase and PM and SG are again in the universality class of critical percolation, with critical exponents summarized in Appendix~\ref{app:details_mapping_percolation} and Table~\ref{table:critical_exponents}.


\begin{figure*}[t]
    \centering
    \includegraphics[width=\columnwidth]{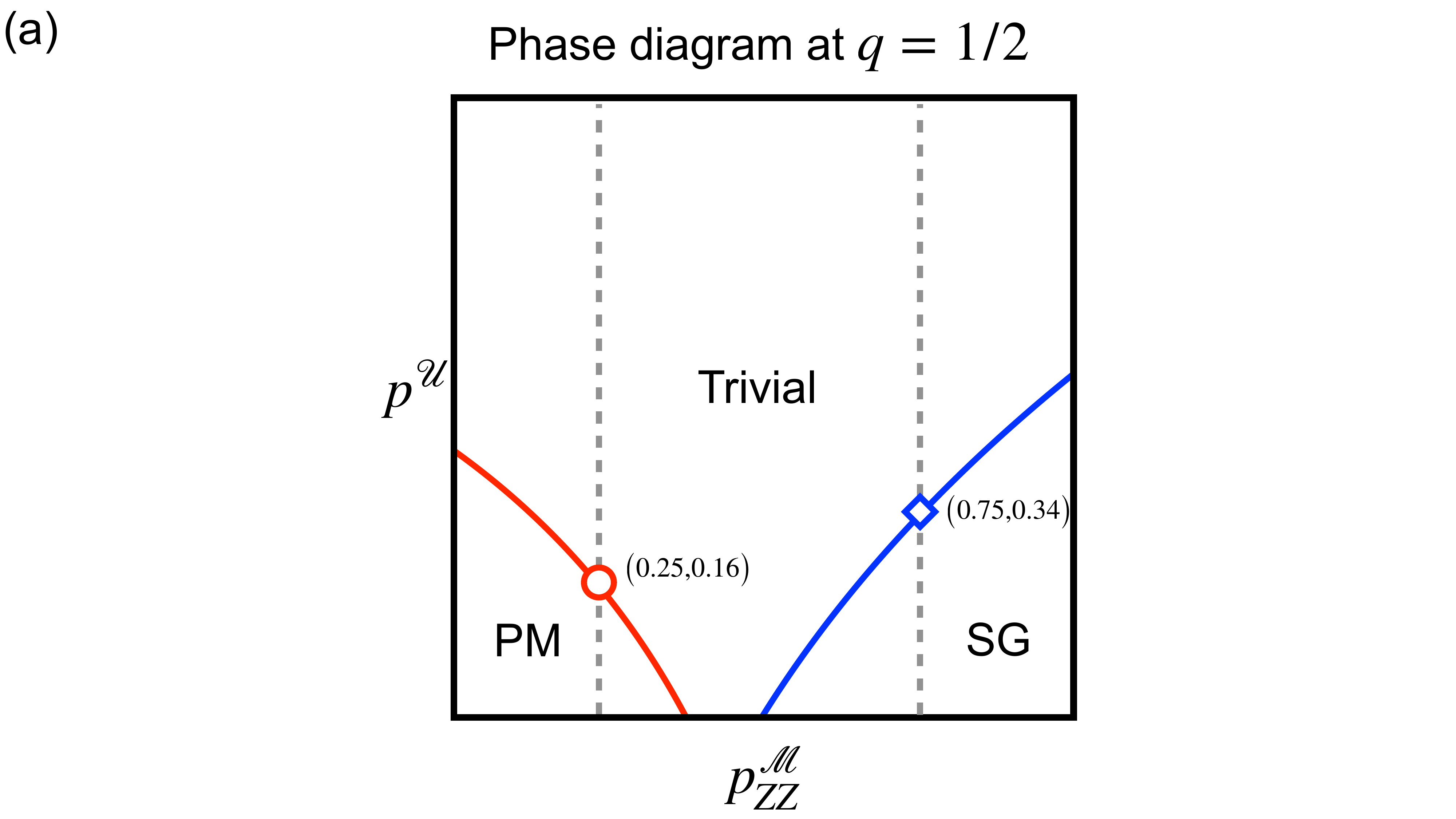}
    \includegraphics[width=\columnwidth]{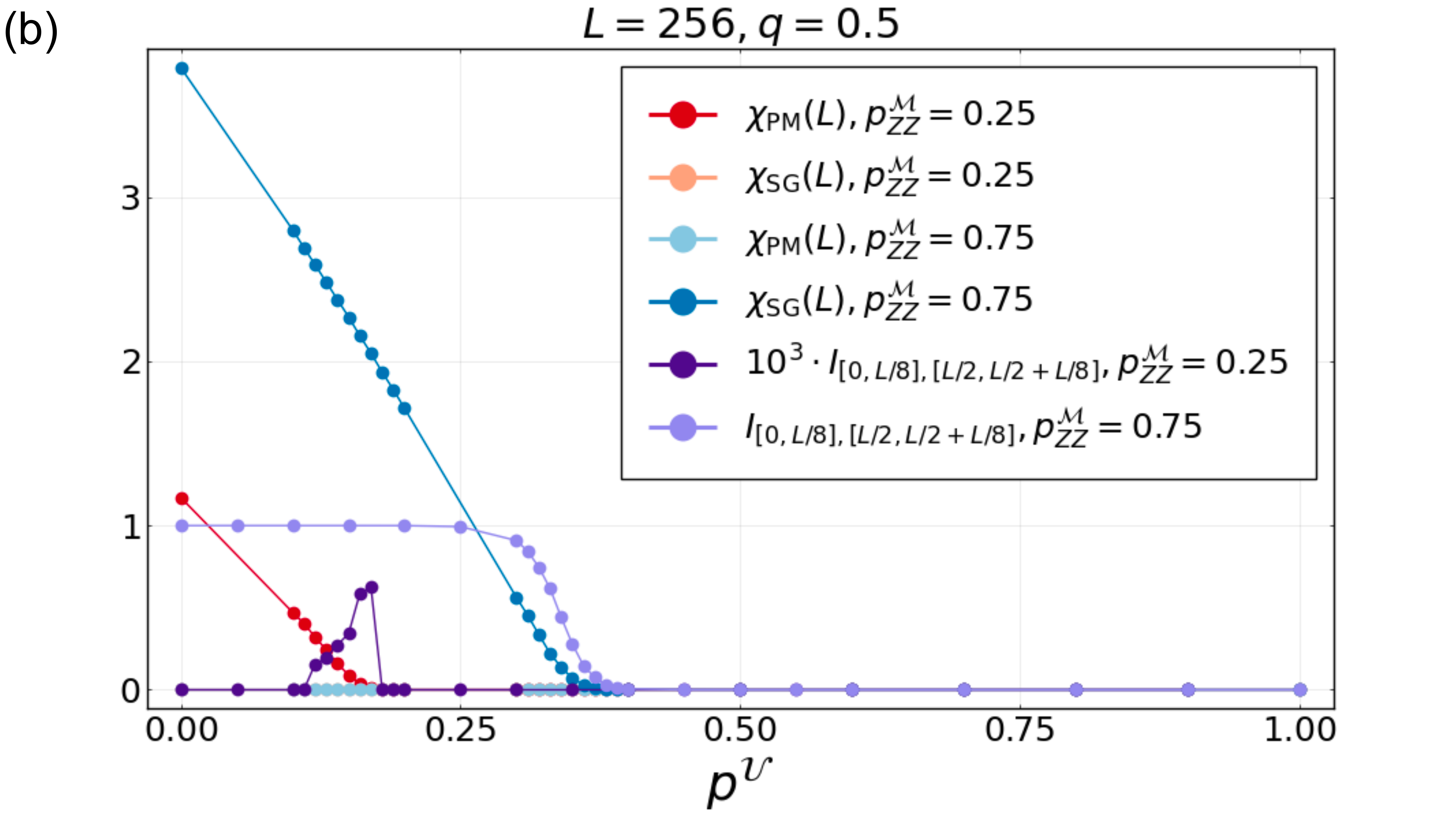}
    \includegraphics[width=\columnwidth]{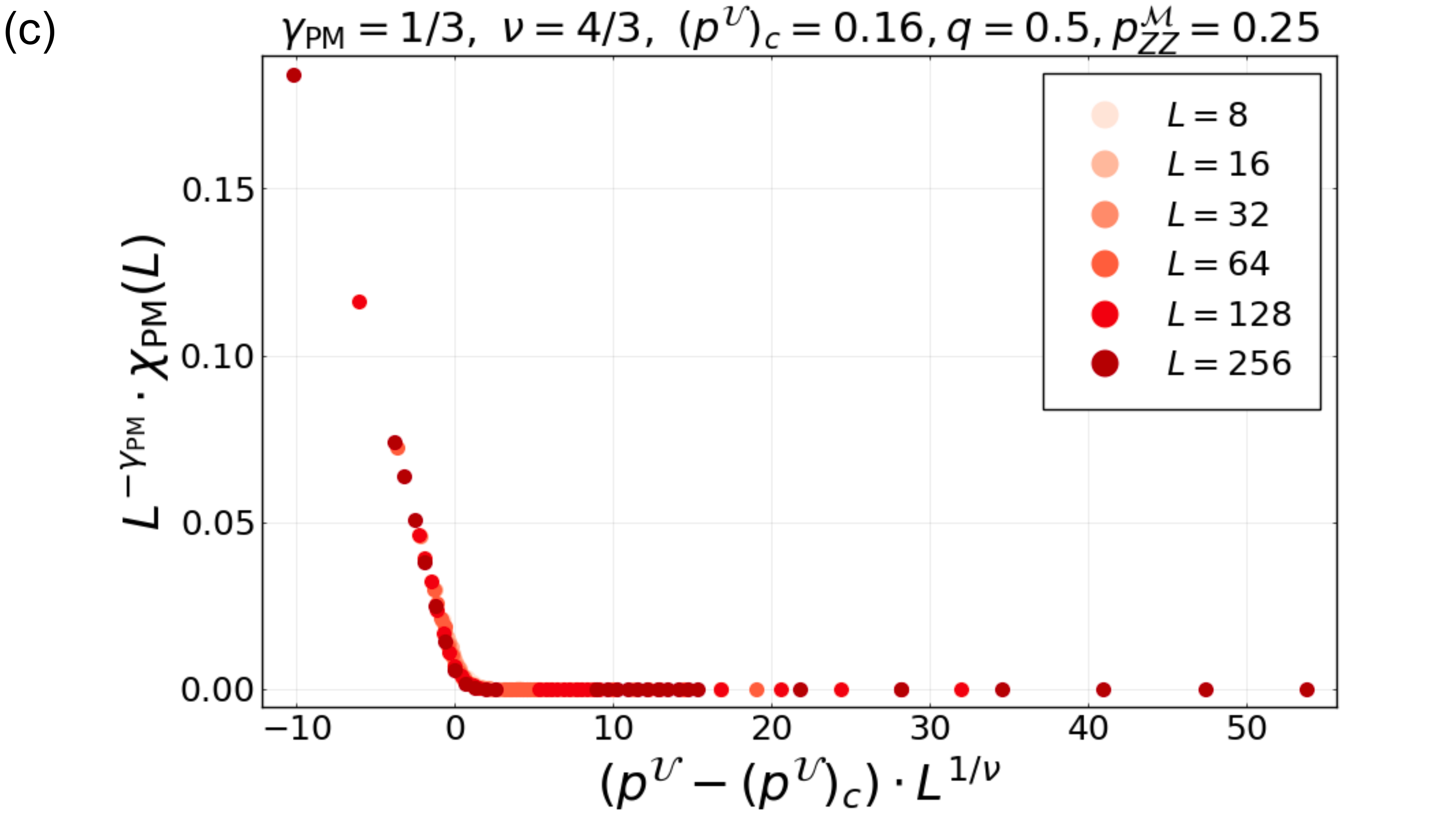}
    \includegraphics[width=\columnwidth]{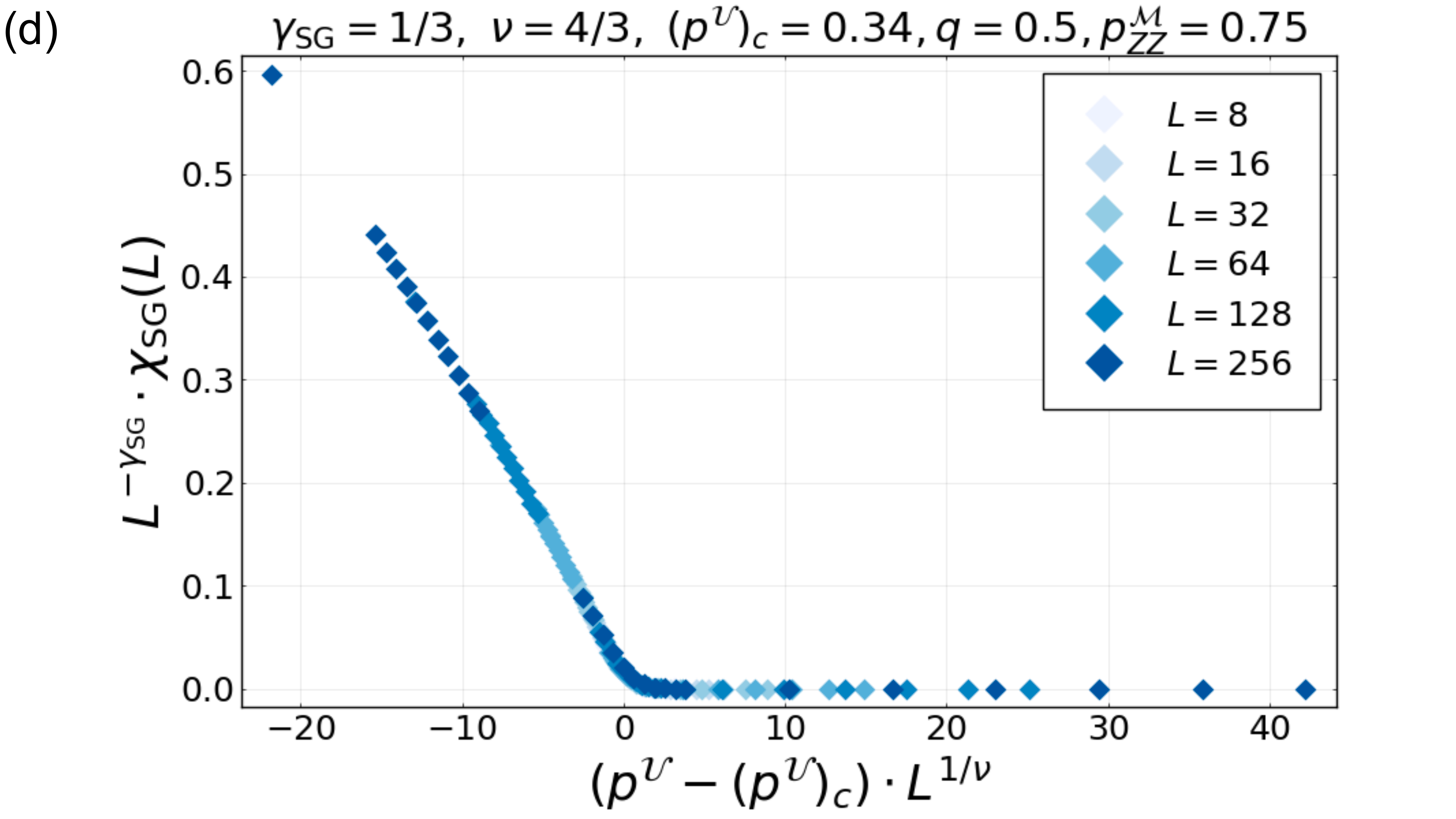}
    \includegraphics[width=\columnwidth]{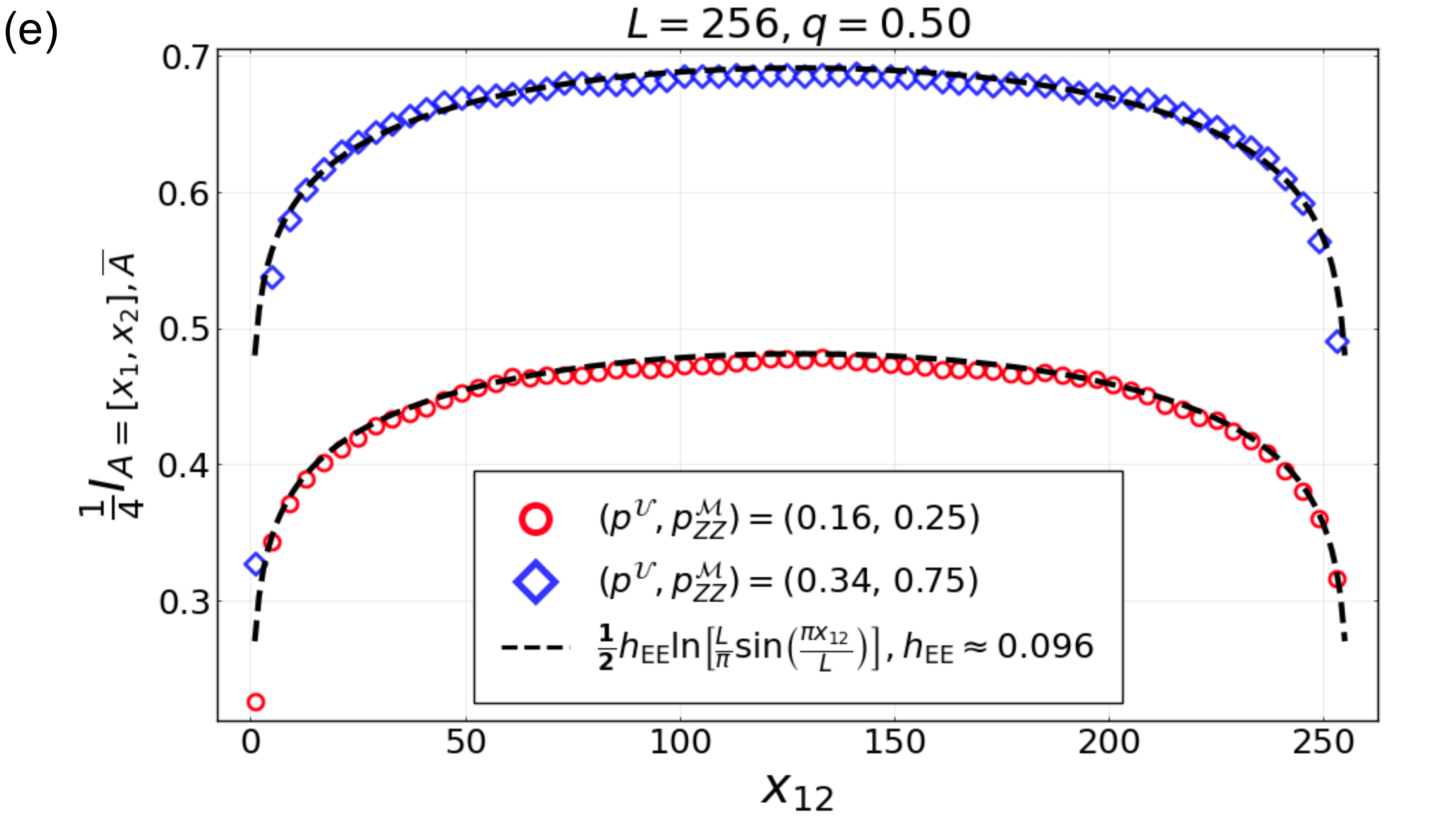}
    \includegraphics[width=\columnwidth]{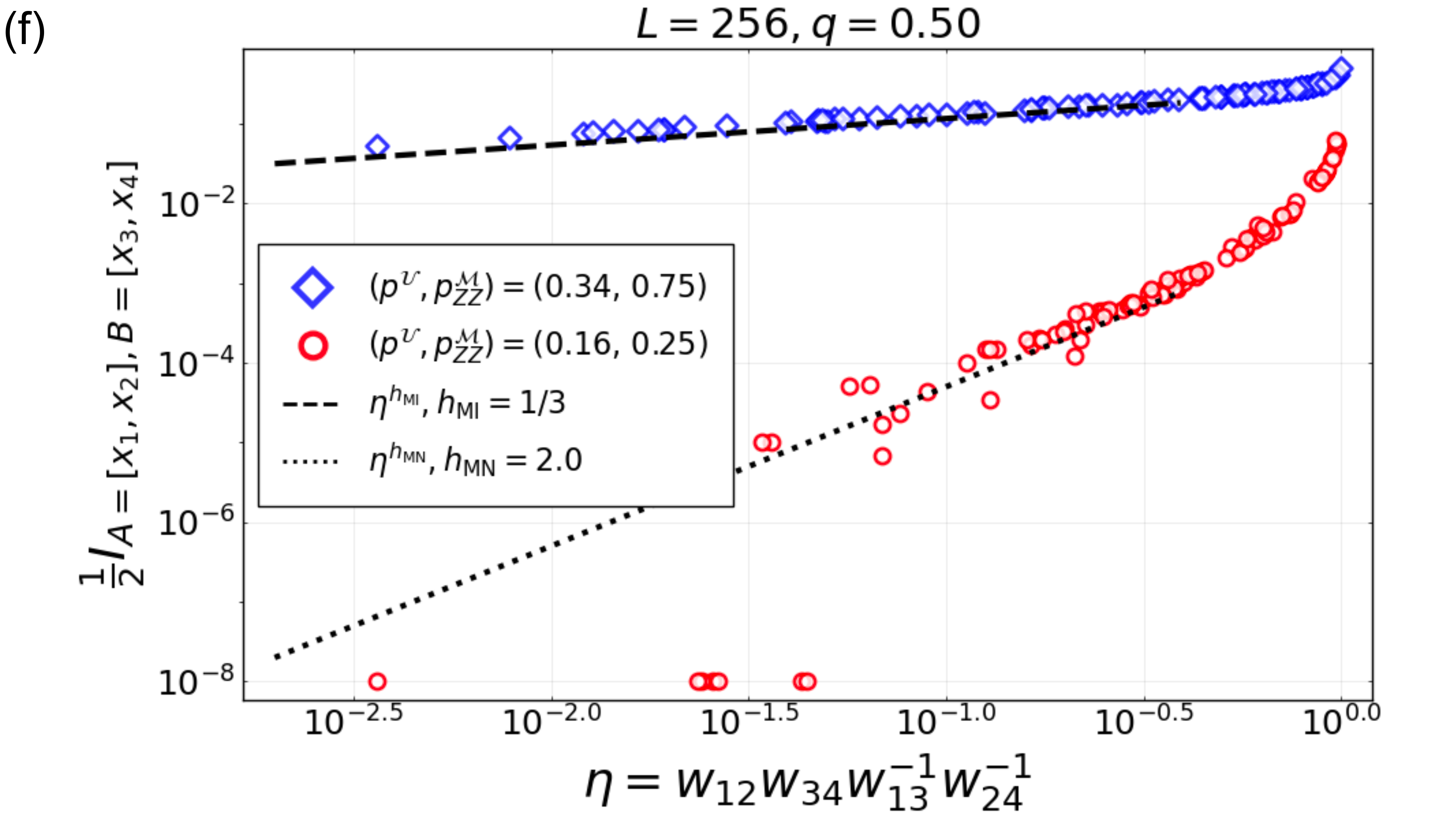}
    \caption{Numerical results for the model in Fig.~\ref{fig:z2_circuit_plusU}(a), where we introduce $\mb{Z}_2$ symmetric unitaries into the baseline circuit.
    Focusing on $q=1/2$, we find the phase diagram in (a), and confirm that the phase transitions are in the universality class of critical percolation, whose
    critical exponents are summarized Table~\ref{table:critical_exponents}.
    }
    \label{fig:plusU_numerics}
\end{figure*}

Below, we numerically confirm the phase diagram in Fig.~\ref{fig:z2_circuit_plusU}(c).
We choose $q=1/2$, and take two cuts of the phase diagram, at $p_{ZZ}^{\mc M} = 0.25$ and at $p_{ZZ}^\mc{M} = 0.75$, as shown in  Fig.~\ref{fig:plusU_numerics}(a).

In Fig.~\ref{fig:plusU_numerics}(b), we numerically compute the order parameters $\chi_{\rm PM}$ and $\chi_{\rm SG}$.
These are clearly consistent with the phase diagram in Fig.~\ref{fig:plusU_numerics}(a).
We also find the mutual information between two distant regions is $0$ in the PM and Trivial phases, and is $1$ is the SG phase, as consistent with Table~\ref{table:critical_exponents} and Eq.~\eqref{eq:I_AB_exp_summary}.
Evidence for long range correlations are found at the PM-Trivial transition (in the form of a peak in mutual information).

In Fig.~\ref{fig:plusU_numerics}(c,d), we collaspe the $\chi_{\rm PM}$ and $\chi_{\rm SG}$ near the transitions for different system sizes against the scaling forms in Eqs.~(\ref{eq:chi_PM_collapse_perc}, \ref{eq:chi_SG_collapse_perc}), and find consistency with $\gamma_{\rm PM} = \gamma_{\rm SG} = 1/3$, $\nu = 4/3$ (compare Fig.~\ref{fig:phase_diagram}(c,d)).

In Fig.~\ref{fig:plusU_numerics}(e,f), we focus on the scaling of entanglement properties at the two critical points.
The results are consistent with Fig.~\ref{fig:baseline_MI_scaling}.

Thus, {in the presense of decoherence} the unitary gates apparently {do not modify any of the qualitative or quantitative universal physics}.
At a microscopic level, their effects can be accounted for by {generalizing} the (exactly soluble) gate set in the baseline circuit to a generic gate set with $\mb{Z}_2$ symmetry, which (evidently) moves the phase boundaries around, but does not change any of the universal (critical) data.
It follows that the phase diagram in Fig.~\ref{fig:z2_circuit_plusU}(c) and Fig.~\ref{fig:phase_diagram} are \emph{generic} for $\mb{Z}_2$ circuits with decoherence.
In contrast, without decoherence, $p_X^{\mc{E}} = 0$,
inclusion of the unitary gates evidently {modify much of the qualitative physics}~\cite{sang2020protected}. 

\begin{figure*}[t]
    \centering
    \includegraphics[width=\columnwidth]{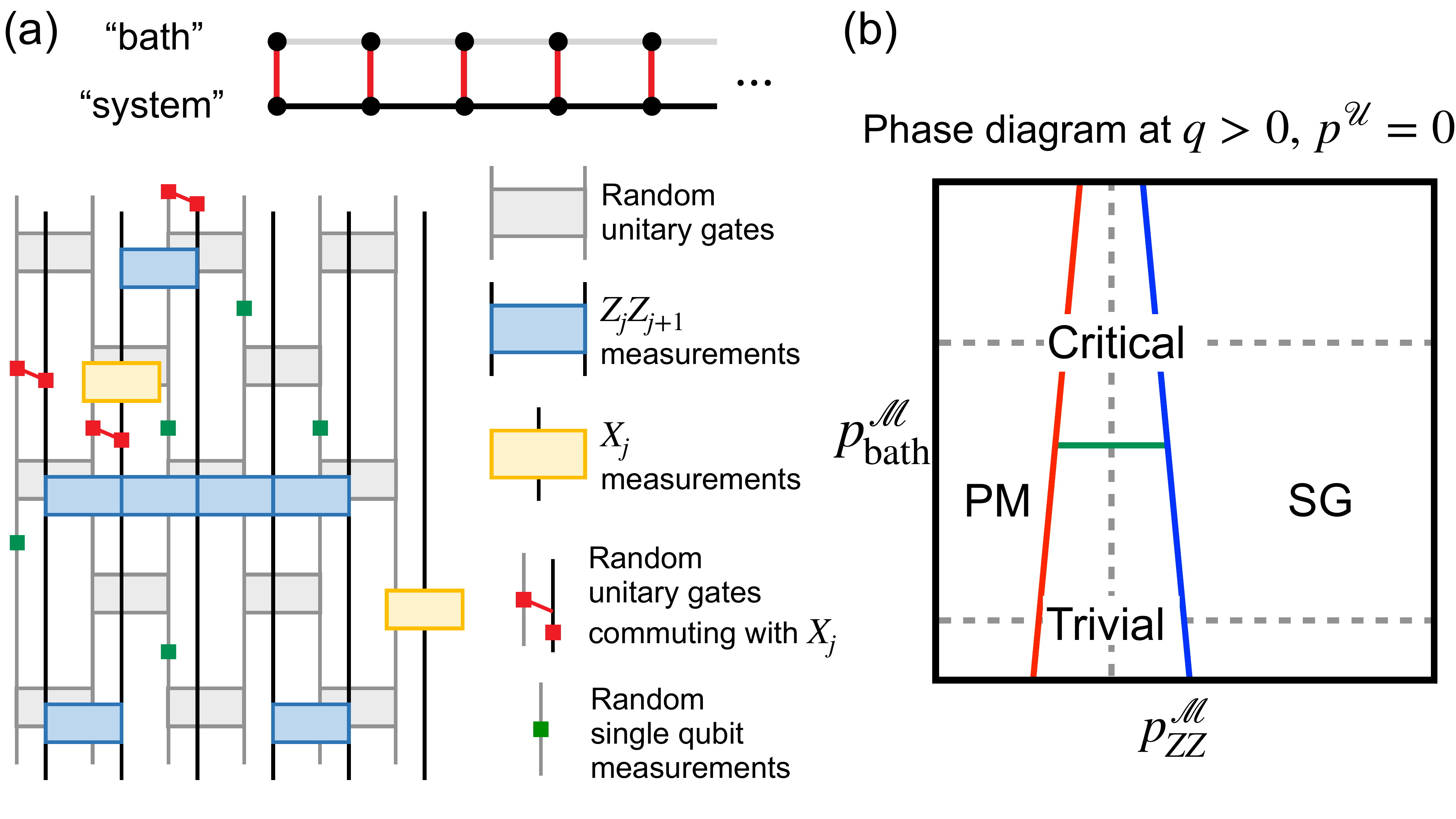}
    \includegraphics[width=\columnwidth]{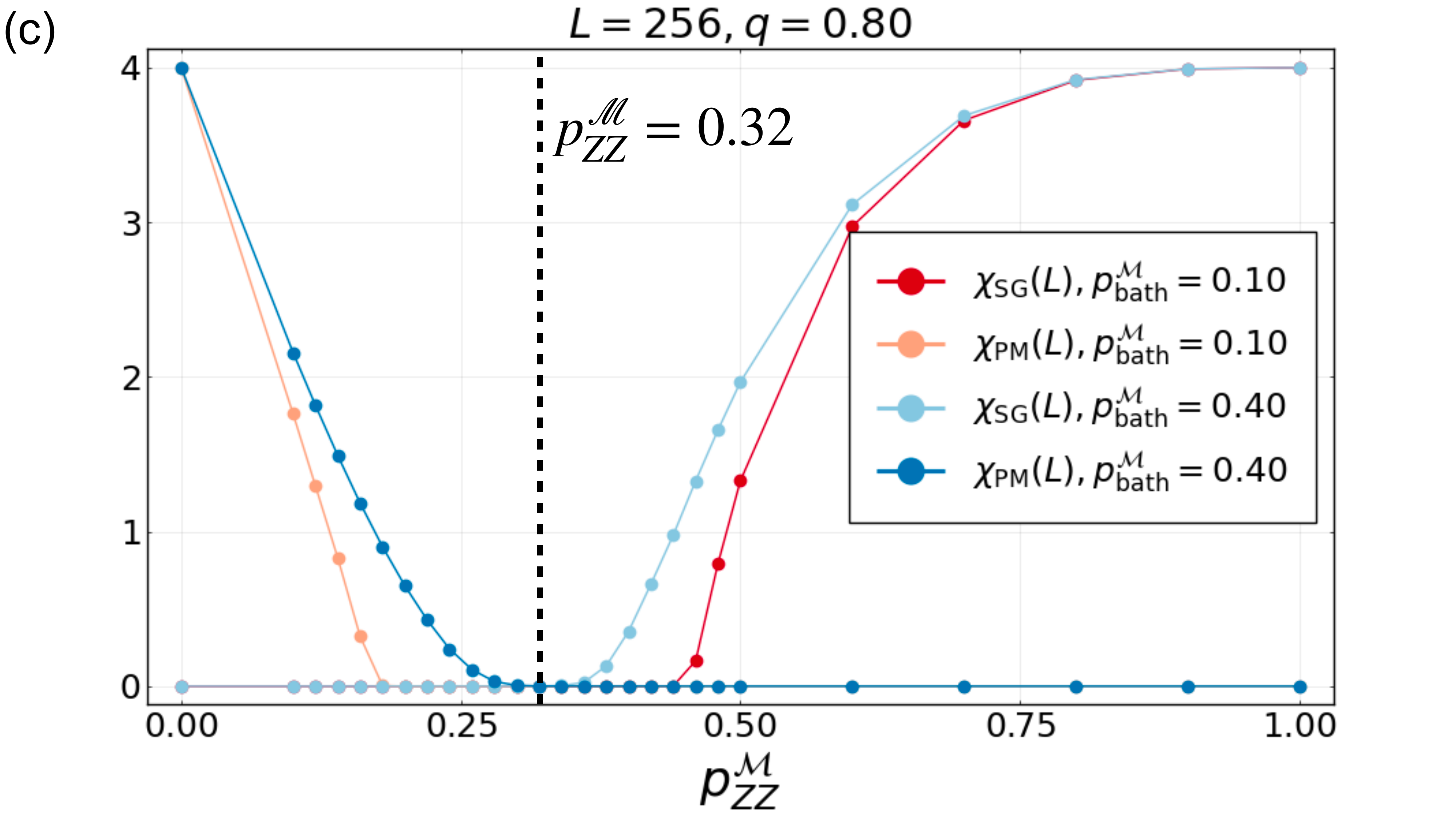}
    \includegraphics[width=\columnwidth]{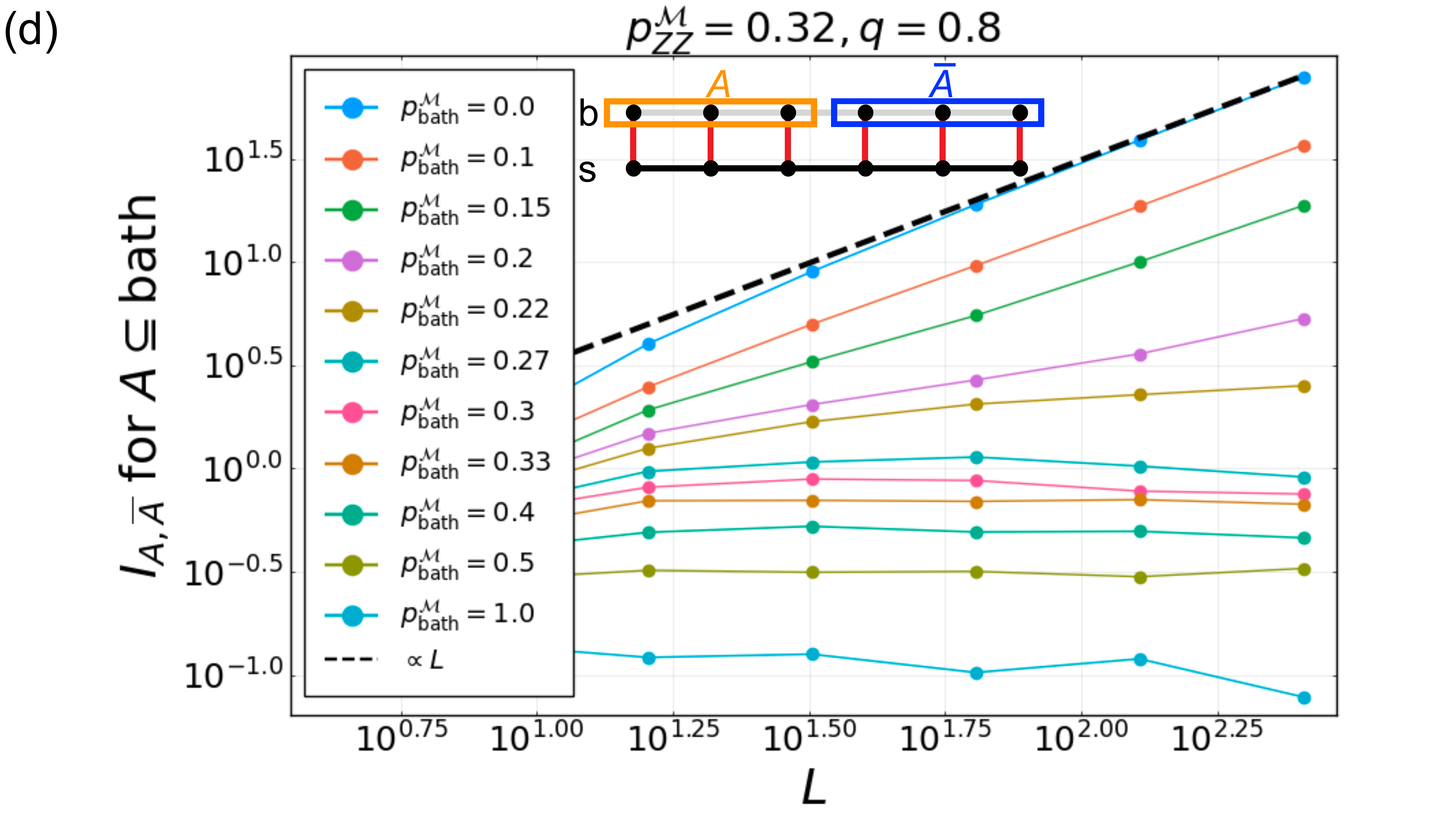}
    \includegraphics[width=\columnwidth]{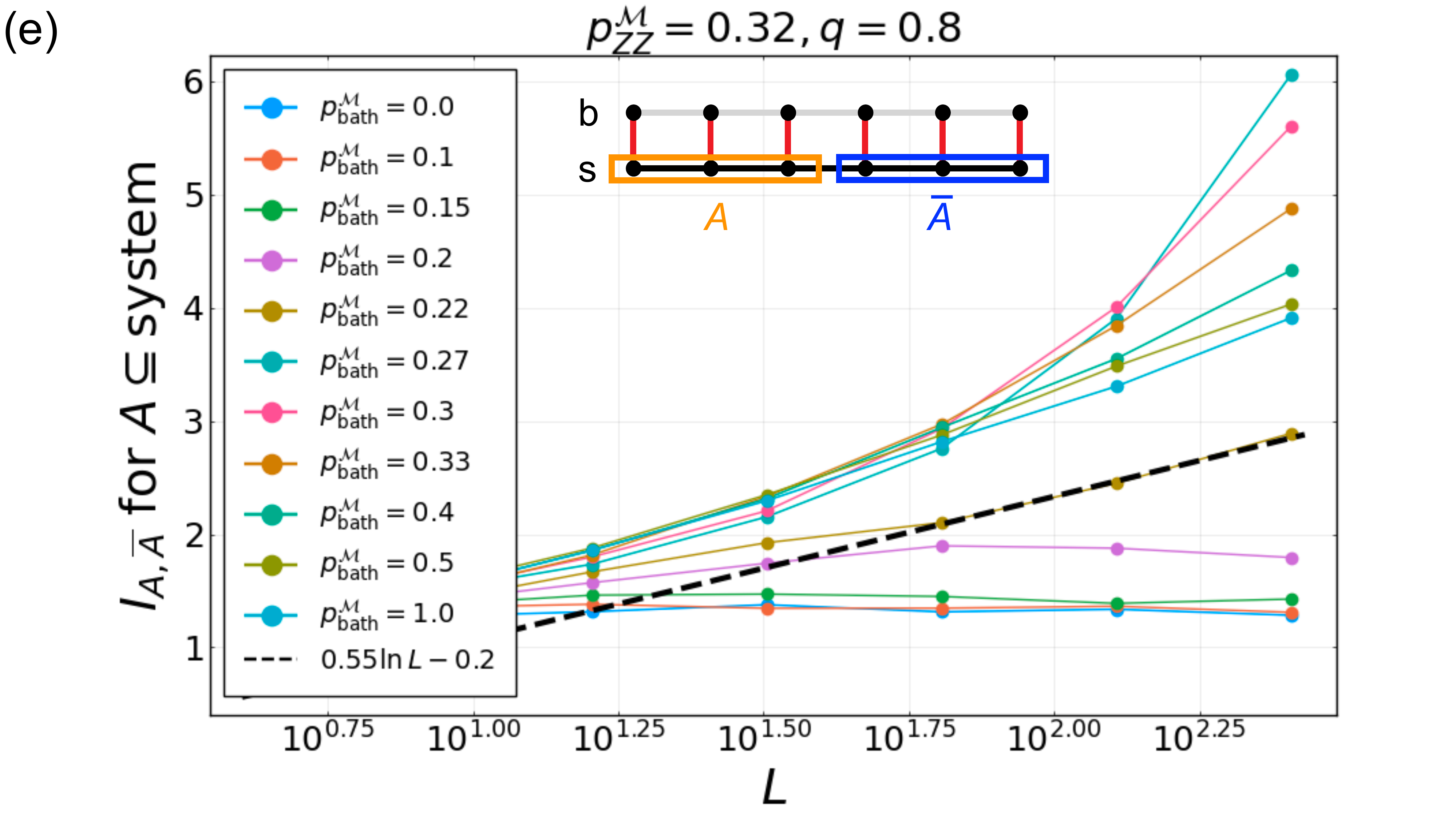}
    \caption{
    (a) The circuit model obtained by introducing local measurements of the bath into the two-leg ladder circuit [Fig.~\ref{fig:phase_diagram_twochain}], and (b) its phase diagram.
    The model has four phases, as we detail in the main text.
    (c) Numerical results on the two horizontal cross sections of the phase diagram in (b), at $p_{\rm bath}^\mc{M} = 0.1$ and $0.4$.
    In both cases we see the SG and PM phases, separated by an intermediate phase where both order parameters vanish.
    (d,e) Numerical results on the vertical cross section of the phase diagram in (b), for an intermediate value of $p_{ZZ}^\mc{M}$ with neither SG nor PM order.
    The half-cut mutual information (d) of the ``bath'' indicates a volume law to area law transition of the bath driven by measurements.
    The half-cut mutual information (e) of the ``system'' indicates a phase transition from a trivial, area law entangled phase, to a ``critically entangled'' phase akin to the one in Fig.~\ref{fig:z2_circuit_plusU}(b).
    }
    \label{fig:bath_transition}
\end{figure*}

\subsection{Entanglement transition of the bath, and its ramifications on the $\mb{Z}_2$ symmetric chain \label{sec:bath_transition}}

We have seen in Sec~\ref{sec:intro_bath} that a random unitary circuit can act as a bath, when satisfying both ``Triviality'' and ``Markovianity'' conditions.
In this section, we explore circumstances where the bath becomes nonthermal.

Starting from the two-leg ladder circuit model~\cite{bao2021enriched}, we introduce single-qubit measurements\footnote{Since the bath dynamics do not have a physical symmetry, and the gates are random, we may take the projective measurements to be single qubit measurements of, say, $Z_j$.}
at a finite rate $p_{\rm bath}^{\mc{M}}$ in the ``bath'' chain, as depicted in Fig.~\ref{fig:bath_transition}(a).
The measurements will drive the bath through a phase transition in entanglement scaling from volume law (at small $p_{\rm bath}^{\mc{M}}$) to area law (at large $p_{\rm bath}^{\mc{M}}$)~\cite{nahum2018hybrid, li2018hybrid}.
The transition occurs at a finite measurement strength, $\big(p_{\rm bath}^{\mc{M}}\big)_c > 0$.
It is plausible that in the volume law phase, the bath is still at ``finite temperature'' with a finite density of entanglement entropy, and the two conditions above are still satisfied;
whereas in the area law phase, the bath looks as if it is at ``zero temperature'', and fails to behave like a good thermal bath.
Instead, in the area law phase most qubits in the bath are measured at each time step, {so that after the bath is traced out}, the system-bath coupling will effectively {induce} (additional) local unitaries and local measurements into the system.
Thus, when the bath passes through the transition, the {system's} intermediate phase induced by the system-bath coupling might also exhibit a transition, from the decoherence-dominated ``trivial'' phase (compare Fig.~\ref{fig:z2_circuit_plusU}(c), where $q > 0$) to the unitary-dominated ``critical'' phase (compare Fig.~\ref{fig:z2_circuit_plusU}(b), where $q = 0$).

The schematic phase diagram that follows from this reasoning is shown in Fig.~\ref{fig:bath_transition}(b).
Here, we vary $p^\mathcal{M}_{ZZ}$ and $p_{\rm bath}^\mathcal{M}$, and parametrize other rates as follows,
\begin{align}
    p_X^\mathcal{M} =&\  (1-q)(1-p^\mathcal{M}_{ZZ}), \\
    p_X^\mathcal{I} =&\ q(1-p^\mathcal{M}_{ZZ}), \\
    p^\mathcal{U} =&\ 0.
\end{align}
The last equation says that we do not introduce any unitary gates into the system explicitly, as in Sec.~\ref{sec:intro_bath}.
Within the phase diagram, we study three cross sections, as highlighted with dashed lines.

We confirm the phase diagram with numerical results in Fig.~\ref{fig:bath_transition}(c,d,e), where we take $q = 0.80$.
First, we choose two values of $p_{\rm bath}^\mc{M} (= 0.10, 0.40)$,
and calculate the order parameters  $\chi_{\rm SG}$ and $\chi_{\rm PM}$ for varying $p_{ZZ}^\mc{M}$.
The results for a finite system size $L = 256$ are shown in Fig.~\ref{fig:bath_transition}(c).
In both cases, there is an intermediate phase where both order parameters vanish.
We note that the width of the intermediate phase decreases with increasing $p_{\rm bath}^\mc{M}$, but remains finite for $p_{\rm bath}^\mc{M} = 1.0$ (see Fig.~\ref{fig:bath_transition_numerics_more} below).

Next, we take $p_{ZZ}^{\mc{M}} = 0.32$, which sits in the intermediate phase for all values of $p_{\rm bath}^{\mc M} \in [0, 1]$ (see Fig.~\ref{fig:bath_transition_numerics_more}), and compare the scaling of the entanglement for varying values of $p_{\rm bath}^{\mc M}$.
To filter out the entanglement between the bath and the system, and focus instead on the internal entanglement of the bath, we consider the ``half-cut mutual information'' $I_{A, \ovl{A}}^{\rm bath}$ for an equal-size bipartition $(A, \ovl{A})$ of the bath, each of $L/2$ qubits.
The results for different values of $p_{\rm bath}^{\mc M}$ are shown in Fig.~\ref{fig:bath_transition}(d).
On a log-log scale, we see a clear transition between a phase where $I_{A, \ovl{A}}^{\rm bath} \propto L$, and another phase where $I_{A, \ovl{A}}^{\rm bath} \propto L^0$.
Thus, the bath itself goes through a volume law to area law transition in its internal entanglement under projective measurements.
The critical point is at $\big(p_{\rm bath}^\mc{M}\big)_c \approx 0.22$.

Similarly, we compute the half-cut mutual information for an equal-size bipartition of the system, denoted $I_{A, \ovl{A}}^{\rm system}$, and plot the results in Fig.~\ref{fig:bath_transition}(e).
When $p_{\rm bath}^\mc{M} < \big(p_{\rm bath}^\mc{M}\big)_c$, $I_{A, \ovl{A}}^{\rm system}$ obeys an area law, as consistent with the trivial phase.
When $p_{\rm bath}^\mc{M} > \big(p_{\rm bath}^\mc{M}\big)_c$, $I_{A, \ovl{A}}^{\rm system}$ grows with the system size, and appears proportional to $\ln L$ as $p_{\rm bath}^\mc{M} > \big(p_{\rm bath}^\mc{M}\big)_c$ approaches $1$.
This scaling behavior is consistent with the critical phase with logarithmic entropy~\cite{sang2020protected} (see also Fig.~\ref{fig:z2_circuit_plusU}(b)).
We note an interesting regime when $p_{\rm bath}^\mc{M}$ is slightly above the critical point, where the entropy grows faster than $\ln L$ for the system sizes accessed.

\begin{figure}
    \centering
    \includegraphics[width=.5\textwidth]{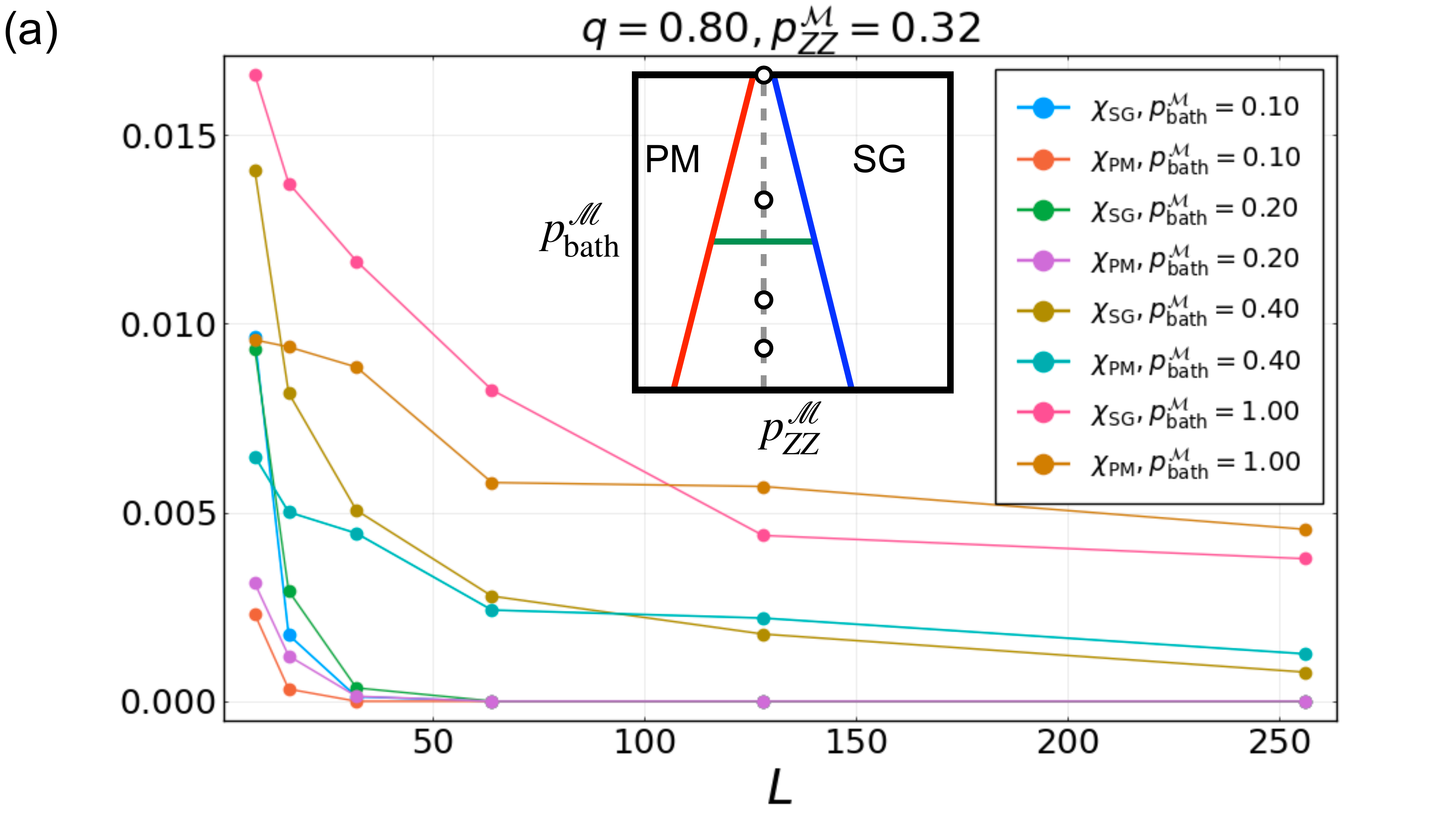}
    \includegraphics[width=.5\textwidth]{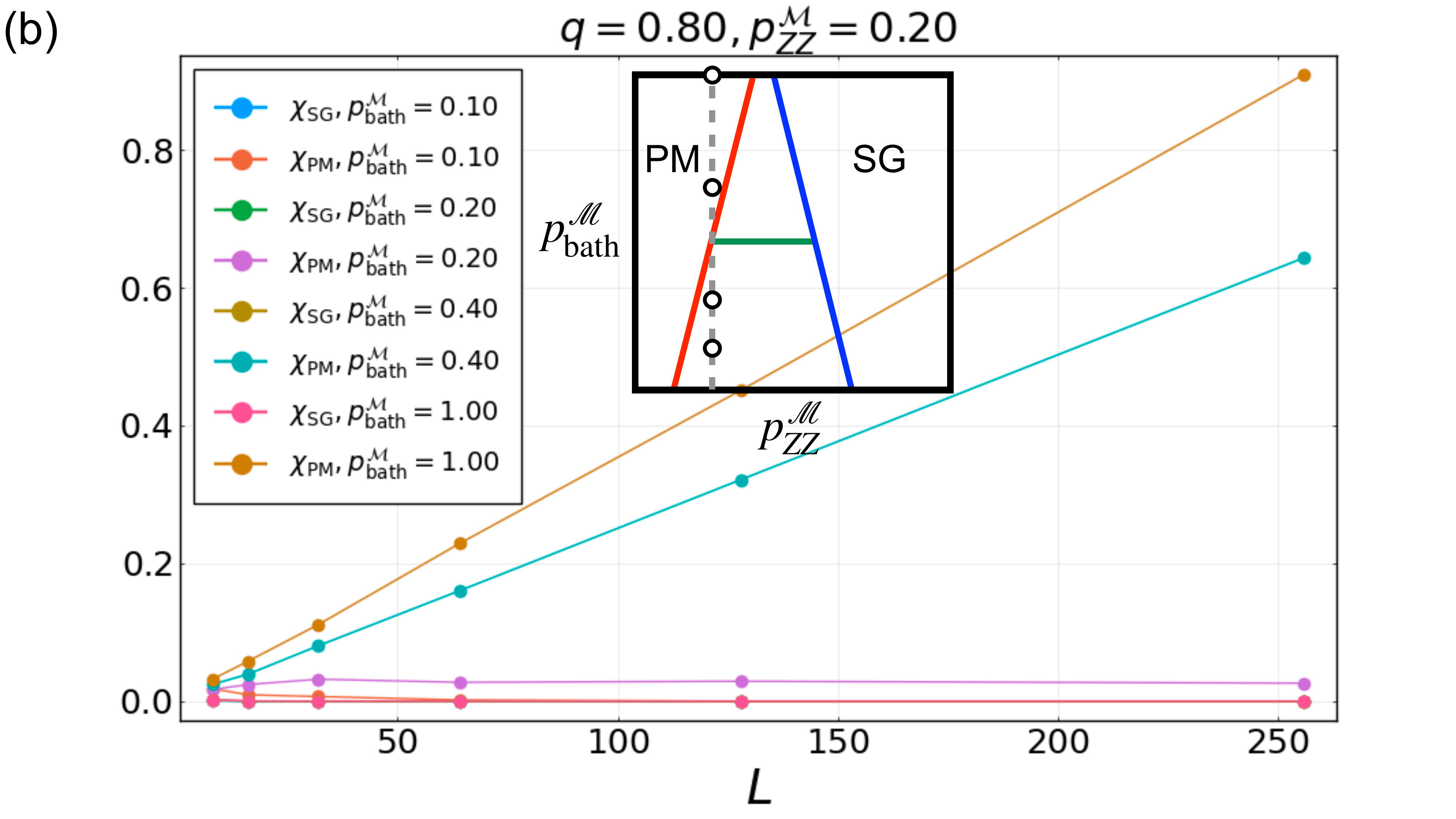}
    \includegraphics[width=.5\textwidth]{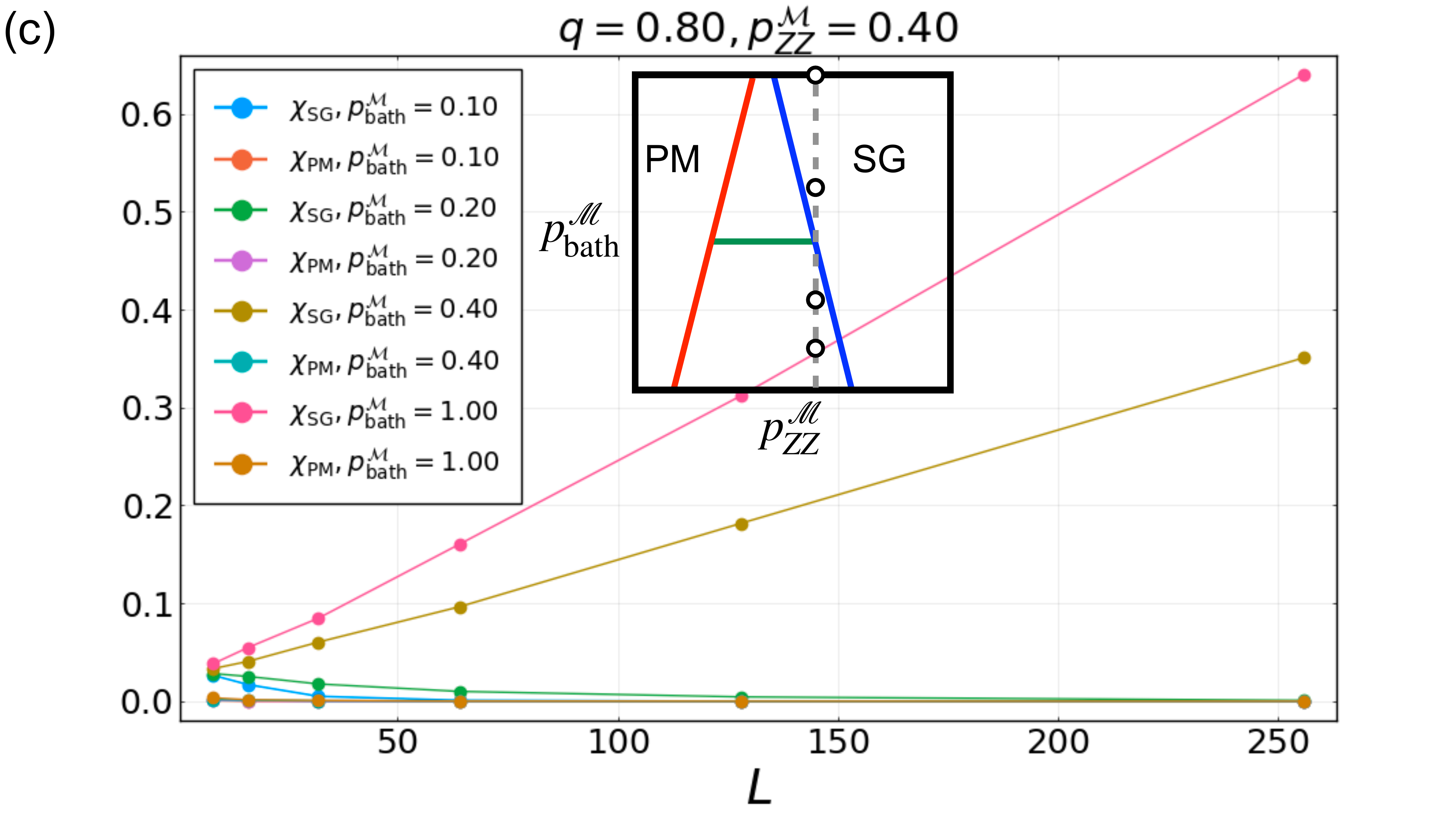}
    \caption{Further numerical results on the circuit model in Fig.~\ref{fig:bath_transition}(a), where the bath undergoes single-qubit projective measurements at rate $p_{\rm bath}^\mc{M}$.
    These results further corroborate the phase diagram [Fig.~\ref{fig:bath_transition}(b)].
    }
    \label{fig:bath_transition_numerics_more}
\end{figure}

In Fig.~\ref{fig:bath_transition_numerics_more}, we present further numerical results of the phase diagram in Fig.~\ref{fig:bath_transition} at $q = 0.8$. 
We take three values of $p_{ZZ}^\mc{M} ( =0.32, 0.20, 0.40)$, and take four values of $p_{\rm bath}^\mc{M} (=0.1, 0.2, 0.4, 1.0)$ along these lines.

Along the line $p_{ZZ}^\mc{M} = 0.32$ [Fig.~\ref{fig:bath_transition_numerics_more}(a)], the system goes from the Trivial phase to the Critical phase with increasing $p_{\rm bath}^\mc{M}$ (see Fig.~\ref{fig:bath_transition}(c,d,e)).
Both phases should have vanishing order parameter $\chi_{\rm SG}$ and $\chi_{\rm PM}$.
This is confirmed by the numerics, where we found that for $p_{\rm bath}^\mc{M} < (p_{\rm bath}^\mc{M})_c \approx 0.22$, both $\chi_{\rm SG}$ and $\chi_{\rm PM}$ decays quickly with increasing system sizes $L$, and for $p_{\rm bath}^\mc{M} > (p_{\rm bath}^\mc{M})_c$, the two order parameters are also decaying with increasing $L$, albeit slowly.
This is perhaps due to the fact that the Critical phase becomes narrower with larger $p_{\rm bath}^\mc{M}$.
In any case, it is clear that these phases are quite different from the SG and PM phases.

In contrary, when we take $p_{ZZ}^\mc{M} = 0.2, 0.4$ (Fig.~\ref{fig:bath_transition_numerics_more}(b,c)), we see a clear transition from the trivial phase to the PM and SG phases, respectively, where the corresponding order parameter scales linearly in $L$.

Thus, we have confirmed the picture outlined above and in Fig.~\ref{fig:bath_transition}(b).
That is, the measurement-driven entanglement transition in the bath is accompanied by a phase transition for intermediate values of $p_{ZZ}^{\mc{M}}$ in the system, as reflected in the entanglement scaling.

We also find that at the Critical-PM and the Critical-SG transitions, the critical exponents are different from those in Table~\ref{table:critical_exponents}, as expected.
These results are not displayed.

\subsection{Comments \label{sec:bath_transition_discussion}}

In this Appendix, we have focused on the generic phase diagram in (1+1)-dimensions.
It would be nice to understand the critical properties, and the differences between the (super-)logarithmically entangled phase and other ``critical'' phases found in the context of monitored dynamics~\cite{chenxiao2020nonunitary, diehl2020trajectory, jian2020fermionRTN, diehl2021sineGordon, turkeshi2021zeroclicks, tang2021freefermion2+1d}.
It would also be interesting to put all these phases under the classification framework in Ref.~\cite{bao2021enriched}.

In higher dimensions, the phase diagram can be much richer, as we discussed briefly in Sec.~\ref{sec:unlocated_error_2plus1d} for the case without $\mb{Z}_2$ unitaries.
With $\mb{Z}_2$ unitaries, a volume law phase can coexist with the spin glass order, even without a bath~\cite{sang2020protected}.
The ensuing phase diagram will be worth exploring in future works.


\end{document}